\documentclass[a4paper,11pt]{article}

\usepackage{jheppub} 
\usepackage{amsmath}
\usepackage[T1]{fontenc} 
\usepackage{tikz}
\usepackage{caption}
\usepackage{subcaption}
\usepackage{nccmath}
\usepackage{bm}
\usepackage{amssymb}

\usepackage{mathtools}

\DeclarePairedDelimiter\abs{\lvert}{\rvert}%
\DeclarePairedDelimiter\norm{\lVert}{\rVert}%

\makeatletter
\let\oldabs\abs
\def\abs{\@ifstar{\oldabs}{\oldabs*}}
\let\oldnorm\norm
\def\norm{\@ifstar{\oldnorm}{\oldnorm*}}
\makeatother

\renewcommand{\d}{\mathrm{d}}
\newcommand{\p}{\partial}
\newcommand{\e}{\mathrm{e}}

\newcommand{\tr}{{\rm tr }}
\newcommand{\Tr}{{\rm Tr }}

\newcommand{\matr}[1]{\bm{#1}}     

\begin{document} 

\addtocontents{toc}{\protect\setcounter{tocdepth}{2}}

\begingroup\parindent0pt
\begin{flushright}\footnotesize
\end{flushright}
\centering
\begingroup\LARGE
\bf
Analytic and Numerical  Bootstrap for One-Matrix Model and ``Unsolvable'' Two-Matrix Model
\par\endgroup
\vspace{3.5em}
\begingroup\large
{\bf Vladimir Kazakov}$\,$ \it{and} $\,$
{\bf Zechuan Zheng}
\par\endgroup
\vspace{2em}
\begingroup\sffamily\footnotesize
Laboratoire de Physique de l'\'Ecole Normale Sup\'erieure,\\ CNRS,
Universit\'e PSL, Sorbonne Universit\'es,\\
24 rue Lhomond, 75005 Paris, France\\
\vspace{1em}
\par\endgroup
\vspace{2em}

\endgroup

\begin{abstract}

We propose the {\it relaxation} bootstrap method for the numerical solution of  multi-matrix models in the large \(N\) limit, developing and improving the recent proposal of H.Lin. It gives rigorous inequalities on the  single trace moments of the matrices up to a given ``cutoff'' order (length) of the moments. The method combines  usual loop equations on the moments and the positivity constraint on  the correlation matrix of the moments. We have a rigorous proof of applicability of this method in the case of the one-matrix model where the condition of positivity  of the saddle point solution appears  to be equivalent  to  the presence of supports of the eigenvalue distribution only on the real axis and only with positive weight.   We demonstrate the numerical efficiency of our method by solving the analytically ``unsolvable'' two-matrix model with \(\tr[A,B]^2\) interaction and quartic potentials, even for solutions with spontaneously broken discrete symmetry. The region of values for computed moments allowed by inequalities quickly shrinks with the increase of the cutoff, allowing the precision of about 6 digits for generic values of couplings in the case of  \(\mathbb{Z}_2\) symmetric solutions.  Our numerical data are checked against the known analytic results for particular values of parameters. 
\end{abstract}

\thispagestyle{empty}

 \newpage
\tableofcontents
\emailAdd{kazakov@lpt.ens.fr}
\emailAdd{zechuan.zheng@phys.ens.fr}
\flushbottom

\section{Introduction}
\label{sec:intro}

Matrix integrals play an important role in numerous physical and mathematical subjects, such as multi-component quantum field theory~\cite{tHooft:1973alw} (see \cite{Migdal:1983qrz} for the review),  two-dimensional quantum gravity and string theory~\cite{David:1984tx,Kazakov:1985ea,Kazakov:1985ds,Kazakov:1987qg},    mesoscopic physics~\cite{PhysRevLett.52.1}, algebraic geometry~\cite{Dijkgraaf:2002fc,Dijkgraaf:2002pp,Eynard:2007kz,Kontsevich:1992ti}, number theory~\cite{montgomery1973pair}, etc.  A rather general class of matrix integrals has the form
\begin{equation}\label{generalMM}
    Z=\int d^{N^2}A\,d^{N^2}B\,d^{N^2}C\dots\,\e^{-\tr {\cal V}(A,B,C,\dots)}
\end{equation}
where \(A,B,C,\dots\) are Hermitian $N\times N$ matrices with $U(N)$ invariant integration measure and the potential ${\cal V}(x,y,z,\dots)$ is an analytic function (often a polynomial) of the variables $x,y,z,\dots$.  The partition function $Z$ is a function of  parameters (couplings) of the  potential. The typical ``physical'' quantities to study are various correlators  of traces of  ``words'' built out of products of  matrices \(A,B,C,\dots\),  computed w.r.t. the measure represented by the expression under the integral:
\begin{equation}
    \big\langle \frac{\tr}{N}(A^kB^lC^m\dots) \,\frac{\tr}{N}(A^nB^pC^q\dots) \dots\big\rangle.
\end{equation}

The   $N\to\infty$ limit, with the appropriately adjusted parameters of the potential and of the averaged quantity, is of a special importance in multiple applications since it describes the  thermodynamical limit of macroscopically many degrees of freedom for various physical systems. Such a limit deals with the infinite number of integrals, thus the matrix integral becomes a functional integral.

A particularly interesting $N\to\infty$ limit, for the potential scaled as ${\cal V}(x,y,z,\dots)=N\,V(x,y,z,\dots)$, where the function $V(x,y,z,\dots)$  contains only finite, $N$-independent parameters, is usually called the 't~Hooft, or planar limit. Among many important matrix models of this kind there is the so called Eguchi-Kawai $d$-matrix integral equivalent, in the 't~Hooft  limit, to the multicolor Quantum Chromodynamics~\cite{Eguchi:1982nm}. The 't~Hooft limit is characterized by the perturbative expansions given in terms of planar Feynman graphs ($1/N$-expansion appears to be a topological expansion: the ``fat'' graphs of a given genus $g$ are weighted with the factor $N^{2-2g}$). This allows the counting of such planar graphs~\cite{Brezin:1977sv,Itzykson:1979fi,Mehta:1981xt} and enables the introduction and exact solution of statistical mechanical models on random planar dynamical lattices -- Ising model on random triangulations~\cite{Kazakov:1986hy,Boulatov:1986sb} and various generalizations~\cite{Kazakov:1987qg,Kostov:1988fy,Daul:1994qy,Kazakov:1988ch}.

The direct analytic computation of a majority of such multi-matrix  integrals is virtually impossible, apart from some trivial, albeit important, cases, such as the quadratic potential $V(x,y,z,\dots)$ leading to the gaussian integral.\footnote{Expansions w.r.t. parameters around the gaussian point lead in the 't~Hooft limit to the perturbation theory formulated in terms of planar Feynman graphs. It can help to study the model in a specific, narrow domain    of the parameter space.} 

For a  sub-class of such integrals with specific potentials the problem can be reduced to  integrations over a smaller number of variables than $\sim N^2$. For example, sometimes the problem can be reduced to the integrations or summation only over $\sim N$ variables, such as eigenvalues of the matrices. Then, in the large $N$ limit, the problem can be reduced to the saddle point calculation, significantly simplifying the problem of computation of that functional integral.\footnote{An (incomplete) review of such {\it solvable} matrix models can be found in~\cite{Kazakov:2000aq}. } The basic example of such a simplification is the one matrix model 
\begin{equation}
    Z=\int d^{N^2}A\,\e^{-\tr {\cal V}(A)}
\end{equation}
solvable for any potential ${\cal V}(x)$. Once we have two or more matrix integration variables in \eqref{generalMM} the problem usually gets much more complicated. Generically, such models are unsolvable, i.e. the number of degrees of freedom cannot be efficiently reduced, and the saddle point approximation is inappropriate since the characteristic ``energy'' and entropy of the integration variables are both of the order $\sim N^2$. Here comes the question whether we can study these integrals at least numerically.  

Virtually the only universal general method of numerical computation of functional integrals is the Monte-Carlo method. It has been applied to some matrix integrals with more or less of success. Its main drawbacks are well known: i) the result comes with a statistical error; ii) it is sometimes difficult to reach the numerical equilibrium state in a reasonable time; iii) MC is bad for the systems with sign-changing Boltzmann weights or non-local interactions; iv) the size of the system (the number of integrals) is limited by computational facilities v) The precision is usually rather modest, maximum about 3-4 digits.  

Do we have any alternative?  

In the 1980's, in a series of papers \cite{Jevicki:1982jj, Jevicki:1983wu, Rodrigues:1985aq}, the authors formulated the problem of large N matrix integral and large N quantum mechanics in the loop space (space of moments). The authors attempted the numerical study for the loop variables by  minimizing  an effective action. They were the first to stress the importance of positive semi-definiteness conditions for certain matrices of loop variables in getting physically meaningful results.

Recently, an important progress has been made in the computations of multi-point correlators in conformal field theories in various dimensions, due to the conformal bootstrap method~\cite{2008JHEP...12..031R}. The method  uses various properties of correlators, such as crossing and positivity, to ``bootstrap'' numerically their values and the values of the critical exponents. It appeared to be far more efficient and precise then other numerical approaches, giving the critical exponents of 3d Ising model with the record 6-digits precision~\cite{2017JHEP...03..086S}. An appealing property of this method is the absence of any statistical error in the results, which are given within rigorously established margins. 

Inspired by this success a few authors applied the philosophy of the numerical bootstrap to the computations of various matrix integrals~\cite{2020JHEP...06..090L,2020PhRvL.125d1601H} and even of the lattice multi-color QCD and ${\cal N}=4$ SYM theory~\cite{2017NuPhB.921..702A}. Instead of the direct study of the matrix integrals they proposed to study the large $N$ Schwinger-Dyson equations which are often also called loop equations, in analogy with their applications to QCD~\cite{Makeenko:1979pb}. They are easily obtained by the obvious Ward identities resulting from insertion of the full matrix derivative under  the matrix integral:
\begin{equation}\label{ScwingerDyson}
    0=\int d^{N^2}A\,d^{N^2}B\,d^{N^2}C\dots\,\frac{\tr}{N}(\frac{\p}{\p A}A^mB^nC^k\dots)\e^{-\tr {\cal V}(A,B,C,\dots)}
\end{equation}
where the matrix derivative inside the trace $\frac{\p}{\p A}$ acts on all $A$-matrices, including the potential.
All other loop equations correspond to all possible ``words'' of matrices under the trace and to all insertions of various matrix derivatives at any place in the ``words''.\footnote{ In the 't~Hooft limit, the single trace ``words'' are enough due to the factorization property which we will describe in the next section.} Then   the positivity conditions are imposed stating that the inner product\footnote{Here inner product of an operator $\mathcal{O}$ means $\langle\mathrm{tr} \mathcal{O}^\dagger \mathcal{O}\rangle$.}  of any operator with itself is positive. Rigorous bounds on the dynamical quantities of the theory can be derived from these positivity conditions and loop equations. 

This new approach, compared to the previous work of loop variables \cite{Jevicki:1982jj, Jevicki:1983wu, Rodrigues:1985aq}, imposes the large $N$ Schwinger-Dyson equations (loop equations) explicitly, rather than getting loop equations as a result of effective action minimization. In parallel with the philosophy of conformal bootstrap, this approach focuses more on the geometry of the space of loop variables under sensible physical constraints, which guarantees the rigorousness of the bounds on physical quantities.

In the inspiring work of Lin~\cite{2020JHEP...06..090L} the method was rather successfully applied to the one-matrix model mentioned above, to the exactly solvable two-matrix model with $\tr(AB)$ interaction~\cite{Mehta:1981xt,Itzykson:1979fi} describing the Ising model on planar graphs~\cite{Kazakov:1986hy} as well as to the model with $\tr(AB^2+A^2B)$  interaction, presented there as a case of ``unsolvable'' matrix model~\footnote{We will demonstrate in the Appendix~\ref{sec:linsolve} that, in fact, all two-matrix models with cubic interactions, including this one, are solvable in the above-mentioned sense. }. Lin uses the non-linear equations~\eqref{ScwingerDyson} to bootstrap the loop averages up to the positive semi-definite matrix of size $45$.

This new approach has, in our opinion, a great potential for the precision computations  of physically important matrix integrals in the 't~Hooft limit. But at the same time it is very much perfectible at this stage.

Firstly, the numerical matrix bootstrap approach of \cite{2017NuPhB.921..702A,2020JHEP...06..090L,2020PhRvL.125d1601H}, based on the loop equations and positivity constraint, is not well understood analytically. Its efficiency, and the power of positivity, still looks quite mysterious. It is not even fully understood why we need the positivity condition. Secondly, the matrix bootstrap has a very distinguished feature comparing to most of the other bootstrap problems we dealt with so far: it is in general non-convex. The non-convexity comes from the quadratic terms in the loop equation, which is a result of large $N$ factorization. In optimization theory, this is called Nonlinear SDP (semi-definite programing) and all the solvers for it are not mature enough compared with the highly developed SDP solvers dealing with linear problems. In \cite{2017NuPhB.921..702A,2020JHEP...06..090L,2020PhRvL.125d1601H}, the authors tried to bootstrap the matrix models by the Nonlinear SDP directly, and this non-linearity limited the bootstrap capabilities to very simple models, or to more complex models but only up to  very small lengths of operators.

\subsection{Main results}

In this work, we advance the matrix bootstrap approach trying, on the one hand, to understand analytically the role of positivity conditions, and on the other hand,  to overcome, at least partially, the above-mentioned limitations of the method.

First, we derive a necessary and sufficient condition for the positivity  of bootstrap for large $N$ one-matrix model, to clarify how this method is working. Namely, we show that the positivity  is equivalent to the condition for   the resolvent to have the cuts only on the real axis,  with the positive imaginary part corresponding to the positive density of eigenvalue distribution. This condition actually enables us,  in principle,  to analytically solve the bootstrap problem for any one-matrix model. For the illustrative purposes, we will apply the new positivity condition to the  one-matrix model with quartic potential:
\begin{equation}
    V'(x)=\frac{\mu}{2} x^2+\frac{g}{4} x^4,\qquad (\mu=\pm 1)
\end{equation}
where we normalized the quadratic term to $\pm 1$. We will use the analytic bootstrap to completely classify the admissible set of solutions of the loop equations and positivity conditions, and to locate the critical value of $\mathbb{Z}_2$ symmetry breaking.

So far, we could solve exactly a very limited set of bootstrap problems, and most of them correspond to very simple theories, such as Sine-Gordon theory in S-matrix bootstrap~\cite{2017JHEP...11..143P} and 1d mean field theory in conformal bootstrap~\cite{2019JHEP...02..162M}. Since this method appears to be applicable to any one-matrix model and generalized to some solvable multi-matrix models,  it provides us with a big new family of exactly solvable bootstrap problems. Hopefully these solvable bootstrap models will give us  more  of intuition about the way the bootstrap method works.

The other new result of this work is a new bootstrap scheme for the study of non-linear SDP for multi-matrix integrals, which appears to be numerically much more efficient than those proposed in the past. The main ingredient of the method is the introduction of relaxation matrix in the place of  non-linearity of the loop equation. Namely, we treat the quadratic terms as  independent variables and impose the positivity condition on these variables. Surprisingly, it seems enough to bootstrap the region of admissible values of the computed quantity that is quickly shrinking with the increase of the ``cutoff'' -- the maximal length of ``words'' in the involved operators.  

As a particular example of analytically unsolvable matrix integral we will study by this method the following two-matrix model
\begin{equation}\label{2MMcom2}
    Z=\lim_{N\rightarrow \infty}\int d^{N^2}A\,d^{N^2}B\,\e^{-N\tr\left( -h[A,B]^2/2+A^2/2+g A^4/4+B^2/2+g B^4/4\right)}.
\end{equation}
Various versions of this model have been studied in the past in connection to certain ${\cal N}=1$ supersymmetric Yang-Mills theories~\cite{1999NuPhB.557..413K}. In the particular case $g=0$ the model is solvable and it will serve us as an important check of  applicability and efficiency of our relaxation bootstrap method.  Our results show a very good precision: up to 6 digits with the maximal cutoff  equal to 22 for the words under averages. We were also able to establish with a reasonable accuracy the phase structure of the model in the $g,h$ coupling space, i.e. the positions of critical lines corresponding to the convergence radius of planar perturbative expansion, as well as to the spontaneous $\mathbb{Z}_2$  symmetry breaking.


The two-matrix model \eqref{2MMcom2} considered in this paper serves mostly for the illustration of the power of our method, though it could have in principal some physical applications, such as the statistical mechanics on dynamical planar graphs, in the spirit of~\cite{Kazakov:1998qw,Kostov:1999qx,Zinn-Justin:1999chi}.

This article is organized as follows. The next  Section~\ref{sec:review} serves as a retrospect of the Hemitian matrix integral and the numerical bootstrap technique developed for it so far. Then in Section~\ref{sec:equiv} we propose our equivalent condition for the positivity condition described in Section~\ref{sec:review}. This condition will justify the numerical bootstrap method and enable us to analytically solve the corresponding bootstrap problem. in Section~\ref{sec:relax}, we will describe the way our  relaxation method works for  analytically unsolvable large \(N\) multi-matrix integrals. We test this relaxation method in Section~\ref{QCSM} on the concrete unsolvable model~\eqref{2MMcom2}. We will see that our relaxation method is able to largely meet our expectations, with remarkable precision.
In the last section, after short conclusions, we will briefly discuss possible applications of our method to some more physical problems, such as the multicolor lattice Yang-Mills theory.

\emph{Note}: The main results of this work are  compared with the later Monte Carlo(MC) results~ \cite{Jha:2021exo}. This comparison convinces us that the bootstrap method is more efficient than MC regarding the large N two-matrix model calculation.

\section{Hermitian one-matrix model bootstrap}\label{sec:review}

In this section we will revisit several basic facts about large \(N\) limit Hermitian one-matrix model and the related numerical bootstrap proposed in \cite{2020JHEP...06..090L}. We will be mainly focused here on the aspects of this model which are crucial for the theoretical development in the next section and provide us with important intuition. The reader can refer to numerous works and reviews, some already  cited above (see e.g. \cite{Eynard:2004mh} for a good state-of-art description of results on Hermitian one-matrix model).

\subsection{Hermitian one-matrix model in the planar limit and loop equations}\label{sec:revisit}

The Hermitian one-matrix model is defined by matrix integral:
\begin{equation}\label{MMint}
    Z_{N}=\int d^{N^2}M\,\e^{-N\tr { V}(M)}
\end{equation}
where the invariant Hermitian measure is   $d^{N^2}M=\prod_{ i,j 1}^{N}dM_{ij}$. The potential is usually taken polynomial~\footnote{We believe that our final conclusion can be generalized to non-polynomial potentials, but there may be some subtleties.}:  
\begin{equation}\label{pot}
   V(x)=\sum_{k=2}^{d+1} \frac{g_k}{k} M^k.
\end{equation}
The main ``physical observable'' is the \(k\)-th moment:
\begin{equation}\label{moment}
    \mathcal{W}_k=\langle \mathrm{Tr }M^k\rangle=\int \frac{d^{N^2}M}{Z_{N}}\,\frac{1}{N}\tr M^k\e^{-N\tr { V}(M)}.
\end{equation}
 This model is solvable in the planar limit for arbitrary polynomial potentials~\cite{Brezin:1977sv}. There exist several methods for that: direct recursion relations for planar graphs, orthogonal polynomials, saddle point approximation for the eigenvalue distribution, loop equations (see \cite{Migdal:1983qrz,DiFrancesco:1993cyw} for a review). The loop equations  will play the crucial role in our bootstrap method. 
 
 To derive them we simply use the Schwinger-Dyson method by writing
\begin{equation}\label{ScwingerDyson1}
    0=\int d^{N^2}M\,\frac{\tr}{N} \left(\frac{\partial}{\partial M} M^k\right)\e^{-N\tr { V}(M)}
\end{equation}
since the expression under the integral is a total derivative.  The   boundary terms are absent assuming that the highest power $d+1$ of the potential is even and its coefficient is positive $t_{d+1}>0$.~\footnote{For the ``unstable'' potentials, which do not satisfy one of these conditions, the matrix integral might still exist with appropriate deformation of the integration contour. The large $N$ solutions can exist even independently of the contour deformation since they correspond to local minima of the effective potential for the eigenvalues.}

Applying explicitly the matrix derivative in \eqref{ScwingerDyson1} we write the loop equation in terms of the moments:\footnote{Here for the conciseness, we introduce the normalized trace $\mathrm{Tr}=\frac{1}{N}\tr$, so that $\mathrm{Tr}I=1$}
\begin{equation}\label{loopN}
    \langle \mathrm{Tr}V'(M)M^k\rangle=\sum_{l=0}^{k-1} \langle \mathrm{Tr}M^l  \mathrm{Tr}M^{k-l-1}\rangle.
\end{equation}
In the \(N\rightarrow \infty\) limit we can  use the factorization property:
\begin{equation}\label{factorN}
    \langle \mathrm{Tr}M^l \mathrm{Tr}M^{m}\rangle= \langle \mathrm{Tr}M^l\rangle \langle \mathrm{Tr}M^{m}\rangle+\mathcal{O}(1/N^2).
\end{equation}
Then the loop equation reduces to
\begin{equation}\label{loop}
    \sum_{j=1}^{d} g_j\mathcal{\,\,W}_{k+j}=\sum_{l=0}^{k-1} \mathcal{W}_{l} \mathcal{\,\,W}_{k-l+1}.
\end{equation}  

The simplest way to solve~\eqref{loop} is  to introduce the generating function of moments - the resolvent - as a formal power series in terms of $z^{-1}$:
\begin{equation}\label{reso}
    G(z)=\sum_{k=0}^{\infty} z^{-k-1}\mathcal{W}_k .
\end{equation}
We have not yet assumed anything about the convergence of the series.  Multiplying~\eqref{loop} by $z^{-k}$ and summing from $k=1$ to $\infty$ we represent the loop equation in a compact form as a quadratic equation for the resolvent. 
\begin{equation}\label{Riccati}
    G(z)^2+P(z)=V'(z) G(z).
\end{equation}
The function $P(z)$ comes from carefully collecting in the summation the terms with small $k$'s . It can be written compactly as:
\begin{equation}
    P(z)=\langle \mathrm{tr} \frac{V'(z)-V'(M)}{z-M}\rangle.
\end{equation}
This is a polynomial of $z$ and a linear function of $\mathcal{W}_k,(k=1,...,d-1)$. For example, if $V'(z)=z+g z^3$, then we have
    $P(z)=1 + g z^2 + g \mathcal{W}_2 + g z \mathcal{W}_1$. We can solve ~\eqref{Riccati}, picking the relevant branch of the root which reproduces the leading $z^{-1}$ behavior of the resolvent $G(z)=\frac{1}{z}+{\cal O}(\frac{1}{z^2})$ at infinity:
\begin{equation}\label{root}
   G(z)=\frac{1}{2}(V'(z)-\sqrt{V'(z)^2-4P(z)}) .
\end{equation}
This result will play an important role in our work, so we make several comments on it:
\begin{enumerate}
    \item By~\eqref{root}, the resolvent is understood as a genuine analytic function, at least in the neighborhood of infinity point. So the formal series defined in~\eqref{reso} has a finite radius of convergence. As a consequence, there must be an exponential bound for the moments\footnote{Strictly speaking the radius of convergence is the inverse of the module of largest root of the polynomial under the square root in~\eqref{root}, unless two of such roots merge. Here it is enough for us that it is bounded exponentially.}:
    \begin{equation}\label{bound}
    \mathcal{W}_n\leq C_0 R(g_k,\mathcal{W}_1,...\mathcal{W}_{d-1})^n .
    \end{equation}
    
    \item It is clear from this formula, that all the moments are determined by several low-order moments in $P(z)$ and the couplings:
    \begin{equation}
        \mathcal{W}_n=\mathcal{W}_n (g_k, \mathcal{W}_1,..., \mathcal{W}_{d-1}).
    \end{equation}
    This is what the loop equation tells us. But the loop equation doesn't tells us how to fix the low-order moments involved in $P(z)$. Those can be fixed only by additional assumptions on the solution, such as for example the single support solution for the eigenvalues (single cut on the physical sheet of  $G(z)$). We will see  how to classify the solutions which are picked up by the bootstrap method. 
\end{enumerate}

\begin{figure}[ht]
    \centering
    \includegraphics[width=.8\textwidth]{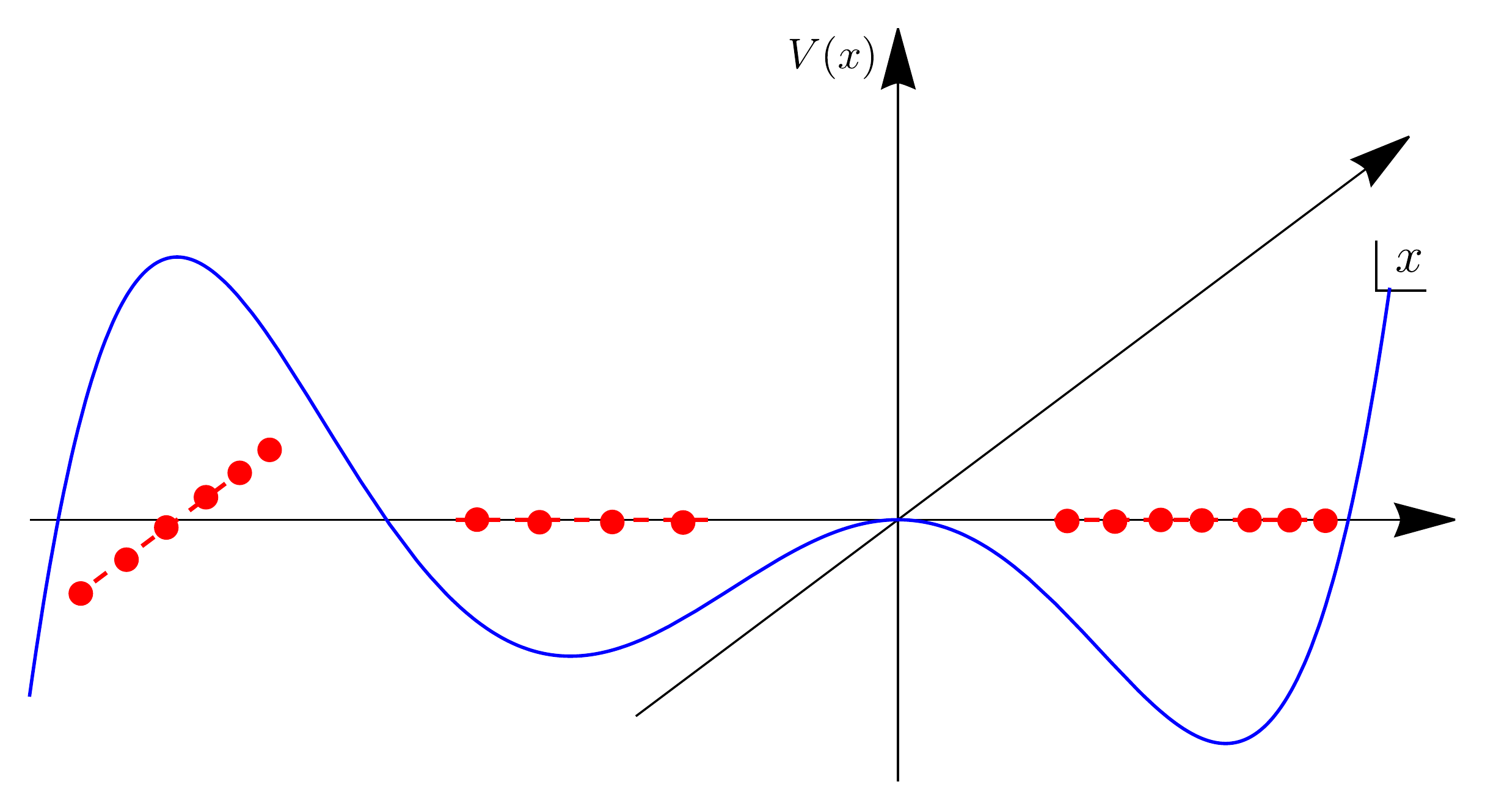}
    \caption{Coulomb gas interpretation for the eigenvalue configurations and a typical cut configuration. For a general solution of the SDP~
    \eqref{SPE}, we can have a complex cut at the maximum of the potential.}
    \label{fig:gas}
\end{figure}

To have more intuitive ideas of possible large \(N \) solutions it is useful to reduce the matrix integration~\eqref{MMint} to the integration over the eigenvalues of the Hermitian matrix. Namely, if we represent it as \(M=\Omega^\dagger X \Omega\) where \(X=\text{diag}(x_1,x_2,\dots,x_N)\) is the diagonal matrix of eigenvalues and \(\Omega\) is the diagonalizing unitary matrix, the matrix integral reduces to only \(N\)  integrations over the eigenvalues~\cite{Brezin:1977sv}: \begin{equation}\label{MMev}
    Z_{N}=\int \prod_{j=1}^{N}\,\left(dx_j\,\e^{-NV(x_j)}\right)\,\Delta^2(x_1,\dots,z_N)
\end{equation}
where the square of the  Vandermonde determinant \(\Delta(x_1,\dots,z_N)=\prod_{i>j}(x_i-x_j)\) represents the Jacobian of the change of integration variables (Dyson measure).
Here the integrand  is of the order \(e^{N^2(\dots)}\) whereas the number of variables is reduced to \(N\). This allows for the application of the saddle point approximation, giving the BIPZ saddle point equations (SPE)~\cite{Brezin:1977sv}
\begin{align}\label{SPE}
V'(x_j)=\sum_{k(\ne j)}\frac{2}{x_j-x_k},\qquad j=1,2,\dots N.
\end{align}
It looks as the condition of electrostatic equilibrium of two-dimensional point-like electric charges (of the same sign)  with coordinates \(x_j\) on a line, locked in the potential \(V(x) \) (see the Fig~\ref{fig:gas}. )

The possible physical solutions correspond to filling \(n\) minima of such a potential with fractions \(N_1,N_2,\dots,N_n\) of these charges, such that \(\sum_{l=1}^n N_k=N\).  The eigenvalues then form a continuous distribution with \(n\) finite supports along the real axis. 

The SPE describes all extrema of the effective potential
 \begin{equation}\label{Veff}
 V_{\mathrm{eff}}=N\sum_{k}V(x_k)+\log \Delta^2(x_1,\dots,z_N),
 \end{equation}
  not only the minima but also the maxima.  For the solutions with the  filling of some maxima of the potential the linear supports of distributions around the maxima should inevitably turn into the complex plane, with the complex conjugate endpoints,  as shown in Fig~\ref{fig:gas}. We will call such solutions ``unphysical''. The values of fractions \(\nu_j=N_j/N,\,\,\,j=1,2,\dots,d\) are in one-to-one correspondence
with the values of first \(d-1\)  moments \(\mathcal{W}_1,...\mathcal{W}_{d-1}\) and they completely fix the algebraic curve of the solution~\cite{Dijkgraaf:2002fc,Dijkgraaf:2002pp}. The solutions where we fill only the minima of the effective potential will be called "physical". The supports for such solutions will be located only on the real axis, with positive weight for the distribution of the eigenvalues. 

In the large \(N\) limit, the distribution of eigenvalues converges to a continuous function \(\rho(x)\) and the corresponding SPE actually becomes the quadratic equation for resolvent~\eqref{Riccati}. In this limit, the resolvent function, the eigenvalue distribution and the series of moments are closely related.

The moments can be computed via the resolvent~\eqref{reso} by a simple contour integration formula:
\begin{equation}\label{WGint}
\mathcal{W}_n=\frac{1}{2\pi i}\oint_{\Gamma} z^n G(z) \mathrm{d} z=-\frac{1}{4\pi i}\oint_{\Gamma} z^n C(z) \mathrm{d} z
\end{equation}
where the contour \(\Gamma\) must encircle all branch points  of \(G(z)\)~(see  \cite{Dijkgraaf:2002fc,Dijkgraaf:2002pp} for the details). We introduced here the ``cut-function'' -- the square root of the discriminant -- by the formula 
\begin{equation}\label{eq:cutfunc}
    C(x)=\sqrt{V'(x)^2-4P(x)}=\sqrt{D(x)}=g_{d+1}\sqrt{\prod_{k=1}^{d}(x- a_{k})(x- b_{k})}.
\end{equation}
The roots \(\{ a_k, b_k\}\) of the discriminant become the branch points of the cut-function. Their number is always even. For real couplings in the potential the branch points lay only on the real axis or come in complex conjugate pairs. Note that it is easy to relate the eigenvalue fractions to these branch points: \begin{equation}\label{NjGint}
\nu_j=\frac{1}{2\pi i}\oint_{\Gamma_{j}}  G(z) \mathrm{d} z=-\frac{1}{4\pi i}\oint_{\Gamma_{j}} C(z) \mathrm{d} z
\end{equation} 
where the contour \(\Gamma_{j}\) encircles anticlockwise only the branch points \(\{a_k, b_k\}\) , and \(\nu_j\) can be in principal of either sign. 

The eigenvalue density is expressed as the discontinuity of the cut function \(C(x)\):
\begin{equation}\label{eigd}
\rho(x)=\frac{1}{4\pi i}(C(x+i0)-C(x-i0))=\frac{1}{2\pi} \Im C(x+i0)
\end{equation}
and the moments can be expressed by the eigenvalue density:
\begin{equation}\label{peigen}
\mathcal{W}_n=\int_{-\infty}^{\infty} x^n \rho(x) \mathrm{d} x.\qquad 
\end{equation}

\subsection{ Bootstrap method for  the large \texorpdfstring{$N$}{N} one matrix model }

As we stated above, the loop equation~\eqref{loop},\eqref{Riccati} has in general a continuum of solutions of the form~\eqref{root} labeled by a finite number of parameters -- the lowest moments  \(\mathcal{W}_1,...\mathcal{W}_{d-1}\) which can take a priori arbitrary values. But not all of these solutions are ``physical'', i.e. rendering all moments \(\mathcal{W}_{k},\,\,\,k\in \mathbb{Z}_+\) real and compatible with the finite \(N\) Hermitian matrix ensemble. For example, the physical even moments should be positive, but this condition is not the only one.        

A more general physical condition on a solution is the positivity of inner product for the matrix integral. This condition states that, for any operator of the form $\mathcal{O} = \sum_{i=0}^{n-1} \alpha_i M^i\, \mathrm{s.t.}\, \matr \alpha\in \mathbb{R}^n,$ and for any \(n\in \mathbb{Z}_+\), we have the positive semi-definite quadratic form\footnote{Here we  assume that all the expectation values we study are real. This is actually a non-trivial result from the \(M\rightarrow M^{\mathrm{T}}\) symmetry of the potential. Since this symmetry is always present for all the models considered in this article, we will implicitly assume this to be always true.}:
\begin{equation}\label{posit1}
    \langle\mathrm{Tr}\mathcal{O}^\dagger \mathcal{O}\rangle=\matr \alpha^{\mathrm{T}} \mathcal{\mathbb{W}} \matr \alpha\geq 0\quad \forall \matr \alpha\in \mathbb{R}^n.\, \,\,\,
\end{equation}
Here we introduced the matrix \(\mathbb{W}_{ij}=\mathcal{W}_{i+j-2}\) which  will be called below for convenience the correlation matrix. The above condition is equivalent to the positive definiteness of correlation matrix\footnote{Here we slightly abused the notations: $\mathbb{W}$ sometimes means the matrix with finite size, involving only the moment up to a certain order \(\Lambda\) (``cutoff''), and sometimes it means the infinite dimensional matrix. But the positive semi-definiteness is always well defined as a positivity of the corresponding quadratic form.}:
\begin{equation}\label{eq:corpos}
    \mathbb{W}\succeq 0\,.
\end{equation}
The condition~\eqref{posit1} is obvious for a finite $N$ matrix model with converging integral~\eqref{MMint}, i.e. when \(d+1 \) is even and \(g_{d+1}>0\) in the potential~\eqref{pot}, since the moments are just given by the integration of positive definite functions  with a positive measure. In addition, at finite \(N \) there exists only one solution for the moments~\footnote{To define the matrix integral for  unstable potentials, when \(d+1\) odd or/and \(g_{d+1}<0\) one usually deforms appropriately the integration contours. Then the the questions of positivity become less obvious for finite \(N\). But we will see on the example of quartic potential that at infinite \(N\)  we can still have positivity for certain solutions, even for such, globally unstable, potentials.      }. But it is far from trivial in large $N$ limit, where we have to understand what solutions from the continuum are really physical.

\begin{figure}[ht]
    \centering
    \includegraphics[width=.6\textwidth]{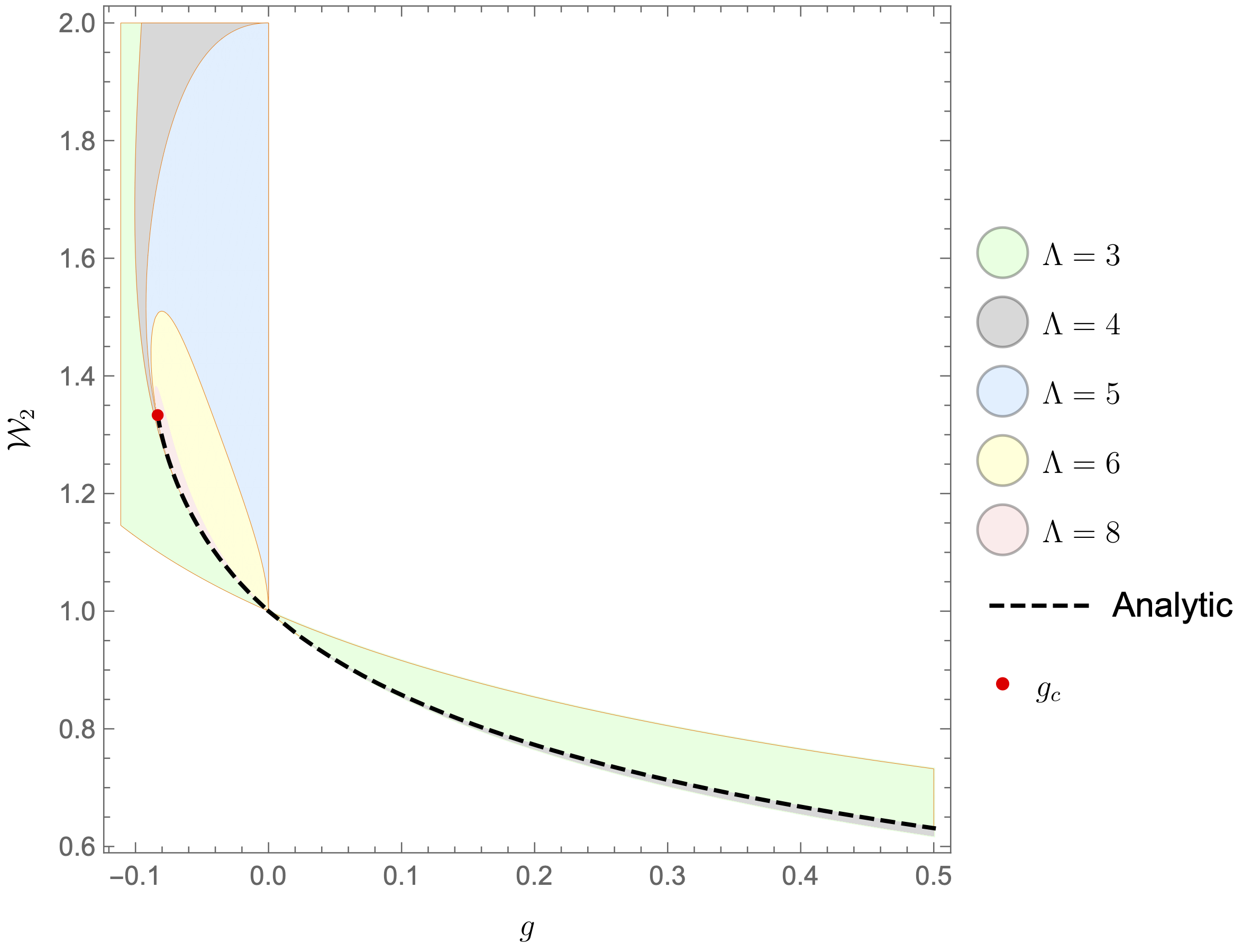}
    \caption{The allowed region for quartic model $V(z)=\frac{1}{2}z^2+\frac{g}{4} z^4$ for different cutoff $\Lambda$, compared with the analytic solution. Here we assuming $\mathcal{W}_1=0$, we also note that for \(g>0\), we didn't plot $\Lambda\geq5$ since they are almost indistinguishable from the analytic solution on the figure.}
    \label{fig:re}
\end{figure}

To do numerical bootstrap, we set a finite cutoff $\Lambda$,  i.e. the highest moment in the correlation matrix \(\mathbb{W}\) and loop equations is $\mathcal{W}_{2\Lambda}$ and the size of the correlation matrix is $(\Lambda+1)\times (\Lambda+1)$. We can use the loop equations to express the higher moments through a certain number of the lower moments and  substitute them  into the correlation matrix as functions of lower moments. Then the positivity of the correlation matrix provides us with algebraic inequality on these lower moments. In general, we expect to get the inequalities for each of the lower moments both from above and from below, for example:
\begin{equation}
    \mathcal{W}_{2\Lambda }^\mathrm{min}\leq \mathcal{W}_2\leq \mathcal{W}_{2\Lambda }^\mathrm{max}
\end{equation}
at a cutoff $\Lambda$. In practice, the allowed region $[\mathcal{W}_{2\Lambda }^\mathrm{min}, \mathcal{W}_{2\Lambda }^\mathrm{max}]$ shrinks fast as we increase $\Lambda$, giving us tight bounds on $\mathcal{W}_2$.

We exemplify this approach on the case of quartic potential $V(z)=\frac{1}{2}z^2+\frac{g}{4} z^4$. We plot in Fig~\ref{fig:re} the  region for the allowed values of function \(\mathcal{W}_2(g)\) under the assumption of \(M\Leftrightarrow-M\) symmetry of the solution, i.e. $\mathcal{W}_{2l+1}=0$. Under this assumption, all the higher moments are polynomials in terms of $\mathcal{W}_2$ and $g$. The positivity of the correlation matrix reduces to a list of algebraic inequalities on $\mathcal{W}_2$.\footnote{To depict the allowed region, maybe the simplest way is the Mathematica \textbf{RegionPlot} function. Although not really numerically efficient, it is already good enough for the simplest quartic one-matrix model.}

It is a bit surprising that the bootstrap scheme described above for Hermitian one-matrix model is generally analytically solvable, considering that it is usually non-trivial to solve an infinite series of algebraic inequalities. In the next section we will propose a necessary and sufficient condition for the positivity constraint~\eqref{eq:corpos} by virtue of a result of solution of Hamburger moment problem. By this condition we can not only generally solve the bootstrap problem analytically but also justify why the numerical bootstrap process excludes the unphysical solutions of SPE described in Section~\ref{sec:revisit}.

\section{Hamburger moment problem and positivity of resolvent}\label{sec:equiv}

As it should be clear from the previous section, the bootstrap method for one-matrix model has two main ingredients: 1. The Schwinger-Dyson loop equations for the moments of the random matrix variables; 2. The positive definiteness of the correlation matrix of these moments. We will rigorously prove that the second ingredient, in virtue of  the Hamburger moment problem \cite{reed1975ii}, picks up in the planar limit the solutions of loop equations only with real-positive supports of the matrix eigenvalue distribution. We will employ this condition to analytically solve the bootstrap condition.

\subsection{Hamburger problem versus the positivity condition on resolvent}\label{sec:Hamburger}

The loop equation \eqref{loop} renders all possible large $N$, saddle point  solutions of the Hermitian one matrix model. Some of them ``look''  physical, i.e. corresponding to the stable equilibrium in the Coulomb gas picture for the eigenvalues locked in the effective potential \eqref{Veff}. For the other, the stable solution corresponds to ``unphysical'' picture when the supports of eigenvalue distribution become complex. What are the solutions captured by our bootstrap procedure?

As we just reviewed, numerical bootstrap of one matrix model consists of two ingredients: loop equation and positivity of correlation matrix. It sets a cutoff on both constraints and gets a rigorous bound on the physical quantities we are interested in. Analytically all the information contained in the loop equations is encoded in the quadratic equation of resolvent~\eqref{Riccati}, which has a simple solution~\eqref{root}. It describes the hyper-elliptic algebraic curve parameterized by complex variable \(x\). It is natural to ask the question: can the positivity of correlation matrix  also be expressed as a simple condition on the resolvent? The answer is, luckily and a bit surprisingly, yes.

For convenience, we make the following definition: a resolvent satisfies the positivity condition if the corresponding eigenvalue density~\eqref{eigd} is supported on the real axis and is positive on its support.
Our main conclusion of this section will be:
\begin{equation}\label{equiv}
    \textit{Positivity of correlation matrix}\Leftrightarrow\textit{Positivity of Resolvent}
\end{equation}
We prove the necessity first. The proof is based on a well-known mathematical conclusion that will play an important role in our demonstration -- the result of the solution of the Hamburger moment problem~\cite{reed1975ii}:

\textit{For a given series of real numbers $\{m_n\}_{n=0}^\infty$, there exists a positive Borel measure $\mu$ such that:
\begin{equation}\label{ham}
    m_n=\int x^n \mathrm{d}\mu
\end{equation}if and only if the matrix $H_{ij}=m_{i+j-2}$ is positive semi-definite. Moreover, if there exist the constants $C$ and $D$, such that $|m_n|\leq C D^n n!$, the measure is unique.}

Applying the result of Hamburger momentum problem~\eqref{ham}, we have for each moment $\mathcal{W}_n=\int_{\mathbb{R}} x^n \mathrm{d}\mu(x)$. We notice from the exponential bound condition~\eqref{bound}  that $\mu$ must be supported in a finite region $[-R,R]$, since otherwise, if we have $\epsilon>0$, such that $\mu\big((-\infty, -(R+\epsilon))\bigcup (R+\epsilon,\infty)\big)=\mu_0>0$, then
\begin{equation}
    \mathcal{W}_{2n}=\int_{\mathbb{R}} x^{2n} \mathrm{d}\mu(x)>(R+\epsilon)^{2n} \mu_0,
\end{equation}
which contradicts the exponential bound~\eqref{bound}.

Consequently for $|z|>R$:
\begin{equation}\label{analy}
    G(z)=\sum_{k=0}^{\infty} z^{-k-1}\mathcal{W}_k=\sum_{k=0}^{\infty} z^{-k-1}\int_{[-R,R]} x^k \mathrm{d}\mu(x)=\int_{[-R,R]} \frac{\mathrm{d}\mu(x)}{z-x}\,.
\end{equation}
The exchange of infinite sum and integration is justified by Fubini's theorem. Due to this equation, $G(z)$ is analytic in the region outside of the disk $|z|>R$. The last equality  in~\eqref{analy} enables us to analytically continue $G(z)$ to the whole region $\mathbb{C}\backslash[-R,R]$. So the function $G(z)$ must be analytical away from the real line, which  eliminates the possibility of cuts between complex branch points. Comparing with~\eqref{root} and~\eqref{eigd}, we come to the conclusion that all  supports of \(\rho(x)\) function must be located on the real line. For the positivity of eigenvalue density, we note that we can extract  by contour deformation the coefficient of the series in~\eqref{reso}:
\begin{equation}\label{peigen1}
\mathcal{W}_n=\frac{1}{2\pi i}\oint z^n G(z) \mathrm{d} z=\int_{[-R,R]} x^n \rho(x) \mathrm{d} x.
\end{equation}

By the uniqueness of  solution of the  Hamburger momentum problem, we must have $\rho(x) \mathrm{d} x=\mathrm{d} \mu$ i.e. they are equal in terms of positive measure.\footnote{We note that the uniqueness is not strictly necessary here. To see this, the reader can combine Stone-Weierstrass theorem and the fact that compactly supported continuous function is dense in $L_p,\, 1\leq p< \infty$. Then if $\rho(x) \mathrm{d} x$ is not positive almost everywhere then the positivity of correlation matrix is violated. } So we have $\rho(x)$ real supported and positive. This concludes our proof of necessity for~\eqref{equiv}.

The proof of sufficiency is straightforward. It is already true because the sufficiency is a part of the result of the solution of Hamburger moment problem. For a more direct argument, suppose we have a resolvent that satisfies the positivity condition, i.e.~\eqref{peigen} with \(\rho(x)>0\). We notice that the matrix $(\matr X)_{ij}=x^{i+j-2}$ is trivially positive semi-definite for real $x$, so that if we integrate the matrix $\matr X$ w.r.t. the positive measure $\rho(x) \mathrm{d} x$, it stays positive semi-definite as well. The result of the integration is actually our correlation matrix $\mathbb{W}$. This concludes the proof of sufficiency and hence  of the equivalence~\eqref{equiv}.

It is easy to demonstrate by the direct computation that in the presence of  complex branch points in \(G(x)\) the corresponding
correlation matrix is{\it\ not} positive definite. A simple example is the resolvent for the matrix model with the unstable potential \(V(M)=-\frac{1}{2}\Tr~M^2\), which is \(G(x)=\sqrt{x^2+2}-x=\frac{1}{x}-\frac{1}{2 x^3}+O\left(\left(\frac{1}{x}\right)^4\right)\). We see that \(\langle\Tr M^2\rangle=-\frac{1}{2}\) so that the correlation matrix is not positive definite.

This suggests the validity of the numerical bootstrap approach at least in the case of the one-matrix model: by imposing the positive semi-definiteness condition on the solutions of loop equations, at least for a finite cutoff \(\Lambda\),  we exclude the ``unphysical'' large \(N \) solutions with the eigenvalue distributions having complex or negative supports, i.e. violating the hermiticity  of the matrix measure. 

The result of the present section actually enables us to analytically solve the bootstrap constraints, since the infinite series of inequalities from the positivity of the correlation matrix have been proven to be equivalent to the positivity property of resolvent. As an example, in the next subsection we will present the analytic result of solving the bootstrap problem of quartic one-matrix model.

The application of numerical bootstrap to the multi-matrix models, such as the one studied in Section~\ref{QCSM}, has not as strong theoretical basis as the one presented in this section for the one-matrix model. However the arguments presented here give a good intuition why the numerical bootstrap can work even in the multi-matrix model case. In the following sections we will demonstrate its viability empirically, by showing its numerical efficiency for a specific, ``unsolvable'' matrix model.

\subsection{Classification of physical solution of quartic one-matrix model}\label{sec:ana}

In Fig~\ref{fig:re}, we saw that as we increase $\Lambda$ the allowed region converges to the analytic solution. One may ask whether the allowed region will ultimately exclude all other solutions as \(\Lambda\) increases, or it will stabilize to a very tiny island  which will not shrink further. The results of the current section will support the first of these options. 

In this section, we will  apply the positivity of resolvent to the one-matrix model with quartic potential 
\begin{equation}\label{qpotential}
V(x)=\frac{1}{2}\mu x^2 +\frac{1}{4}g x^4
\end{equation}
in order to fully classify all physical solutions. This is equivalent to solving the positivity condition of one-matrix bootstrap analytically. Since in the previous subsection we have already formulated this problem as a precise mathematical theorem, we will not  present the formal mathematical derivation here.  For the details the reader can refer to the Appendix~\ref{Quartic}. 

For the bootstrap problem we are trying to solve, we will not assume the $\mathbb{Z}_2$ symmetry of the solutions. This symmetry  would  mean $\mathcal{W}_{2k+1}=0$. We will see that there exist solutions that break this symmetry. In fact, for  solutions we find numerically  the breakdown or preservation of $\mathbb{Z}_2$ symmetry will be established dynamically and not necessarily imposed  as an input. Alternatively, if we  assume $\mathbb{Z}_2$ symmetry  from the beginning, the numerical efficiency for such  solutions considerably increases.

For the specific potential~\eqref{qpotential}
the positivity condition for the resolvent 
\begin{equation}
\begin{split}
    G(x)&=\frac{1}{2}(V'(x)-\sqrt{V'(x)^2-4P(x)})\\
    &=\frac{1}{2} \left(-\sqrt{-4 g (\mathcal{W}_2+x (\mathcal{W}_1+x))+\left(g x^3+\mu  x\right)^2-4}+g
   x^3+\mu  x\right)
\end{split}
\end{equation}
 translates into the condition that it has a only real positive eigenvalue distribution. This condition can be solve rigorously, namely:
\begin{enumerate}
    \item $\mu=1$ and $g\geq-\frac{1}{12}$: \(\mathcal{W}_1=0, \,\mathcal{W}_2=\frac{(12 g+1)^{3/2}-18 g-1}{54 g^2}\).
    \item $\mu=1$ and $g<-\frac{1}{12}$, there is no possible solution.
    \item $\mu=-1$ and $g\leq0$, there is no possible solution.
    \item $\mu=-1$ and $g\geq\frac{1}{4}$: \(\mathcal{W}_1=0,\, \mathcal{W}_2=\frac{(12 g+1)^{3/2}+18 g+1}{54 g^2}\).
    \item $\mu=-1$ and $0< g<\frac{1}{4}$: This situation is a bit involved. The bootstrap solution is a curve segment parametrized by \(\mathcal{W}_1\). Explicitly, the solution is a branch of the algebraic equation:
    \begin{equation}
\begin{medsize}
\begin{split}
    &0=11664 g^6 \mathcal{W}_2^5+\left(-27216 g^5-864 g^4\right) \mathcal{W}_2^4+\mathcal{W}_2^3 \left(-16200 g^5 \mathcal{W}_1^2-13824 g^5+19872 g^4+1440 g^3+16 g^2\right)+\\
    &\mathcal{W}_2^2 \left(\left(43200 g^5+33480 g^4+888 g^3\right) \mathcal{W}_1^2+23040
   g^4-3232 g^3-544 g^2-16 g\right)+\\
   &\mathcal{W}_2 \left(\left(4125 g^4-22500 g^5\right) \mathcal{W}_1^4+\left(-65280 g^4-24568 g^3-1480 g^2-16 g\right) \mathcal{W}_1^2+4096
   g^4-8704 g^3-1072 g^2-32 g\right)\\
   &+3125 g^5 \mathcal{W}_1^6+\left(18500 g^4-3925 g^3-16 g^2\right) \mathcal{W}_1^4+\\
   &\left(-1024 g^4+22848 g^3+7096 g^2+608 g+16\right)
   \mathcal{W}_1^2-4096 g^3-512 g^2-16 g.
\end{split}
\end{medsize}
\end{equation}
The physical branch of solution is selected by the one passing through \(\mathcal{W}_1=0,\,\mathcal{W}_2=\frac{1}{g}\),\footnote{Actually for \(\mu=-1\) and \(0<g\leq \frac{1}{4}\), the $\mathbb{Z}_2$ symmetry preserving solution is just \(\mathcal{W}_1=0,\,\mathcal{W}_2=\frac{1}{g}\). So the first discontinuity of \(\mathcal{W}_2(g)\) at \(g=1/4\) happens for second  derivative. } with \(-\mathcal{W}_{1c}\leq \mathcal{W}_{1}\leq \mathcal{W}_{1c}\). 

For \(0<g\leq \frac{1}{15}\),
\begin{equation}
    \mathcal{W}_{1c}=\frac{2 \sqrt{4500 g^2+75 g-2 (1-15 g)^{3/2} (60 g+1)+2}}{75 \sqrt{5} g^{3/2}},
\end{equation}
and for \(\frac{1}{15}<g< \frac{1}{4}\),
\begin{equation}
    \mathcal{W}_{1c}=\frac{2 \sqrt{12000 g^2+1200 g-\sqrt{3} (20 g+7)^{3/2} (60 g+1)+102}}{75 \sqrt{5} g^{3/2}}.
\end{equation}
\end{enumerate}

\begin{figure}[ht]
\centering
     \begin{subfigure}[b]{0.48\textwidth}
         \centering
         \includegraphics[width=\textwidth]{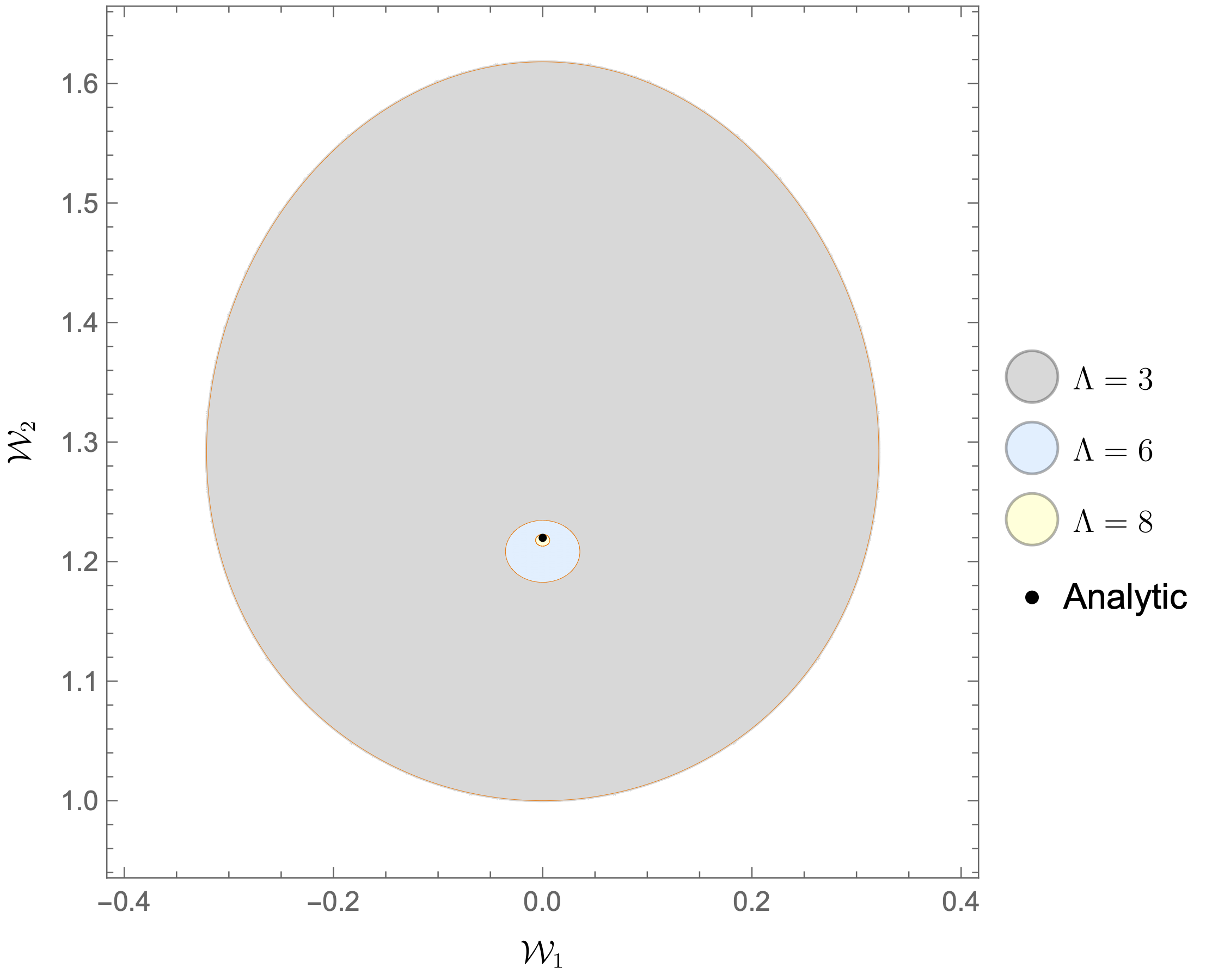}
         
     \end{subfigure}
     
     \begin{subfigure}[b]{.48\textwidth}
         \centering
         \includegraphics[width=\textwidth]{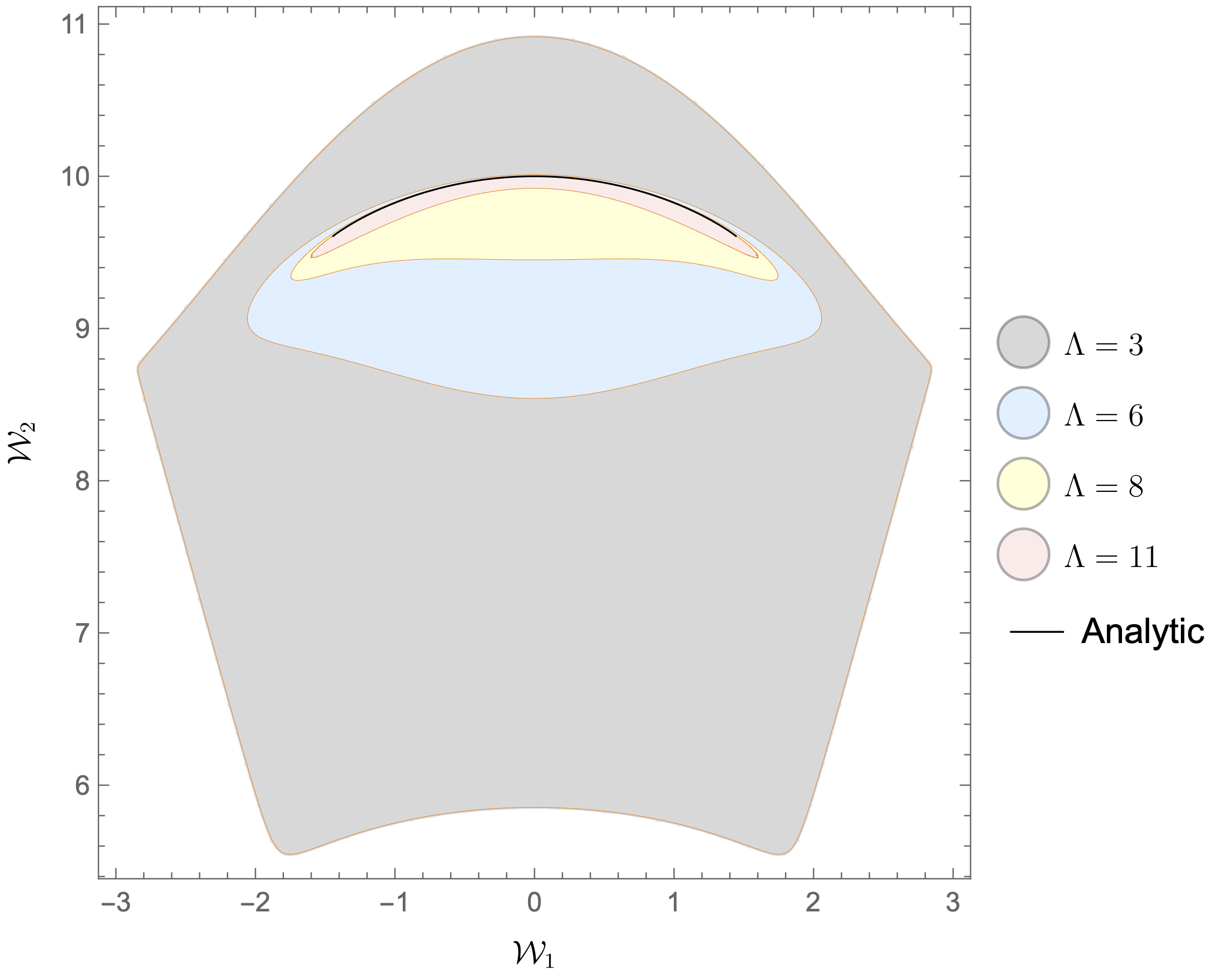}
         
     \end{subfigure}
     
     \begin{subfigure}[b]{0.48\textwidth}
         \centering
         \includegraphics[width=\textwidth]{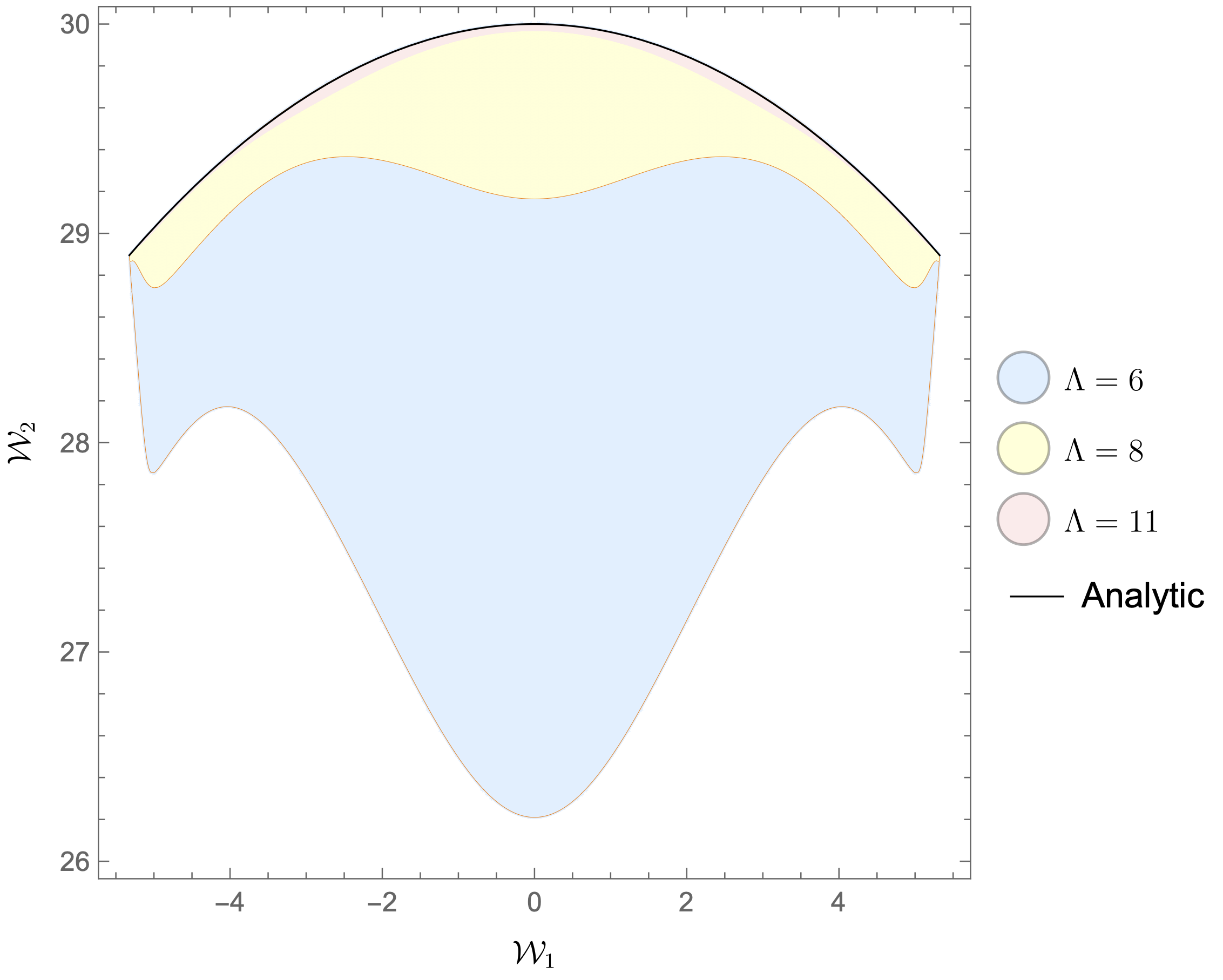}
         
     \end{subfigure}
        \caption{The comparison of numerical bootstrap with our analytic bootstrap results (the black point or the black curve on the figures) in Section~\ref{sec:ana} with \(\mu=-1\) in potential~\eqref{qpotential}. They are representatives from different phases of the model, with \(g=1,\, \frac{1}{10},\,\frac{1}{30}\) (for the figures from above to below, respectively). We notice even visually that for \(g<\frac{1}{4}\) i.e. when the symmetry breaks, the exact solution is a non-convex set.}
        \label{fig:num}
\end{figure}

This reproduces the exact solution of quartic one-matrix model.  In Fig~\ref{fig:re} we have already compared the exact solution and the numerical bootstrap result for \(\mu=1\). A typical comparison for \(\mu=-1\) case is Fig~\ref{fig:num}.

In Fig~\ref{fig:num} we take a representative from each phase and compare it with the above analytic solutions. We see that the numerical bootstrap results converge quickly to the analytic result. A distinguishable feature of these figures is that the allowed region is not guaranteed to be convex. This is very different from the convex optimization problems which we encountered in CFT bootstrap and S-matrix bootstrap. Generally, the large-scale non-convex problem is hard and usually unsolvable. We will discuss  in the next section how to overcome this difficulty.

\subsection{Comments}
Here we present several comments on the results of this section:
\begin{enumerate}
\item There may exist certain doubts on particular choices of the positivity condition of the correlation matrix in numerical bootstrap. In the work~\cite{2017NuPhB.921..702A,2020JHEP...06..090L}, the authors showed that in some cases one only needs the positivity of even moments $\mathcal{W}_{2k}\geq0$ to make bootstrap converging to the analytically known solution. But in general one should be careful about the choices of the positivity condition. For example,  consider the model with \(V(x)=-\frac{1}{2}x^2+\frac{1}{4}x^4\). Under the assumption of the \(\mathbb{Z}_2\) symmetry, the loop equations of this model read:
\begin{equation}
    \mathcal{W}_{2k}=\mathcal{W}_{2k-2}+\sum_{l=0}^{k-2}\mathcal{W}_{2l}\mathcal{W}_{2k-4-2l}\quad k=2,3,4,....
\end{equation}
We see that the positivity condition on even moments only provides us with the constraint  \(\mathcal{W}_2\geq 0\), evident by induction in loop equations. In this situation we can bootstrap the physical solution only with the  positivity condition on the full correlation matrix. This fact  explains to some extent why the convergence in Fig~\ref{fig:num} is not as fast as for the model with positive quadratic coefficient.
\item For the one-matrix integral with  integration over the unitary matrix instead of the Hermitian matrix, we can establish and justify a similar bootstrap method. This enables us with the analytic solution of such  bootstrap problems. The main difference in this case comparing to the Hermitian integral is that the correlation matrix is of the form \(\mathbb{W}_{ij}=\mathcal{W}_{i-j}\). It is called the Toeplitz matrix in linear algebra\footnote{For the Hermitian integral the correlation matrix is of the form of  the Hankel matrix.}. For this correlation matrix, we have the following result of solution, this time for trigonometric moment problem:

\textit{For a given series of real numbers $\{m_n\}_{n=-\infty}^\infty$ such that \(m_{-k}=m_k^*\) , there exists a positive Borel measure $\mu$ on \([0,2\pi]\) such that:
\begin{equation}\label{ham}
    m_n=\frac{1}{2\pi}\int_0^{2\pi} \exp (-i n t) \mathrm{d}\mu(t)
\end{equation}if and only if the matrix $T_{ij}=m_{i-j}$ is positive semi-definite. }

Applying this result to our unitary matrix integral, we come to the conclusion that the positivity of correlation matrix for large \(N\) unitary matrix integral is equivalent to  the  positivity of the eigenvalue density which is supported on the unit circle in the complex plain. 
\end{enumerate}

\section{Relaxation bootstrap method}
\label{sec:relax}
Now we turn to the discussion of the bootstrap method for multi-matrix models. We will see that a naive generalization of the previous one-matrix model bootstrap will lead to a Non-linear SDP~\footnote{SDP means semi-definite programming}. But it is widely known that a general large-scale Non-linear SDP cannot be solved efficiently.  In this section we will propose a systematic numerical bootstrap procedure to solve the large $N$ multi-matrix models via SDP. 

SDP, unlike the Nonlinear-SDP which is directly applicable  in the case of large $N$ matrix model bootstrap~\cite{2020JHEP...06..090L,2017NuPhB.921..702A}, has a long history in  academic research as well as in  applied sciences.  The standard primal form of SDP is\footnote{There  exists  also the dual form of these problems, which will be discussed in Appendix~\ref{rdual}. We also note that in some literature  different conventions for dual and primal for SDP are used.}:
\begin{equation}\label{SDPp}
    \begin{split}
        &\mathrm{minimize}\qquad \sum_{i=1}^m c_i x_i\quad \text{w.r.t.}\,\, \{x_1,x_2,\dots,x_m\}\in\mathbb{R},\\
        & \mathrm{subject \,\,to} \quad \matr \sum_{i=1}^m \matr F_i x_i -\matr F_0\succeq 0,\quad \matr \quad \matr F_i \in \mathcal{S}^n.
    \end{split}
\end{equation}
Here $\mathcal{S}^n$ denotes the space of  $n\times n$ real symmetric matrices. As long as we can transform our bootstrap problem to the form~\eqref{SDPp}, we can  get rigorous bounds on the physical quantities of interest --  linear functions of  \(\{x_1,x_2,\dots,x_m\}\) --  by efficiently solving  the SDP problem~\eqref{SDPp}. 

So the problem reduces to the question how to efficiently transform our matrix integral problem  into   the constraints of  the form~\eqref{SDPp}. Then the original physical problem is transformed into a purely numerical   SDP  problem. 

In this section we will describe our relaxation bootstrap method on the example of single trace moments in a large $N$ two-matrix model with the partition function\footnote{The generalization to multi-matrix models with more matrices is straightforward.}:

\begin{equation}\label{general2MM}
    Z=\lim_{N\rightarrow \infty}\int d^{N^2}A\,d^{N^2}B \e^{-N\tr { V}(A,B)}
\end{equation}
where $V(A,B)$ is assumed to be a so far general polynomial in $A$ and $B$, to make the loop equations more tractable. In the next section we will apply it to a model with a concrete potential, generally unsolvable by the known analytic methods. We will see that our method has four types of constraints: loop equations, global symmetries, positivity of correlation matrix and positivity of relaxation matrix (which will be explained later).

\subsection{Physical constraints}

To make this section as self-contained as possible, we briefly review here the terminology already introduced in the previous sections and show how the constraints of the type~\eqref{SDPp} are specified in the two-matrix model. 

The positivity of correlation matrix is still  at the heart of our method. Since we are doing numerical analysis, we set the cutoff $2\Lambda$ to the length of operators that we are considering, i.e. to the length of ``words'' built from two ``letters'' --  the matrices $A$ and $B$:  \(\mathcal{O}=ABBAAAB\dots\). For any word $\mathcal{O}$ of the length$\le\Lambda$, we assume:
\begin{equation}\label{posit}
    \langle\mathrm{Tr} \mathcal{O}^\dagger \mathcal{O}\rangle\geq 0\,.
\end{equation}

The set of words with length$\le\Lambda$ is a vector space spanned by all the words constructed from two letters with the length cutoff  $\Lambda$. This is a set of $L=2^{\Lambda+1}-1$ elements which we denote  as $\mathcal{O}_{i}$, where $i$ runs from $1$ to $L=2^{\Lambda+1}-1$. For example, when $\Lambda=2$ the basis of this vector space reads:
\begin{equation}
I,\,A,\,B,\,A^2,\,AB,\,BA,\,B^2.
\end{equation}
We can expand the equation~\eqref{posit} w.r.t. this base:
\begin{equation}\label{qform}
    \langle\mathrm{Tr} (\sum_{i=1}^{L}\alpha_i \mathcal{O}_i)^\dagger (\sum_{i=1}^{L}\alpha_i \mathcal{O}_i)\rangle=\matr \alpha^{\mathrm{T}} \mathcal{M}_L \matr \alpha\geq 0.
\end{equation}
Let us  introduce the correlation matrix $\mathcal{M}_{Lij}=\langle\mathrm{Tr}  \mathcal{O}_i^\dagger  \mathcal{O}_j\rangle$ which  consists of expectation values of operators with the lengths up to $2\Lambda$. Since~\eqref{posit} is true for all operators, the condition \eqref{qform} holds for all $\matr \alpha\in \mathbb{R}^L$, i.e. the semi-definite positivity of correlation matrix is ensured:
\begin{equation}
    \mathcal{M}_L\succeq 0.
\end{equation}
This  correlation matrix condition can be directly applied to the two-matrix model. We see that the main difference with the one-matrix model is that the dimension of correlation matrix grows exponentially with $\Lambda$.

Another important ingredient for our bootstrap method is the loop equations.  For the two-matrix model it can be schematically represented as:
\begin{equation}\label{dMWord}
    \int d^{N^2}A\,d^{N^2}B \,\,\tr (\p_{M}(\mathrm{Word}\times\e^{-N\tr\, V(A,B)})=0,\qquad  M=\{A,B\}
\end{equation} 
where ``Word'' means the matrix word built by arbitrary finite product of matrices \(A\)  and \(B\).~\footnote{Note that ``word'' is not yet traced, so that generically it is not cyclically symmetric: a cyclic transformation gives in general a new word.}  The differentiation \(\p_{M}\) can be either w.r.t. the matrix \(A\) or w.r.t. the matrix \(B\).

 The loop equations for large $N$ multi-matrix model in general close on{\it\ all}
words.\footnote{Here we mean that there is generally no infinite subset of loop equations and operators closed among themselves. This fact will be explored further in Section~\ref{sec:structure}.} Schematically, they have the following quadratic form:
\begin{equation}\label{loopAB}
   \langle\Tr\left (\text{Word}_l\,\times\p_MV(A,B)\right)\rangle=\sum_{l_1=1}^{l} \langle\Tr\, \text{Word}_{l_1-1}^{(M)}\rangle\cdot\langle \Tr\, \text{Word}_{l-l_1}^{(M)}
\rangle
\end{equation}
which is a direct generalization of \eqref{loop} of the one-matrix model.
Here \(\text{Word}_{l_1-1}\,\,\text{and}\,\,\, \text{Word}_{l-l_1}\) are the words
obtained by cutting the word \(\text{Word}_l\,\) in two words whenever one has the matrix \(M\) on the \(l_1\)-th place in \(\text{Word}_{l}\).     The matrix factor \(\p_MV(A,B)\) in the l.h.s. comes from the derivative of the exponential factor in~\eqref{dMWord}, which  generically renders a sum over single trace operators with lengths from $l_1$ to $l_1+d$    (the degree of polynomial $V(A,B)$ is assumed to be $d+1$). So we expect that  a loop equation of length $l_1$ involves quadratic relations of operators with lengths up to $l_1+d$.
 In the next section we will precise all these steps on a particular example of the two-matrix model.

The set of all loop equations can be efficiently  generated by applying the derivatives in \(M=\{A,B\}\) to  any word of the length less than a certain cutoff\footnote{We will discuss the detail of the choice of the cutoff in the Appendix~\ref{imple}.}. However, the loop equations obtained in this way are not all independent, which means that there may exist linear dependence and/or algebraic dependence among them. It turns out that these redundancies are numerically crucial when applying the SDP solver to the constraints of our system, but they are not important at this stage of explanation. We will discuss these technicalities in Appendix~\ref{imple}.

If the model has some discrete symmetries, such as \(M\to -M\) or \(A\leftrightarrow B\),  it is not necessary to  assume them from the beginning in our bootstrap scheme, but factoring  it out will significantly increase our numerical efficiency if we are only interested in the symmetry preserving solution. Generally, the symmetry assumptions not only simplify the loop equation by reducing the number of operators\footnote{For example, if the potential has $\mathbb{Z}_2$ symmetry $A\leftrightarrow B$ , we could identify all the operators identical  by $A\leftrightarrow B$  transformation.}  but in certain cases they make the correlation matrix block diagonal, thus greatly simplifying our problem. We will encounter this situation in the next section for a concrete model.

At last,  we identify all the operators related by cyclicity of trace and the reversion of the word. These transformations also 
reduce considerably the number of unknowns in our scheme.

In summary, for the two-matrix integral~\eqref{general2MM}, assuming the global symmetries or not, we set up all the physical constraints. A natural question is what is the solution of these constraints. But this is not a good question since generally, apart from some solvable models where the loop equations close on a very limited subclass of operators (like in a  two-matrix model~\cite{Kazakov:1989bc,Staudacher:1993xy} or some \(n\)-matrix models~\cite{Kazakov:1987qg,Daul:1994qy,Kostov:1988fy}), the number of operators grows faster than the number of constraints, which means that the solution is a region in an extremely high dimensional space. A constructive question at this stage  is:  given a cutoff to the length \(\le 2\Lambda\), what is the minimal or maximal possible value of a physical quantity? This amounts  to asking what is the allowed interval when the region allowed by the constraints is projected on the linear subspace corresponding to the specific physical quantity. 

Rephrasing it in the language of optimization theory, we  deal with the problem of the form:
\begin{equation}\label{NSDP}
\begin{aligned}
\text{minimize} \quad & c^{\mathrm{T}}x\\
\textrm{subject\,\,to} \quad & x^{\mathrm{T}} \mathcal{A}_i x +b_i^{\mathrm{T}} x+a_i=0\quad (\text{\(i\)'th loop equation)},\\
  \text{and}\quad &M_0+\sum_{j=1}^L M_j x_j\succeq 0
\end{aligned}
\end{equation}
where $c$ is a vector defining the dynamical quantity we want to optimize and $x$ is the column vector of all our operator expectations $x_i=\langle\Tr \mathcal{O}_i\rangle$, up to the length $2\Lambda$. The quadratic loop equation (in the middle) is written in the vector form, where \(\mathcal{A}_i\) is the quadratic form encountered in the \(i\)th equation; linear and constant terms are represented accordingly.   The matrix inequality is the expansion of the  correlation matrix in terms of the operator expectations. This is certainly  not equivalent to the standard SDP which we introduced by~\eqref{SDPp} since the quadratic equations represent non-convex conditions. One of the conventional methods to deal with it is relaxation.

\subsection{Relaxation matrix}

The constraints discussed in the last section define a problem which is called Non-Linear SDP in optimization theory. There are indeed some solvers specialized for it but, from our limited trials, they are not mature enough to solve large-scale problems such as the ones we encountered in matrix bootstrap. To improve the situation, we propose to modify  the problem \eqref{NSDP}  by relaxing the non-convex conditions involving the non-linear loop equations, into  convex ones. Our intuition here is that we don't really need all of the loop equation constraints for our bootstrap method to converge as $\Lambda$ increases. 

To see how our method works, let us begin with a simple example which will provide us with a heuristic argument. Suppose we have only three quadratic ``loop equations'':
\begin{equation}
\begin{cases}
x^2=T_1\\y^2=T_2\\xy=T_3
\end{cases}
\end{equation}
Here $T_i=\sum_{j}\,q_i^j\,w_j,\, (i=1,2,3)$ denote linear combinations of some other variables $w_1,w_2,\dots$. These equations are of course non-convex. But we can relax them to make them convex by replacing $x^2=T_1$ with $x^2\leq T_1$ or, in the positive semi-definite matrix form,
\begin{equation}\label{trial1}
\begin{pmatrix}
1 & x \\
x & T_1 
\end{pmatrix}\succeq 0\,.
\end{equation}
We can do the same thing with the second equation \(y^2=T_2\), to relax it to a convex condition. But the same operation cannot be reproduced for equation $xy=T_3$, since neither $xy\leq T_3$ nor $xy\geq T_3$ is  convex~\footnote{Because the bilinear form $xy$ is not positive semi-definite.}. It is tempting to consider the positive semi-definite combinations:
\begin{equation}\label{trial}
(x+\alpha y)^2\leq T_1+\alpha^2 T_2 +2\alpha T_3,\, \forall \alpha \in \mathbb{R}\,.
\end{equation}
It is not very elegant to implement~\eqref{trial} by  introducing extra parameters like $\alpha$, although  numerically this is viable. Can we write instead of \eqref{trial} a condition that does not contains explicitly $\alpha$? In fact yes. Since $T_1\geq x^2 \vee T_2\geq y^2$ , we only need the discriminant of \eqref{trial} w.r.t. \(\alpha\) to be non-positive, to exclude the existence of real solution for \(\alpha\) when \eqref{trial} becomes an equality.  That means
\begin{equation}
\left(T_1 T_2-T_1 y^2-T_2 x^2-T_3^2+2 T_3 x y\right)\geq 0
\end{equation}
is equivalent to \eqref{trial} for all $\alpha\in \mathbb{R}$.
In its turn, it is  equivalent to:
\begin{equation}\label{trial2}
\mathrm{Det}\begin{pmatrix}
1 & x & y\\
x & T_1 &T_3\\
y & T_3 & T_2
\end{pmatrix}\geq 0.
\end{equation}
Combining  \eqref{trial1} and \eqref{trial2} we come to the conclusion that:
\begin{equation}\label{relax1}
\begin{pmatrix}
1 & x & y\\
x & T_1 &T_3\\
y & T_3 & T_2
\end{pmatrix}\succeq 0.
\end{equation}
This is mathematically more elegant and numerically more efficient. 

To apply this relaxation method to the case of our loop equations is a simple generalization of what we just proposed. We make in the loop equation the substitution $\langle\mathrm{Tr} \mathcal{O}_i\rangle\langle\mathrm{Tr} \mathcal{O}_j\rangle=X_{ij}$, or in matrix notations:
\begin{equation}\label{sub}
    X=x x^\mathrm{T}
\end{equation}
where again $x$ is the column vector whose components are $\langle\mathrm{Tr} \mathcal{O}_i\rangle$. Formally, this changes the loop equations in \eqref{NSDP} to a linear form:
\begin{equation}
    \Tr X \mathcal{A}_i+b_i^{\mathrm{T}} x+a_i=0\,.
\end{equation}
  To apply the relaxation method sketched above, we relax~\eqref{sub} by imposing the inequality:
\begin{equation}
    (\alpha^{\mathrm{T}} x)^2\leq \alpha^{\mathrm{T}} X \alpha,\quad \forall\alpha\in \mathbb{R}^L
\end{equation}
which is equivalent to:
\begin{equation}
    X\succeq xx^{\mathrm{T}}\,. 
\end{equation}
By Schur's complement, this can be re-arranged into a more compact form:
\begin{equation}\label{relaxm}
    \mathcal{R}=\begin{pmatrix}
1 & x^{\mathrm{T}}\\
x & X
\end{pmatrix}\succeq 0.
\end{equation}
Here we introduced the relaxation matrix by $\mathcal{R}_{ij}=X_{ij}$ and $\mathcal{R}_{0i}=\mathcal{R}_{i0}=\langle\mathrm{Tr} \mathcal{O}_i\rangle=x_i$. This step concludes our translation of the nonlinear bootstrap  problem into an SDP. This SDP takes now  a numerically much more tractable, convex form:
\begin{equation}\label{relaxp}
\begin{aligned}
\text{minimize} \quad & c^{\mathrm{T}}x\\
\textrm{such that} \quad & \Tr X \mathcal{A}_i +b_i^{\mathrm{T}} x+a_i=0\,,\\
 \text{and}\quad & M_0+\sum_{j=1}^L M_j x_j\succeq 0\,,    \\
\text{and}\quad &\begin{pmatrix}
1 & x^{\mathrm{T}}\\
x & X
\end{pmatrix}\succeq 0\,.
\end{aligned}
\end{equation}
It has now two types of variables to bootstrap: a column vector variable \(x\) and a symmetric matrix variable \(X\).

Several comments are in order:
\begin{itemize}
\item One of the primary questions to the method is: does the relaxed SDP generate the same bounds as the previous Non-linear SDP problem?  Generally, the answer is ``no''. It is obvious that when the optimal solution of the relaxed problem satisfies the constraint of the original problem the relaxed problem will generate the same bound as the original one. From our experience, this is not the case for any finite $\Lambda$. \footnote{More precisely, if the relaxation is saturated for the optimal solution, we expect that the relaxation matrix will only have one non-zero eigenvalue. But practically, we always observe multiple non-zero eigenvalues for the relaxation matrix.} But as we increase the cutoff $\Lambda$, the mismatches for the quadratic conditions are tending to zero. So we are tempted to believe that for infinite $\Lambda$, the relaxed problem and the original problem give  the same result for most of the questions we are interested in. This  indicates that the non-linear constraints in the loop equations are  somehow contained in the positivity conditions for correlation matrix and relaxation matrix.
\item One can regard our relaxation scheme as a numerical compromise:   doing relaxation we replace the nonlinear equalities by linear inequalities but we can thus explore the correlation matrices of a much higher order since we can significantly increase the length cutoff \(\Lambda\). This enable us to embrace more information from correlation matrix. Our numerical results in the next section will show that this is a worthy trade-off. 
\item There is another point of view on our relaxation formulation \eqref{relaxp}. The problem \eqref{relaxp} is actually the dual of the dual of the problem of \eqref{NSDP}. Although this fact is in principle simple  to show its proof  is quite lengthy, so we put it into the Appendix~\ref{rdual}. In that appendix, we also briefly review the definition and basic facts about the dual formulation. As it is known, the dual problem of any general optimization problem is always convex~\cite{boyd_vandenberghe_2004}, so the double dual is guaranteed to be convex. In some sense, this point of view is more general and universal.
\item We believe that the key condition for the relaxation method to work well is that,  under our bootstrap assumption, there is a unique exact solution.\footnote{Here exact solution means bootstrap solution with infinite cutoff.} Then since a single point (corresponding to the $\Lambda=\infty$ solution of bootstrap) is convex, our relaxation procedure leading to convex constraints will not make the results too different even for a finite but sufficiently large $\Lambda$. However, we  observed in Section~\ref{sec:ana} that the set of exact solutions may become non-convex in the presence of a symmetry breaking. In such situation, we need further assumptions to make the exact solution unique. We will further discuss these aspects  in the next section when bootstrapping the symmetry breaking solutions.
\end{itemize}

\section{Bootstrap for ``unsolvable'' two-matrix model with \texorpdfstring{\(\Tr[A,B]^2\)}{commutator square} interaction }\label{QCSM}

In this section, we implement the relaxation bootstrap method described in the previous section to the case of generically unsolvable large $N$ two-matrix model:
\begin{equation}\label{2MMcom21}
    Z=\lim_{N\rightarrow \infty}\int d^{N^2}A\,d^{N^2}B\,\e^{-N\tr\left( -h[A,B]^2/2+A^2/2+g A^4/4+B^2/2+g B^4/4\right)}
\end{equation}
where the integration goes over Hermitian matrices \(A\) and \(B\).
This model  is unsolvable analytically for generic parameters \(h\) and \(g\), at least with the known methods,  such as reduction to eigenvalues or the character expansion. It is still analytically solvable for some particular values:  for $g=0$ it can be reduced to a specific one-matrix model and solved via saddle point method or via the reduction to a KP equation~\cite{1982PhDT........32H,1999NuPhB.557..413K};   for $h=0$ it reduces to two decoupled one-matrix models; for \(h=\infty\) we have \([A,B]=0\) and it reduces again explicitly to another eigenvalue problem. These particular solvable cases are useful to test the power of our numerical method.

The present section is organized as follows: The bootstrap results for the model~\eqref{2MMcom21} with particular choice of parameters  \(g=1,\, h=1\) (which represent a generic  analytically "unsolvable"  example) are shown in Section~\ref{sec:11}. Then in Section~\ref{anacomp} we compare the bootstrap result for the analytically solvable cases \(h=0\) or \(g=0\) with the corresponding analytic solution, to test our method. In Section~\ref{sec:phase}, we explore the phase diagram of the this model and make several comments about the convergence rate in different regions. At last, in Section~\ref{sec:symb}, we investigate the symmetry breaking in the model by our relaxation bootstrap method.

\subsection{Bootstrap solution for a generic choice of  \texorpdfstring{$g,\,h$}{g, h}}\label{sec:11}

In this subsection, we present the results of the bootstrap for the model~\eqref{2MMcom21} where we specify, for  definiteness, the parameters: $g=1, h=1$. We stress that this choice has nothing specific for the properties of the model and it is made mostly for the demonstrative reasons, as an example of generic values of parameters. The method appears to be very efficient almost everywhere in the physical domain of parameters  $g, h$, except when we  approach the critical lines where it is less efficient. We will discuss in the next subsection the phase structure of the model in the \(g-h\) plane.  

The symmetry of this model can be described by the Dihedral Group \(D_4\)\footnote{Actually we implicitly assume the \(A\rightarrow A^{\mathrm{T}},\, B\rightarrow B^{\mathrm{T}}\) which  basically means that  all  the moments are real. We will assume throughout this paper that this symmetry cannot be broken.  At least intuitively, this is unlikely to happen in our model~\ref{2MMcom21}. }, with generator:
\begin{equation}
    \begin{cases}
    A\rightarrow-A\\
    B\rightarrow-B\\
    A\leftrightarrow B
    \end{cases}
\end{equation}
We saw already on the example of the one matrix model that in the large \(N\) limit there could be a multitude  of saddle point solutions, many of them breaking this kind of symmetries. We  begin with the study of \(D_4\)  symmetric large \(N\) solutions. Later we will discuss the solutions with broken symmetries as well. 

In the fully \(D_4\)   symmetric solutions, only the operators with even number of $A$ and even number of $B$ can be non-vanishing, and we should identify the operators under the exchange \(A\leftrightarrow B\). Obviously this assumption of  \(D_4\)  symmetry  of  solution is in principle not necessary for our bootstrap method to work.  However, assuming this symmetry we gain a lot in the efficiency since we are left with approximately $1/8$ of operators comparing to a general non-symmetric 
setup. It also happens that the  symmetry assumption  simplifies the correlation matrix by much. Namely, when constructing the correlation matrix, only the words with the same $\mathbb{Z}_2$ parity in both $A$ and $B$ can appear in the inner product for a non-vanishing correlator. So our correlation matrix break into 4 block-diagonal matrices, corresponding to  $\mathbb{Z}_2$ parities in $A$ and $B$: even-even, even-odd, odd-even, odd-odd. By $A\leftrightarrow B$ symmetry, the even-odd and odd-even blocks are actually the same. So the original correlation matrix can be reduced to three block diagonal matrices: even-even, even-odd, and odd-even.

Here we bootstrap the allowed region for the first two non-vanishing operators $t_2=\langle \mathrm{Tr} A^2\rangle$, $t_4=\langle \mathrm{Tr} A^4\rangle$\footnote{Here we give up the \(\mathcal{W}_k\) notation for the moments we used in one-matrix model since for two-matrix model the moments cannot be characterize by a single positive number.}. According to~\eqref{relaxp}, this corresponds to setting the objective function of the optimization problem as:
\begin{equation}
\text{Minimize: } t_2\cos \theta+ t_4\sin \theta
\end{equation}
 Scanning it  in the interval  \(0\leq\theta< 2\pi\) we can fix the allowed region for these two operators. Using the general method described in the last section we can use the SDP solvers to solve these problem. The readers interested in the details of the implementation can refer to the Appendix~\ref{imple}, where we gather all the technical detail of numerical implementations. We also demonstrated in Appendix~\ref{example} our numerical procedure explicitly,  step by step, on the example of the system with  $\Lambda=4$ cutoff. 

\begin{figure}[ht]
    \centering
    \includegraphics[width=\textwidth]{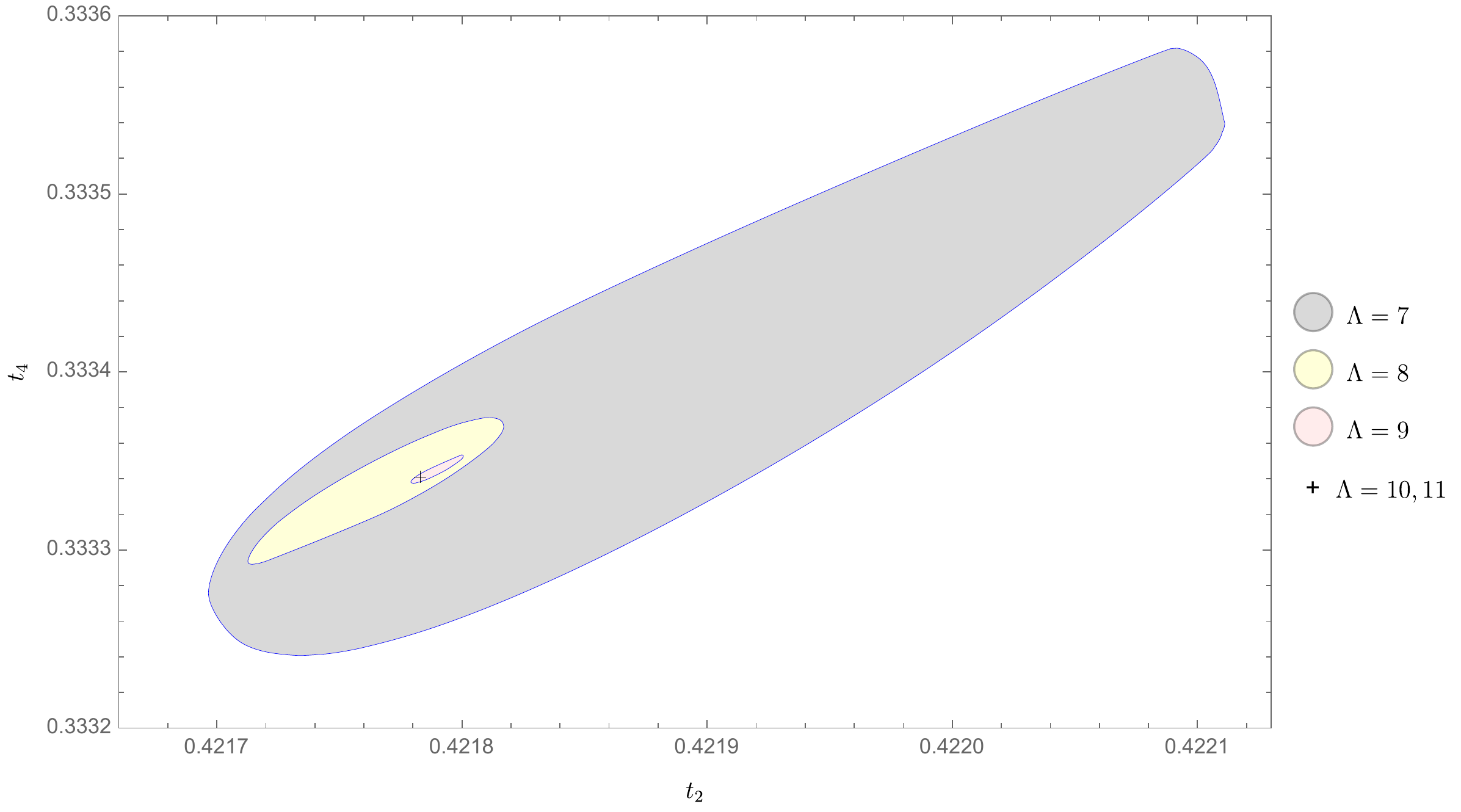}
    \caption{The allowed region of \(t_2-t_4\) of model~\eqref{2MMcom21} with parameter $g=1, h=1$ for the cutoff  \(\Lambda=7,8,9,10,11.\) We recall the definition of \(\Lambda\): the longest operators in the correlation matrix and in the loop equations have the length $2\Lambda$.}
    \label{fig:11region}
\end{figure}

Let us demonstrate our results for various values of the length cutoff  $\Lambda$. We summarized the  allowed regions for the first two correlators \(t_2=\langle \mathrm{Tr} A^2\rangle\) and  \(t_4=\langle \mathrm{Tr} A^4\rangle\)  in Fig~\ref{fig:11region}. The regions for $\Lambda=10$ and $\Lambda=11$ are too small to be plotted on the figure, so we give here the upper and lower bound of $t_2$ and $t_4$. For $\Lambda=10$:
\begin{equation}
    \begin{cases}
    0.421780275\leq t_2 \leq 0.421785491\\
    0.333339083\leq t_4 \leq 0.333343006
    \end{cases}
\end{equation}
and for $\Lambda=11$:
\begin{equation}
    \begin{cases}
    0.421783612\leq t_2 \leq 0.421784687\\
    0.333341358\leq t_4 \leq 0.333342131
    \end{cases}
\end{equation}
We see here that for \(\Lambda=11\) we already  have a six digits precision at $g=h=1$. The $\Lambda=11$ calculation is the largest problem in this work, it is done with SDPA-dd, a solver in SDPA family with the double-double float type. The input to SDPA has $95$ variables, with the correlation matrix size: even-even $683$, odd-odd $682$, even-odd $1365$,  and with relaxation matrix size $8$. We note that this is still within the capability of a single laptop, it only takes $150000s$ CPU time for a single maximization cycle. We also stress that these inequalities, unlike the Monte Carlo methods, are exact: increasing the cutoff \(\Lambda\) we can only improve the margins.

\subsection{Demonstration for analytically solvable cases}\label{anacomp}

It is instructive to apply our numerical method to the analytically solvable cases $g=0$ or $h=0$, which is  a good check for our approach, convincing us that it works well indeed  even for the generic parameters, where we have no analytic data to compare with. In this part we will firstly review the analytic solution for both cases and then compare it with the numerical results of our relaxation bootstrap method.

As we mentioned, for $h=0$ this model reduces to two decoupled one-matrix models -- the case which  we already discussed and studied analytically in Section~\ref{sec:ana}. Integrating out one of the decoupled matrices, we expect the operator containing  only one matrix to have exactly the same expectation value as for the result in Section~\ref{sec:ana}:
\begin{equation}\label{eq:quartic}
    g_c=-\frac{1}{12},\quad t_2=\frac{(12 g+1)^{3/2}-18 g-1}{54 g^2}.
\end{equation}

For $g=0$, this model is already solved analytically in \cite{1982PhDT........32H,1999NuPhB.557..413K}. Here we simply present the analytic solution derived there in our notations and normalization. To have a compact form, we introduce the short-hand notations \(E=E(m)\), \(K=K(m)\), \(\vartheta=E/K\), where K and E are the complete elliptic integrals of first and second kind:
\begin{equation}
    K(m)=\int_0^{\pi/2}\frac{\d\theta}{\sqrt{1-m^2 \sin^2 (\theta)}},\quad E(m)=\int_0^{\pi/2} \sqrt{1-m^2 \sin^2 (\theta)} \d\theta.
\end{equation}
We introduce the new parameter \(m\) related with \(h\) by:
\begin{equation}\label{hm}
    h(m)=\frac{K \left((m-1) -2 (m-2) \vartheta-3 \vartheta^2\right)}{6 \pi ^4},
\end{equation}
and we can express \(t_2\) as:
\begin{equation}\label{KKNeq}
    h(m)t_2(m)=\frac{1}{12}-\frac{K^2 \left(-(m-2) (m-1) +10 (m-2) \vartheta^2+2 ((m-6) m+6)
    \vartheta+10 \vartheta^3\right)}{5 \pi ^2 \left(-(m-1) +2 (m-2) \vartheta+3
   \vartheta^2\right)}.
\end{equation}

 This formula is valid when $h>0$. For $h<0$, we need to analytically continue the solution to the other sheet of Riemann surface of the variable \(m\). For that we introduce the analytic continuation of the elliptical integral \(K(m)\) and \(E(m)\):
 \begin{equation}
\begin{split}
    &K_a=K_a(m)=\frac{K\left(\frac{1}{m}\right)+i K\left(1-\frac{1}{m}\right)}{\sqrt{m}},\\&E_a=E_a(m)=\frac{-(m-1) K\left(\frac{1}{m}\right)+i K\left(\frac{m-1}{m}\right)+m E\left(\frac{1}{m}\right)-i m E\left(\frac{m-1}{m}\right)}{\sqrt{m}},\\
    &\vartheta_a=E_a/K_a.
\end{split}
\end{equation}
To make ~\eqref{hm} and~\eqref{KKNeq} valid for \(h<0\), we simply replace all the \(K,\, E,\, \vartheta\) by \(K_a,\, E_a,\, \vartheta_a\).
 
 The critical point of the smallest possible \(h_c\) for \(h<0\) can be defined as the solution of the equation\footnote{We thank Nikolay Gromov for sharing with us his computation of \(h_c\).}:
\begin{equation}
    \frac{d h(m)}{d m}=0
\end{equation}
which can be numerically solved as:
\begin{equation}
    h_c\approx-0.04965775;\, t_{2c}\approx1.18960475.
\end{equation}

\begin{figure}
\begin{subfigure}[bc]{.9\textwidth}
         \centering
         \includegraphics[width=\textwidth]{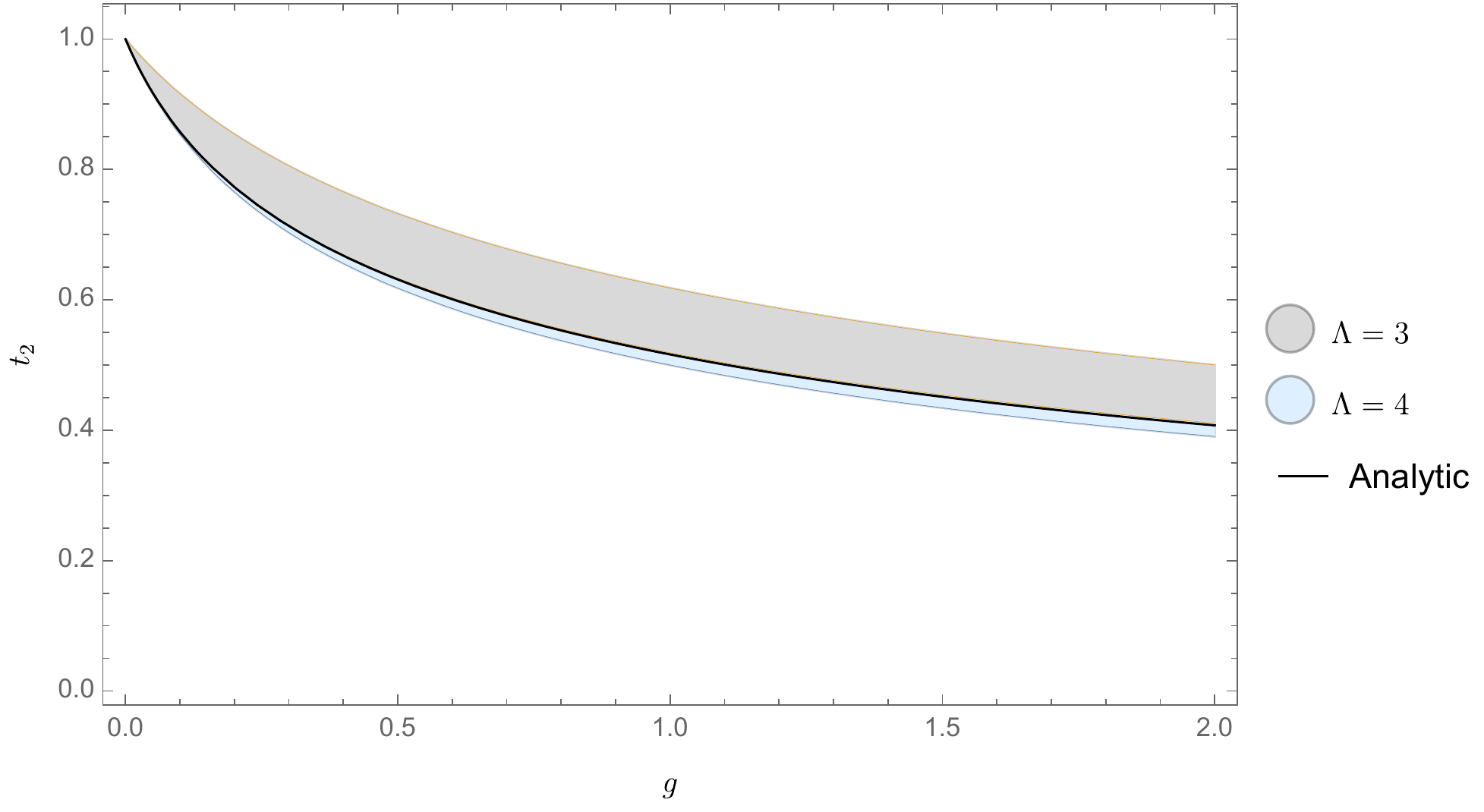}
         \caption{$g>0$}
         \label{fig:ganaplus}
         \end{subfigure}
         \begin{subfigure}[bc]{\textwidth}
         \centering
         \includegraphics[width=.9\textwidth]{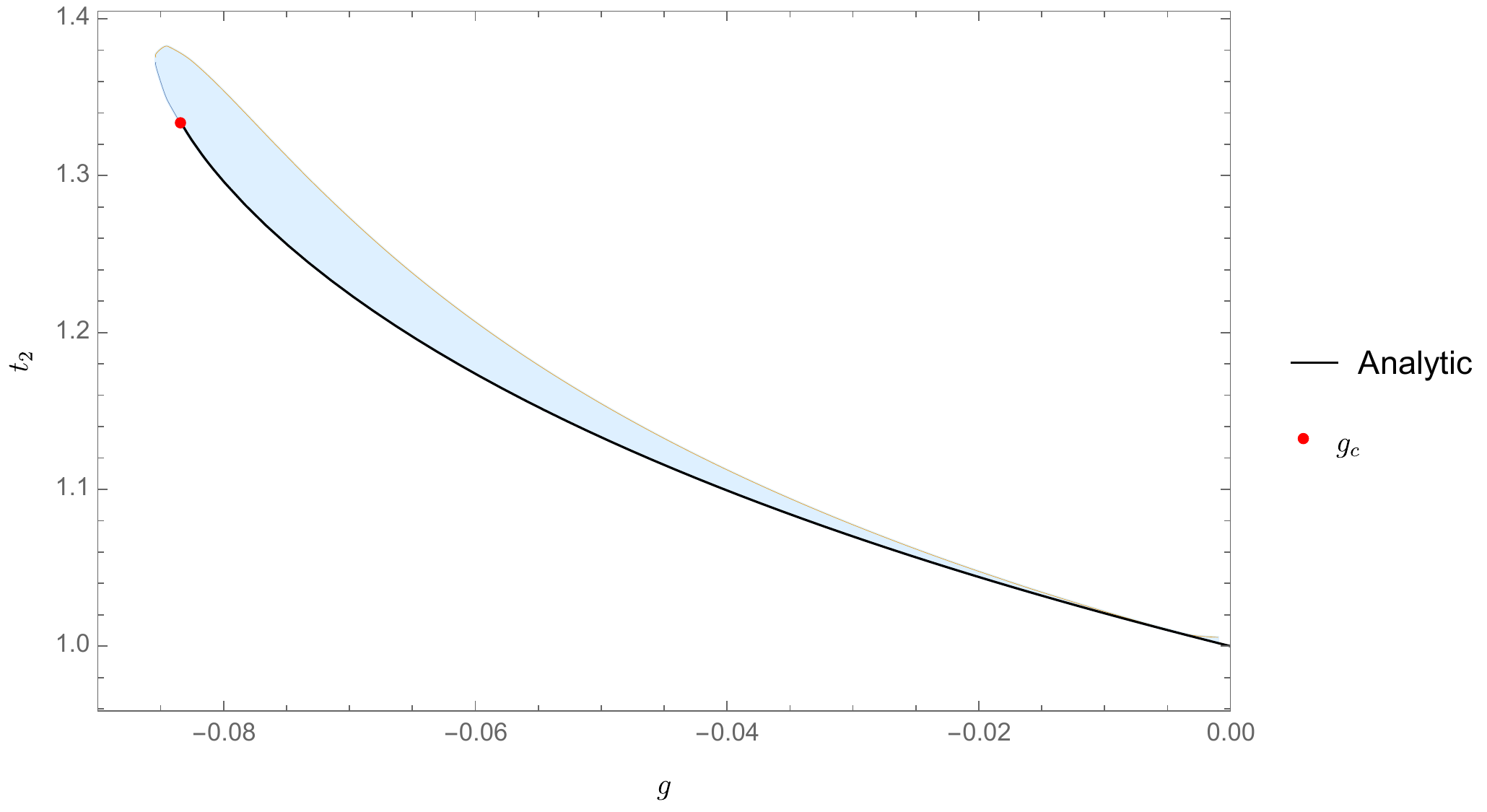}
         \caption{$g<0$}
         \label{fig:ganaminus}
         \end{subfigure}
    \centering
    \caption{Comparison with the exact analytic solution of model~\eqref{2MMcom21} with \(h=0\), i.e. two decoupled quartic one-matrix model. The lower plot is  for $\Lambda=8$.}
    \label{fig:gana}
\end{figure}

\begin{figure}
\begin{subfigure}[bc]{\textwidth}
         \centering
         \includegraphics[width=.9\textwidth]{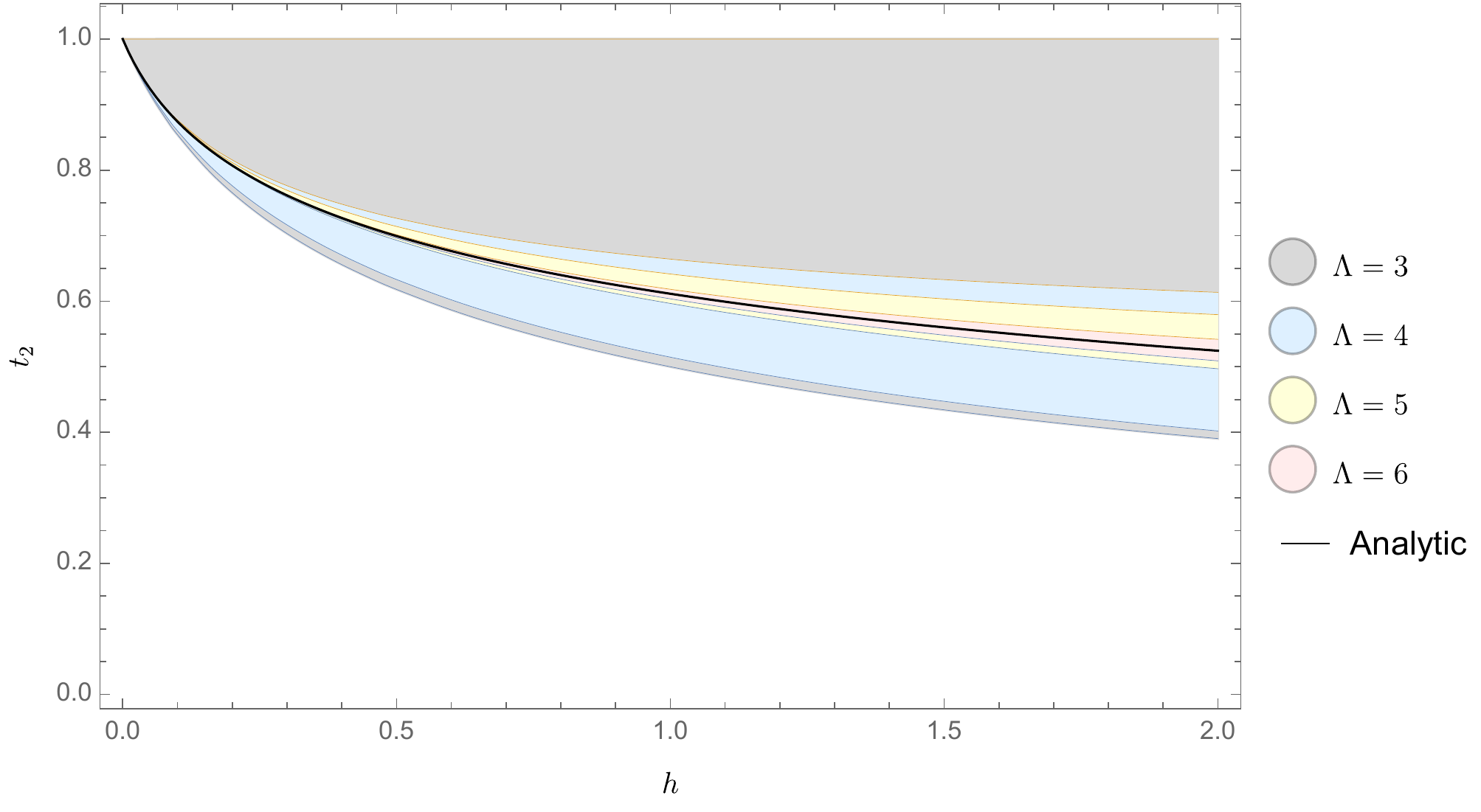}
         \caption{$h>0$}
         \label{fig:hanaplus}
         \end{subfigure}
         \begin{subfigure}[bc]{\textwidth}
         \centering
         \includegraphics[width=.9\textwidth]{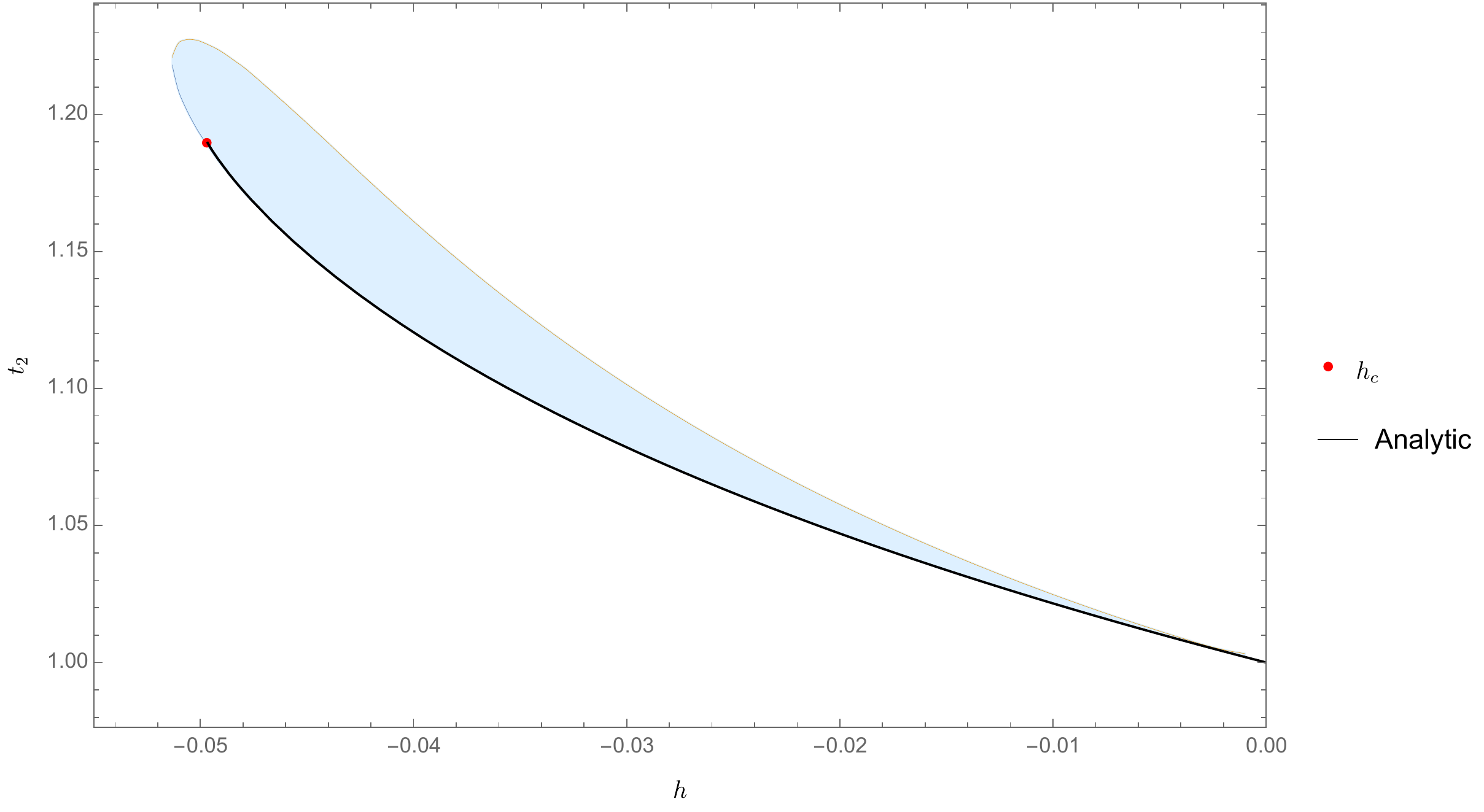}
         \caption{$h<0$}
         \label{fig:hanaminus}
         \end{subfigure}
    \centering
    \caption{Comparison of the numerical bootstrap results with the exact analytic solution of the model~\eqref{2MMcom21} with \(g=0\). The lower plot is  for $\Lambda=8$. }
    \label{fig:hana}
\end{figure}

The comparison of our numerical results with analytic result is presented on Fig~\ref{fig:gana} and Fig~\ref{fig:hana}. Indeed, we see that our numerical results nicely agree with  the analytic formula \eqref{eq:quartic} and \eqref{KKNeq}. An apparent feature of these plots is that when $g<0$ or $h<0$, the allowed region is much larger than the one for the positive coupling case, thus giving less of precision. In general we have the worst convergence  in the neighborhood of critical value. We will discuss this feature in more details in the next subsection. 

There is another fact which is not obvious from the Fig~\ref{fig:gana}. If we compare this figure with Fig~\ref{fig:re} in Section~\ref{sec:review} we will find that for same values of \(\Lambda\), the non-relaxed one-matrix bootstrap bound for \(t_2\) (denoted by \(\mathcal{W}_2\) in that section) and our relaxation bootstrap  bound for $h=0$ case of the 2-matrix model actually coincides within the error bar. This is a very striking feature of our relaxation method since  we relaxed all the quadratic equalities to inequalities but we compensated this with many more mixed operators of two decoupled matrices. So the correlation matrix is much larger in the relaxed case and the final results are basically the same. We will see this feature of relaxation again when we discuss later the bootstrapping of the symmetry breaking solutions. We don't have a very clear explanation for these phenomena in general.

\subsection{Phase Diagram and convergence rate}\label{sec:phase}

In this part, we will discuss the phase diagram of the matrix  model \eqref{2MMcom21} and the corresponding convergence rate in different regions of the diagram. 

In general, for finite $N$ matrix integral the potential $V(A,B)$ must be bounded from below to define a  sensible  integral over Hermitian matrices. But this is not necessary for a large $N$ theory, where we only need deep enough local minima to have a stable saddle point solution. Even for the unstable potentials, the tunnelling effects between the local minima, or to the infinity are suppressed exponentially.  We saw this in Section~\ref{anacomp}, where the bootstrap procedure allowed the existence of solutions with negative values of $g$ and $h$. This provides us with a possibility to study the boundaries of possible $g$ and $h$ values (we will call the region of possible $g$ and $h$ values the feasible region in the following) even when the corresponding potential is not bounded from below.

Before going deeper into the technicalities of bootstrapping the boundaries of the  feasible region, we can get a rough estimate of them by deriving the parameter region of \(g\) and \(h\) which leads to the matrix potential bounded from below. It is obvious that the domain \((h\geq 0,\, g\geq 0)\) is one part of the region we are looking for. Another, less obvious part is \((h<0,\, g\geq -4h)\), as in this case we should have:
\begin{equation}
\begin{medsize}
\begin{split}
\tr V(A,B)&=\tr \left( -h[A,B]^2/2+A^2/2-h (A^4+ B^4)+B^2/2+(g+4h)(A^4/4+B^4/4)\right)\\
&=\tr \left( -h((AB+BA)^2/2+(A^2-B^2)^2)+A^2/2+B^2/2+(g+4h)(A^4/4+B^4/4)\right)\\
&\geq 0\,.
\end{split}
\end{medsize}
\end{equation}
The union of these two domains represents the maximal region where the matrix potential is bounded from below, since for \((h\geq0,\, g<0)\) and \((h<0,\, g< -4h)\) we can always find \(A,\, B\) configurations where the potential is not bounded from below. For \((h\geq0,\, g<0)\), one of these configurations is taking $B=0$ and $A\rightarrow \infty$. For \((h<0,\, g< -4h)\), we simply put $A$ and $B$ to be some constants $\alpha$ times generalized Pauli matrices of dimension \(N\) $\alpha \sigma_1$ and $\alpha \sigma_2$, where $\alpha$ is a large real number. Then we have:
\begin{equation}
    \tr V(\alpha \sigma_1,\alpha \sigma_2)=N(\alpha ^2+(g+4h)\alpha ^4)\,.
\end{equation}
This must be unbounded from below when \((h<0,\, g< -4h)\).

In conclusion,  the region of potential bounded from below is \((g\geq0\bigcap g\geq-4h)\). 
 In addition,  the domain \((g\geq0\bigcap g\geq-4h)\) is guaranteed to lie within the feasible region. But due to the large \(N\) effects, we expect the feasible region to be a little bigger than that. Specifically, for analytically solvable cases, when \(h=0\) we have \(g\geq-\frac{1}{12}\) and when \(g=0\) we have \(h\gtrsim-0.04965775\). These facts give us  an additional information about the location of the boundary of the feasible region.

\begin{figure}[ht]
    \centering
    \includegraphics[width=1.0\textwidth]{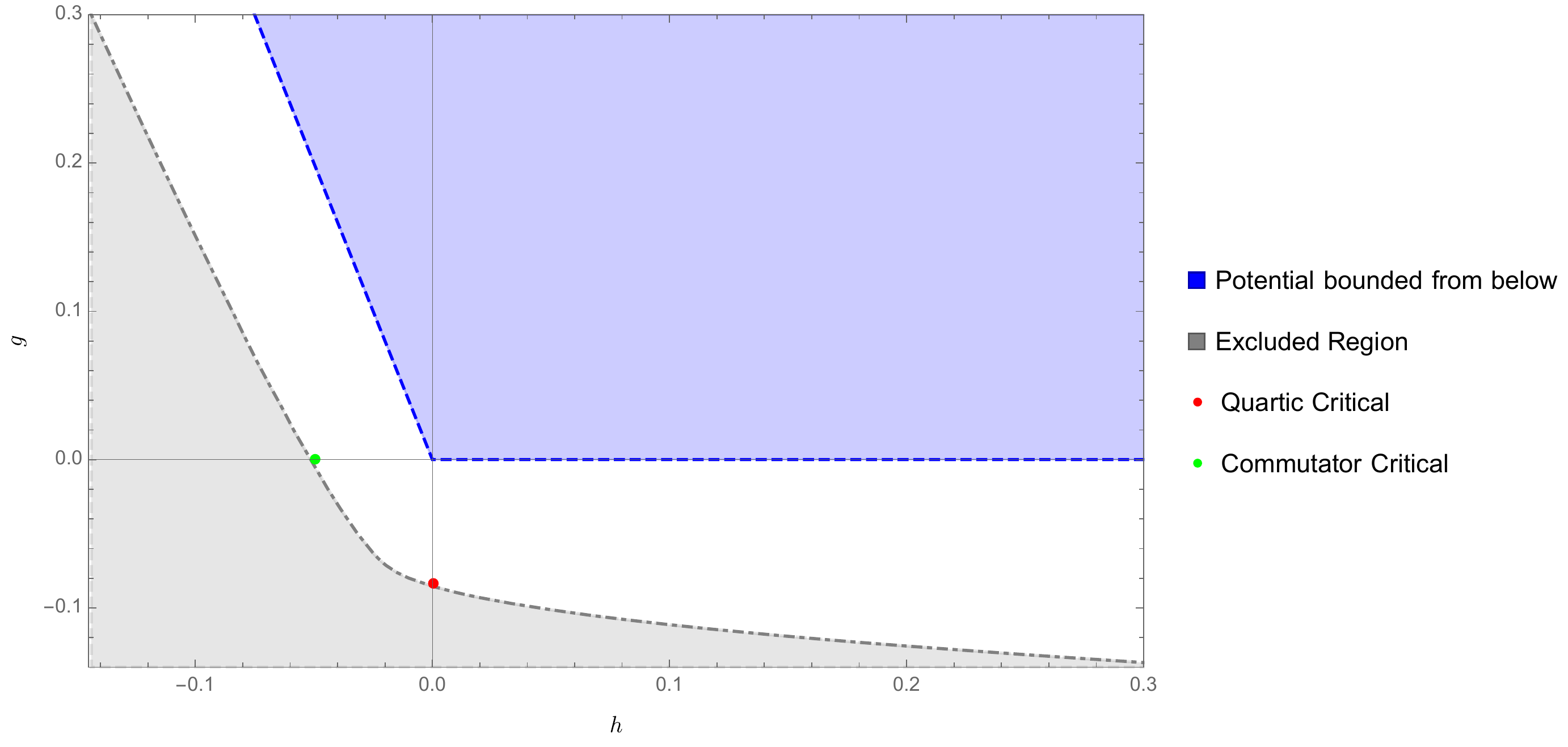}
    \caption{The numerical phase diagram of the model~\eqref{2MMcom21}. The gray region is strictly excluded by our relaxation bootstrap method at $\Lambda=8$. In the  blue region  the matrix potential is bounded from below and its boundary is located above the straight lines \(g=0\) and \(h=-\frac{1}{4}g\). The red and green dots are the critical valus for \(h=0\) and \(g=0\).}
    \label{fig:region}
\end{figure}

To numerically bootstrap the boundary of the feasible region, we can obtain the critical boundary between the allowed and forbidden parameter regions by bisection. Namely, for a given \(\Lambda\) we fix $h$ and take two values of $g$, as $g_1$ and $g_2$. Here $g_1$ is a point that is guaranteed to be forbidden for a given \(h\), and $g_2$ is a point that is guaranteed to be allowed. Then we test the geometric average value \(g_m=\frac{g_1g_2}{g_1+g_2}\). If $g_m$ is allowed, then we make the substitution $g_2=g_m$, otherwise we take $g_1=g_m$. In this way we can recursively approach the maximal forbidden value of \(g\) at fixed $h$. Then we scan over the values of $h$ and get the plot shown in Fig~\ref{fig:region}.

Some  explanations for the plot Fig~\ref{fig:region} are in order. The gray region is  rigorously forbidden as the result of  bootstrap at $\Lambda=8$. On the contrary, the white region is not guaranteed to be allowed for any physical large $N$ solution. As we increase $\Lambda$, the gray region will expand a little. But we  have several hints about the position of the exact boundary line:
\begin{itemize}
    \item We notice the red and green dots on the plot, which are the critical points of the analytic solutions. They are  located on the exact boundary of feasible region, i.e. no matter how large is $\Lambda$, the gray curve cannot go beyond these two points. From this fact we convince ourselves that our numerical curve in Fig~\ref{fig:region} is already very accurate, since the red dot and the green dot are very close to the gray curve.
    \item The blue region is the region where the potential is strictly bounded from below. It is enclosed by the lines $g=-4h$ and $g=0$. Its boundary  can be considered as the exact solution in the ``classical'' limit \(\hbar\to 0\) for this matrix integral, where \(1/\hbar\)  is the coefficient put in front of the potential. In this case we scale  the couplings   as $h\to 1/\hbar,\,\, g\to 1/\hbar$. The boundary of the gray  will coincide for \(\hbar\to 0\) with the boundary of the blue area on Fig~\ref{fig:region}. Then inside the blue area we have a well-defined theory even for finite $N$. At large $N$ and finite $h,g$ there is a  gap between blue region and gray region, as is visible on the Fig~\ref{fig:region}.
\end{itemize}
\subsubsection{Rate of convergence}

\begin{figure}
    \centering
\begin{subfigure}[bc]{\textwidth}
         \centering
         \includegraphics[width=\textwidth]{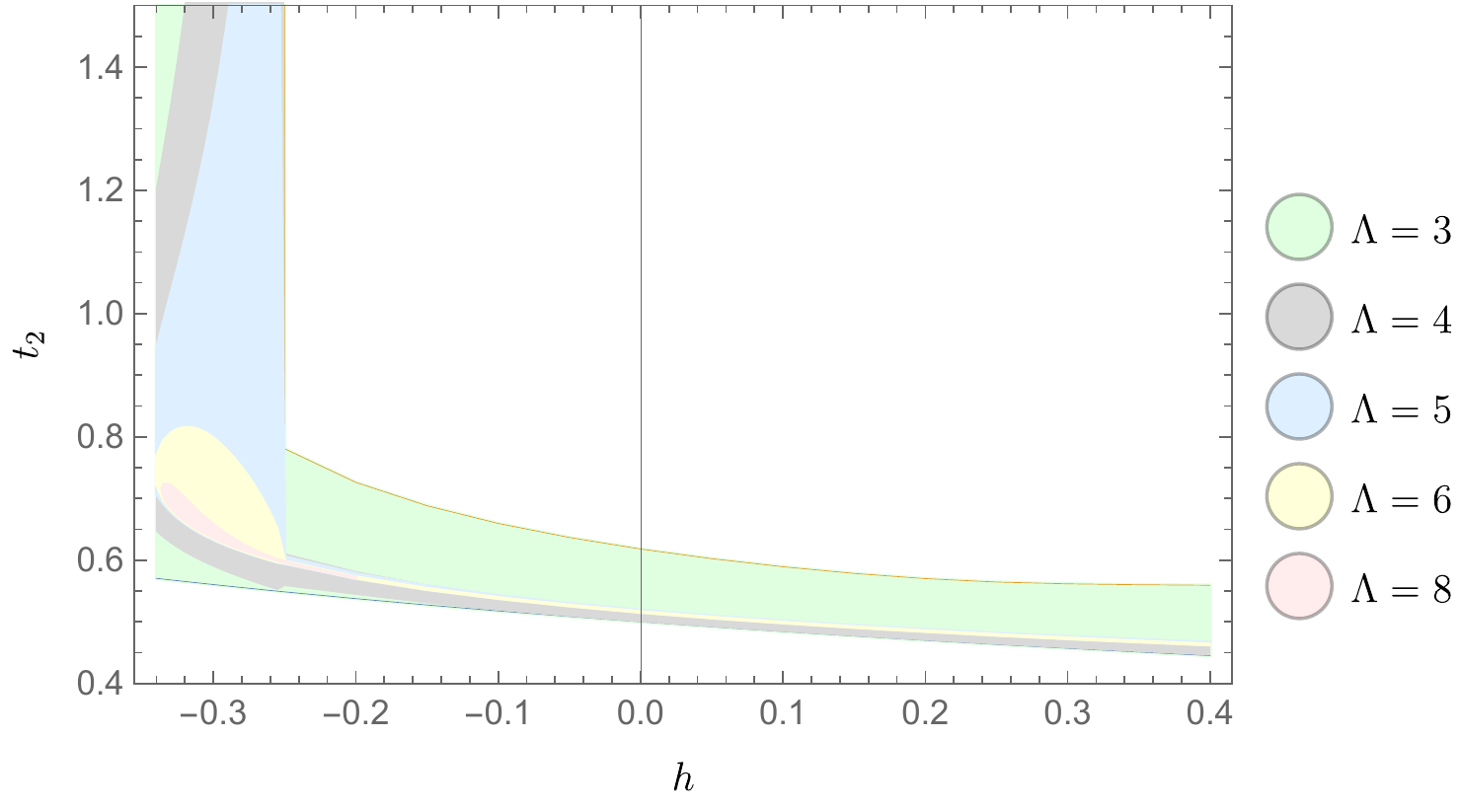}
         \end{subfigure}
    \caption{The allowed region for \(t_2\) when we fix $g=1$ and vary $h$.  In the region where the potential is bounded from below i.e. \(h\geq-\frac{1}{4}\), we have a decent convergence whereas for \(h<-\frac{1}{4}\) the convergence gets much slower.  }
    \label{fig:g1}
\end{figure}

As the reader may have noticed already in Section~\ref{anacomp}, when $g<0$ or $h<0$ the convergence is very bad  compared to the  case \(g>0\) and \(h>0\). From our experience, this is a generic situation when we are outside of the blue region in Fig~\ref{fig:region}, which is defined by the region of parameters yielding a potential bounded from below. For example, Fig~\ref{fig:g1} depicts the allowed region for $t_2$ when we fix $g=1$ and scan over $h$ in the neighborhood of $h=-1/4$. It is clear from this figure that for $h<-1/4$ there is drop in the rate of convergence. Actually, from numerical data, the difference of the upper bound and the lower bound varies between the orders of magnitude from $10^{-4}$ to around $10^{-2}$ when \(h\) varies from $h=-0.25$ to $h=-0.26$.

Nonetheless we can get a rather accurate estimate of physical quantities in the region discussed in the last paragraph. We note that in Fig~\ref{fig:gana} and Fig~\ref{fig:hana}, the analytic solution is very close to the lower bound, comparing to the upper bound\footnote{We believe that the upper bound and the lower bound converge to the same value, but it seems they have rather different convergence behaviors. }. Actually, as we increase $\Lambda$, the lower bound stabilizes already at  rather small \(\Lambda\). Empirically this is a typical behavior in the unbounded region. Under the assumption that there is  a unique solution satisfying the constraint for arbitrarily large \(\Lambda\), we expect that the optimization results for the  maximum and the minimum of $t_2$ will ultimately converge with increasing \(\Lambda\) to the same value. This has been proven for some parameters of the one-matrix model in Section~\ref{sec:ana}, and we have strong numerical evidence to believe it will hold for our model~\eqref{2MMcom21} as well. So we can simply bootstrap the physical quantities by the minimization of \(t_2\) in this region (in the following, we will call this procedure the minimization scheme  as opposed to the maximization scheme). Comparing it to the analytically solvable particular cases we learned that this method can yield especially accurate estimate of  physical quantities. However, we lost the rigorous margin in the region with good convergence  (blue region in Fig~\ref{fig:region}). \footnote{This situation is similar to that of the early days of conformal bootstrap when people used the kink of a plot to estimate the dimension of operators in the \(3d\) Ising model, c.f.~\cite{ElShowk:2012ht}}

\begin{figure}[ht]
    \centering
    \includegraphics[width=1.0\textwidth]{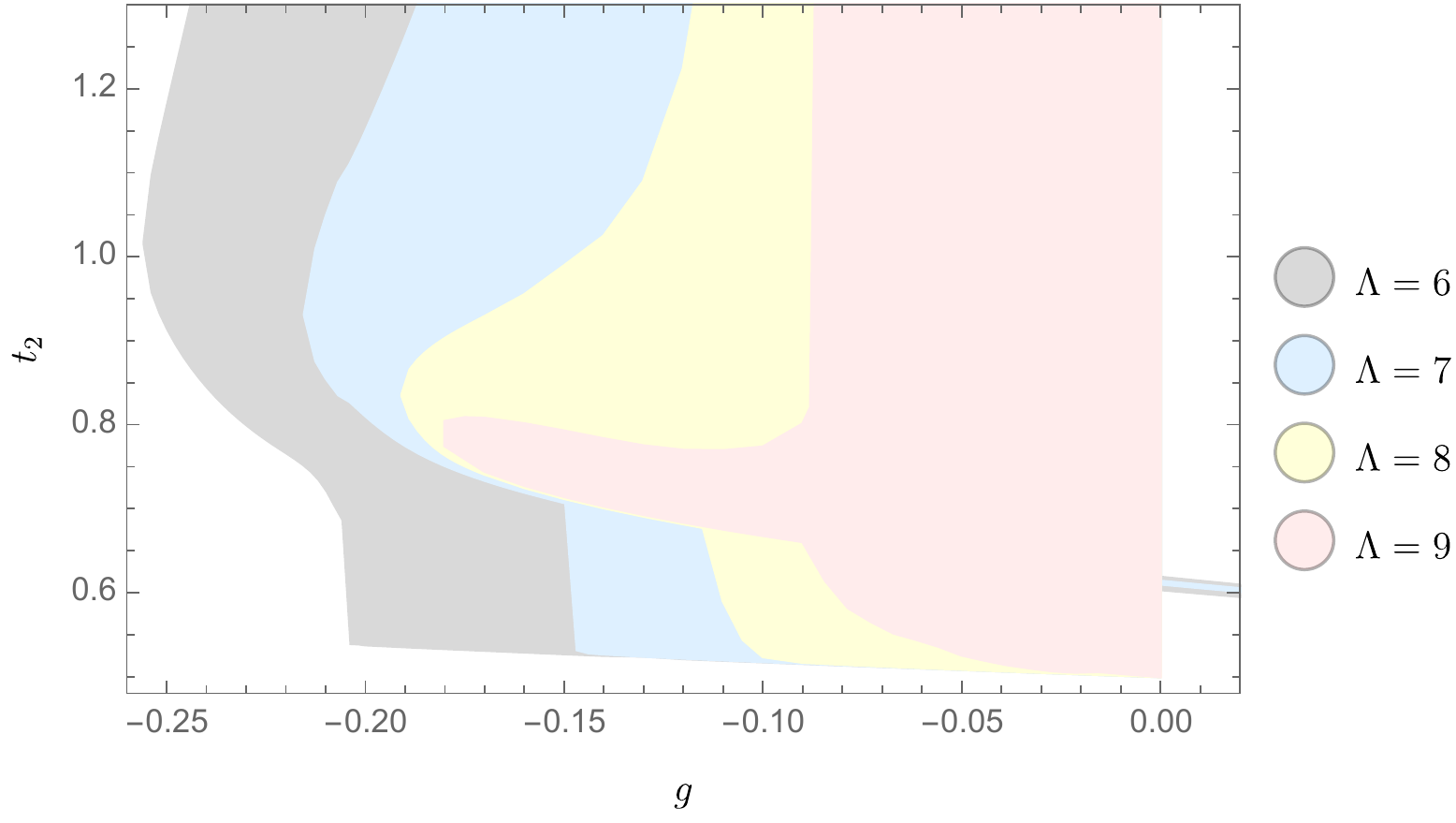}
    \caption{The allowed region for the fixed \(h=1\) and varying \(g\). For the lower bound of the pink region \(\Lambda=9\), there are some numerical instabilities at \(-0.09<g<0\). From careful inspection of our data at various values of \(\Lambda\) it seems that the lower bound at \(\Lambda=9\) should stabilize in this region at the value \(t_2\simeq 0.5\) if no numerical instabilities happened in our SDP solver.}
    \label{fig:bad}
\end{figure}

There exists a region in the phase diagram Fig~\ref{fig:region} where the bootstrap is valid only for very high cutoff \(\Lambda\): it is \(g<0,\, \abs{g}\ll h\). The Fig~\ref{fig:bad} shows the allowed region when we fix \(h=1\) and vary \(g\). For the lower bound of pink region \(\Lambda=9\), there are some numerical instabilities for \(-0.09<g<0\). From careful inspection of our data at various values of \(\Lambda\) it seems that the lower bound at \(\Lambda=9\) should stabilize in this region at the value \(t_2\simeq 0.5\) if no numerical instabilities happened in our SDP solver. We notice a few very distinguishable features of this plot:
\begin{enumerate}
    \item For a fixed \(\Lambda\), there is a region where \(t_2\) is slightly larger than \(0.5\) and not bounded from above. In  other words, the dual SPD problem for the upper bound is infeasible. In this bad region of parameter space, the bootstrap with such \(\Lambda\) essentially tells us nothing about the right physical values. Luckily, the ``bad'' region is shrinking when we increase \(\Lambda\), and hopefully it will disappear when we have a high enough cutoff.
    \item As already stated in the last paragraph,  when we are not in the ``bad region'', the minimization scheme converges much faster than the maximization scheme. So for a reasonable estimate of the operator expectation we should privilege the minimization scheme.
    \item We also notice that for the region \(g>0,\, h>0\), the convergence is excellent as expected, but there is a huge drop in the rate of convergence in the neighborhood of \(h=0\).
\end{enumerate}

\subsection{Bootstrapping the symmetry breaking solution}\label{sec:symb}

In the previous parts of this section, we always assumed the global symmetry, or in other words, we bootstrapped the symmetry preserving solutions. Here in the following, we will make the first attempt to study the symmetry breaking solutions with our relaxation bootstrap method. Consequently, in this subsection we will not make assumptions on a specific global symmetry of operator expectations. For example, we will assume that it is possible to have:
\begin{equation}
    t_1=\langle \mathrm{Tr} A\rangle=  \langle \mathrm{Tr} B\rangle\ne 0
\end{equation}
and any other nonzero expectations containing odd number of letter $A$ or $B$, unlike the solutions with such \(A\rightarrow -A,\, B\rightarrow -B\) symmetry. 

To understand the general features of symmetry breaking solutions, Fig~\ref{fig:num} in Section~\ref{sec:ana} is a good source for our intuition.  We see on that figure that the exact solution is not  unique anymore but there is rather  a continuous family of solutions  parametrized by \(t_1\). This is a non-convex set  of exact solutions, so we don't expect that our relaxation bootstrap method, as  applied in the case of Fig~\ref{fig:11region}, will converge to such a non-convex set as \(\Lambda\) increases. Namely,  if we impose the relaxation bootstrap constraint without the assumption of \(A\rightarrow -A,\, B\rightarrow -B\) symmetry, then  minimize the value of
\begin{equation}
    t_1 \cos \theta +t_2\sin \theta
\end{equation}
and then  scan over \(\theta\) in \([0,2\pi)\), we expect to get a convex set instead of the non-convex one, due to the convex nature of the relaxation method.

So to bootstrap the symmetry breaking solutions, new techniques are needed to tackle the non-convexity. We will describe the general method for bootstrapping solutions and then we apply it to the study of our model~\eqref{2MMcom21}.

\subsubsection{Schemes for symmetry breaking bootstrap}

 The main problems in the study of symmetry breaking solutions in the multi-matrix model of the type considered here are:
\begin{enumerate}
    \item How to identify  the range of parameters for which the model has a possible symmetry breaking solution?
    \item How to numerically bootstrap the symmetry breaking solution?
\end{enumerate}

The answer to the first problem is quiet straightforward. We can establish the relaxed constraint without the symmetry assumption, and bootstrap a dynamical quantity which signals the symmetry breaking. For example, for the \(A\rightarrow -A,\, B\rightarrow -B\) symmetry breaking solution  we take the objective function (\(c^\mathrm{T} x\) in~\eqref{relaxp}) as:
\begin{equation}
    t_1=\langle \mathrm{Tr} A\rangle
\end{equation}
and for the \(A\leftrightarrow B\) symmetry breaking we take the objective function as\footnote{Here in these two situations the dynamical quantity signaling the symmetry breaking is respectively \(\langle \mathrm{Tr} A\rangle\) and \(\langle \Tr A^2\rangle -\langle \Tr B^2\rangle\).}:
\begin{equation}
    \langle \Tr A^2\rangle -\langle \Tr B^2\rangle.
\end{equation}
If the bound of the symmetry breaking expectation is significantly larger than the error bar at the current \(\Lambda\) for a given value of parameters, we believe that this is a strong signal of existence of a symmetry breaking solution.

For the second problem, we propose to transform the non-convex set of exact solutions to a convex one, which means that for the case of Fig~\ref{fig:num}  we fix \(t_1\) by \(t_1=t_{1}^{(0)}\) in our bootstrap procedure. For this  particular value of $t_1$ we should have at infinite cutoff $\Lambda$  a unique exact solution for \(t_2\) and for other higher moments, which is definitely a convex set. Therefore our relaxation bootstrap method with a finite cutoff \(\Lambda\) will yield a rigorous upper bound and lower bound for \(t_2\). Next we scan over \(t_{1}^{(0)}\) until such values that the problem becomes infeasible. In this way we get the allowed region in \(t_1,t_2\) plane.

The above method is easily generalizable to the problem of bootstrapping solutions with the other symmetry breaking patterns. Namely, we establish the bootstrap scheme by fixing the dynamical quantity signaling the symmetry breaking, and then we bootstrap the quantities we are interested in. At this step, we expect that after fixing such dynamical quantity, the exact solution of the bootstrap problem is unique. At the next step, we scan over all possible values of the quantity which was fixed in the previous step. In this way we can bootstrap a non-convex set of solutions.

There is another possibile solution  for the first problem, i.e. to locate the symmetry breaking region. We can assign to the dynamical quantity signaling the symmetry breaking a specific value and then use the method  similar to that of Section~\ref{sec:phase}, i.e. using a bisection to approach the maximal possible value of expectation signaling the symmetry breaking. In principle this bisection method could have given us a tighter bound than our initially proposed method. However, from our test, the two methods yield basically the same numerical result, so we will not bother to use the bisection method in what follows.

\subsubsection{Numerical results for symmetry breaking solution}

Here we apply the method proposed above to the model~\eqref{2MMcom21}. Our results in this part  concern  the breaking of the following symmetries:
\begin{equation}
    A\rightarrow -A,\, B\rightarrow -B
\end{equation}
and 
\begin{equation}
    A\leftrightarrow B.
\end{equation}
In the bootstrap setup, we don't impose the global symmetry assumptions for the corresponding symmetries, i.e. that the non-singlet operator expectations of the \(\mathbb{Z}_2\) symmetry vanish. Then we pick up the dynamical quantities signaling the symmetry breaking as:
\begin{equation}
    \langle \Tr A\rangle
\end{equation}
and 
\begin{equation}
    \langle \Tr A^2\rangle -\langle \Tr B^2\rangle,
\end{equation}
respectively and set them as the objective functions in the corresponding bootstrap problem.

As the result, in the feasible region of Fig~\ref{fig:region} we didn't find any evidence of the existence of a symmetry breaking solution for the model ~\eqref{2MMcom21}. We tried several points in different regions of Fig~\ref{fig:region}. The results show that the maximized values are always lying within the error bar (typically \(10^{-3}\) and \(10^{-4}\), depending on the cutoff \(\Lambda\) and the parameters \(g\) and \(h\)). In particular, for \(\Lambda=8\) and some generic values of \(g\) and \(h\), we have:
\begin{equation}
    -10^{-4} \lesssim \langle \mathrm{Tr} A\rangle, \langle \mathrm{Tr} A^2\rangle -\langle \mathrm{Tr} B^2\rangle\lesssim 10^{-4} \,.
\end{equation}

 We believe this to be a strong evidence that the two symmetries we investigated are not spontaneously broken for all the regions in Fig~\ref{fig:region}.

Some other interesting facts:
\begin{enumerate}
    \item For \(g=0\), i.e. when the quartic coefficient vanishes, the preservation of symmetry is automatic from the loop equation. This fact provides us with the intuition that the commutator square interaction is to some extent not a symmetry-breaking interaction. Regarding that at \(h=0\) the model is not in symmetry breaking phase, since it reduces to two decoupled one-matrix models, intuitively it points on the absence of symmetry breaking phase the for model~\eqref{2MMcom21} (with positive coefficients in front of quadratic terms).
    \item For the region \(h>0\) and \(g\) slightly smaller than zero, we have a very large upper bound for the exposed quantities, sometimes of order \(10\), which might signal the symmetry breaking. But we note that the bootstrap convergence is really bad in this region where some bootstrap results for symmetry preserving solution are presented on Fig~\ref{fig:bad}, and the error bar here is almost infinitely big. So we believe this cannot be a reliable evidence that there  a symmetry breaking takes place in this region.
\end{enumerate}

As we don't find evidence for the existence of symmetry breaking solutions for the model~\eqref{2MMcom21}, we consider the same model but with negative coefficients in front of quadratic terms:
\begin{equation}\label{2MMcom23}
    Z=\lim_{N\rightarrow \infty}\int d^{N^2}A\,d^{N^2}B\,\e^{-N\tr\left( -h[A,B]^2/2-A^2/2+g A^4/4-B^2/2+g B^4/4\right)}.
\end{equation}We know from the Section~\ref{sec:ana} that for \(h=0\) where we have just two decoupled one-matrix models, we have a symmetry breaking phase for \(g<1/4\). At such values of \(g\) we can test our method for bootstrapping the symmetry breaking solutions. 

\begin{figure}
\begin{subfigure}[bc]{.48\textwidth}
         \centering
         \includegraphics[width=\textwidth]{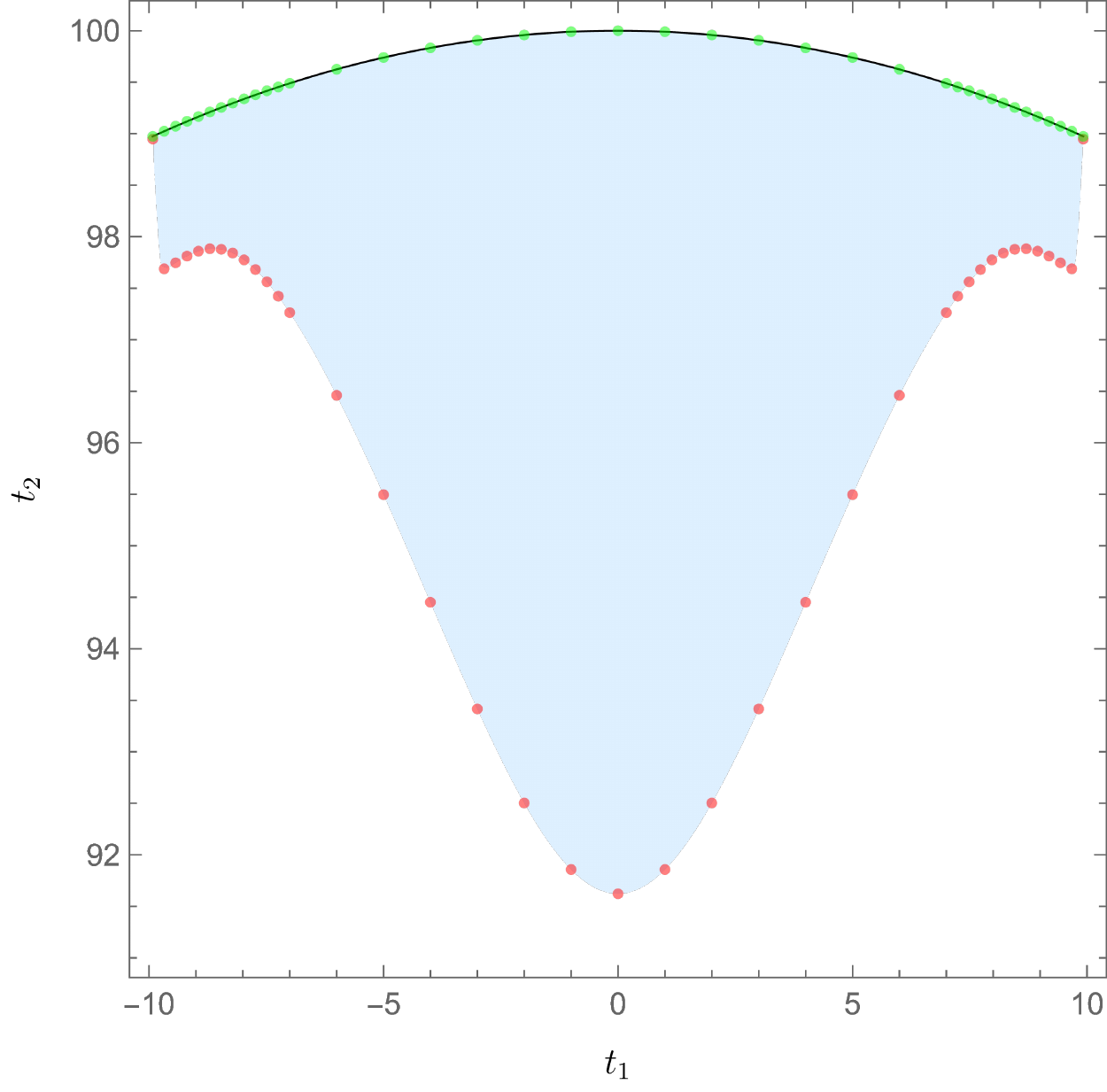}
         \caption{\(\Lambda=6\)}
         \label{fig:sym16}
         \end{subfigure}
         \begin{subfigure}[bc]{.48\textwidth}
         \centering
         \includegraphics[width=\textwidth]{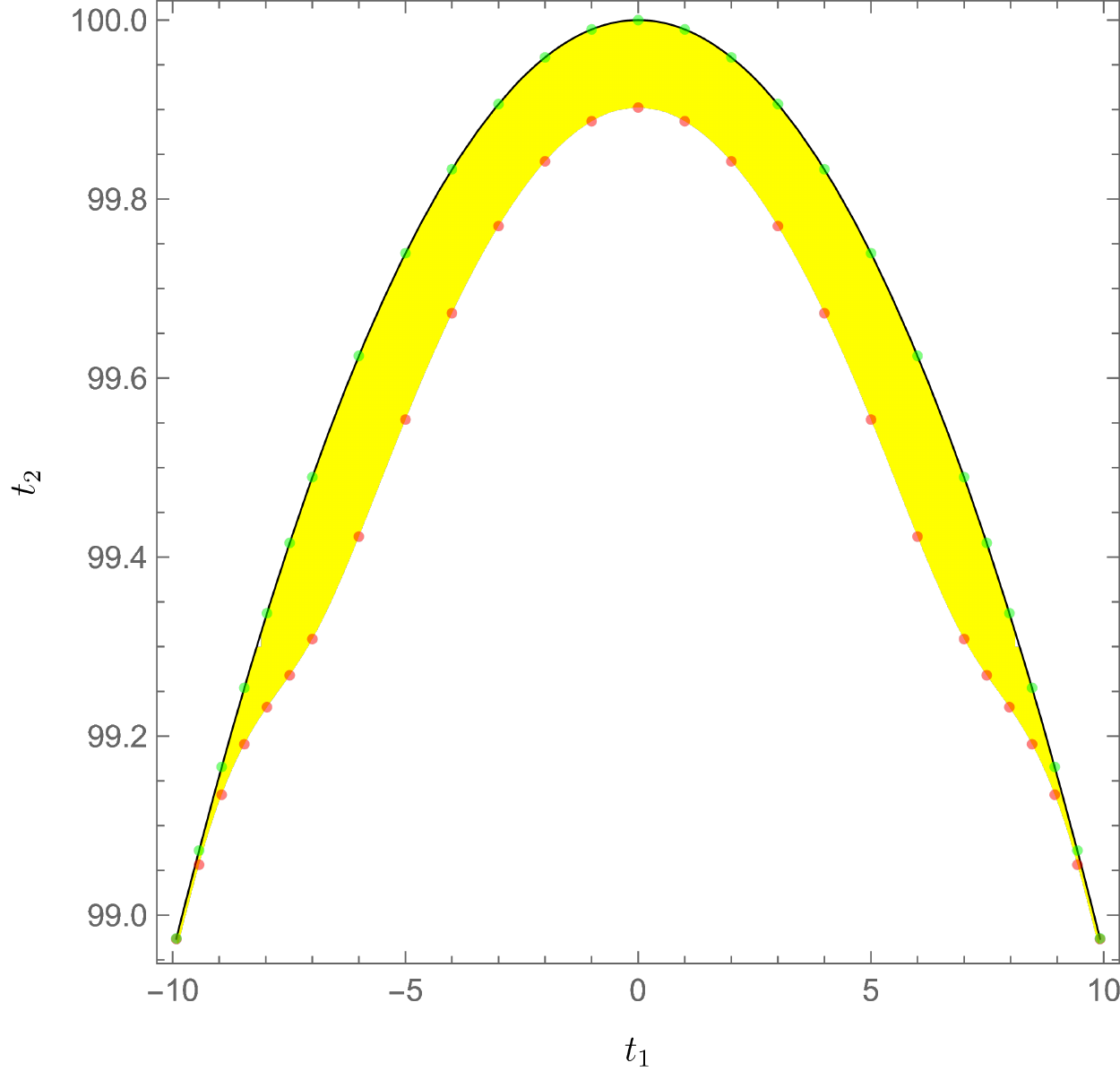}
         \caption{\(\Lambda=9\)}
         \label{fig:sym19}
         \end{subfigure}
    \centering
    \caption{The allowed \(t_1-t_2\) region for \(\Lambda=6\) and \(\Lambda=9\). The corresponding parameters  in the model~\eqref{2MMcom23} are \(g=\frac{1}{100},\,h=0\). The shaded region is the result of one-matrix bootstrap. The black line is the exact analytic  solution described in Section~\ref{sec:ana}. The green and red dots are the upper bounds and lower bounds of our relaxation method from scanning over \(t_1\).}
    \label{fig:sym1}
\end{figure}

In Fig~\ref{fig:sym1} we compare the results of our  relaxation method described above  with the exact results and the one-matrix bootstrap plot at the same cutoff  and  the parameters \(g=\frac{1}{100},\,h=0\). We see that our method is indeed able to bootstrap the symmetry breaking solution, even though it is non-convex. It is especially striking that not only our relaxation method  converges to the highly non-convex exact solution, but it even coincides with the one-matrix bootstrap at each cutoff  \(\Lambda\) within the error bars.  It seems that, in spite of some loss of information when applying the relaxation method, we recover this information  by considering the positivity condition of the mixed operators containing both matrices, such as  \(\langle\Tr ABAB\rangle\). We don't have yet a good explanation why these two approaches give equal or very close results. We also note that in this case the maximization scheme converges faster than the minimization scheme. Namely,  the upper bound (green dots) in the plot is much closer to the exact solution than the lower bound (red dots). This suggests that if we are looking for a good approximation for the exact solution, we should use the upper bound solution as the best approximation. 

\begin{figure}[ht]
    \centering
    \includegraphics[width=.7\textwidth]{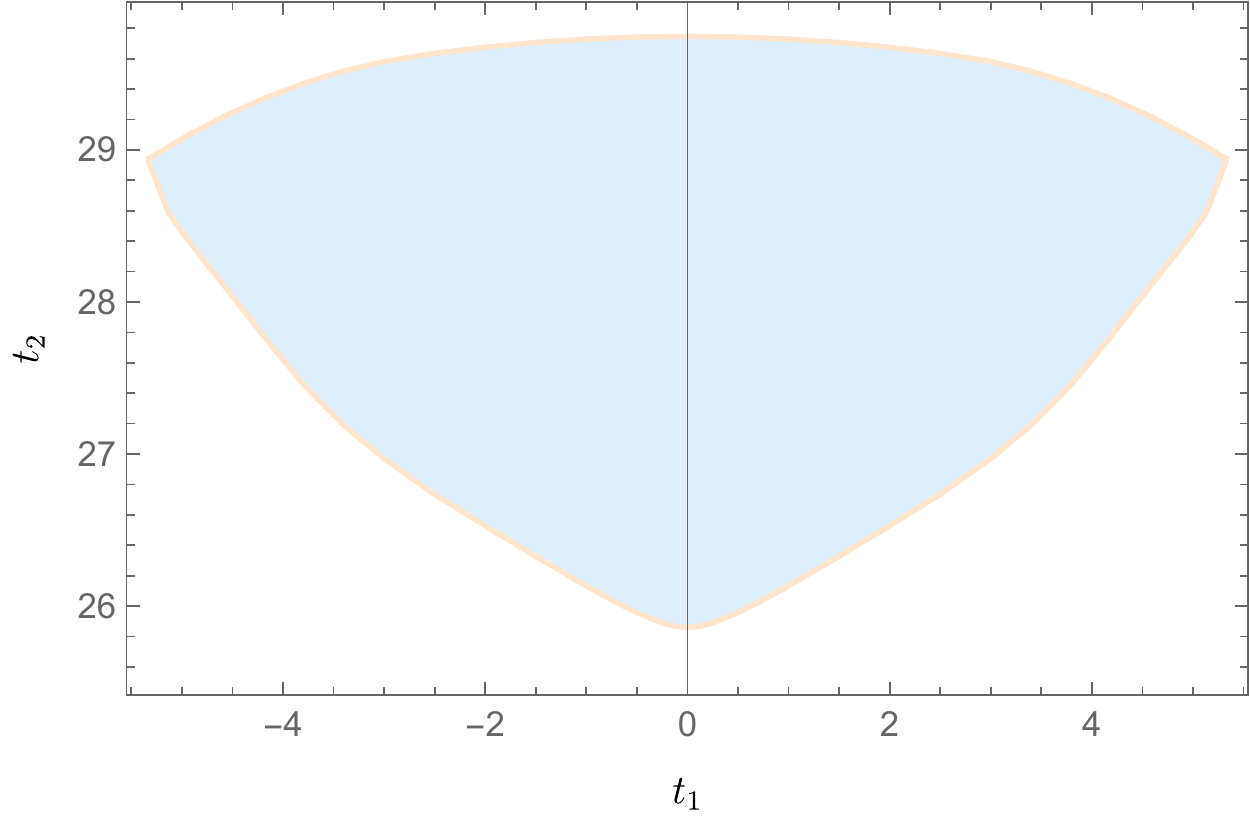}
    \caption{The allowed region for \(g=\frac{1}{30}\) and \(h=\frac{1}{15}\). These numerical results are obtained for \(\Lambda=8\).}
    \label{fig:symb1}
\end{figure}

For generic values of \(h\) and \(g\) for the model~\eqref{2MMcom23}, the convergence is slower than in the analytically  solvable particular case. It would be good to understand whether such a situation for solvable versus unsolvable models is typical. In Fig~\ref{fig:symb1} we plot the bootstrap result for \(g=\frac{1}{30},\,h=\frac{1}{15}\). Obviously,  it is still a symmetry breaking solution. We expect that taking the upper bound we can get a very accurate estimation of the physical quantities. We didn't try to further  increase the value of \(\Lambda\), being already satisfied to see that the proposed method works for  rather generic values of parameters. 

\section{Conclusion and discussion}

In this work, we develop further the matrix bootstrap method pioneered in the papers \cite{2017NuPhB.921..702A,2020JHEP...06..090L} and propose a crucial improvement -- the relaxation procedure -- applicable to a large class of multi-matrix problems and allowing to bootstrap them with a much higher precision. The relaxation transforms a Non-linear  SDP, with the non-linearity due to the structure of loop equations, to the usual, linear SDP. We demonstrate  the efficiency of our approach on the analytically unsolvable two-matrix model and establish its phase structure with rather high precision. The method appears to work well even for the discrete symmetry breaking large \(N\) solutions.

Actually, the efficiency of the matrix bootstrap, based on the positivity of the correlation matrix, remains an enigma. Nevertheless we managed to theoretically study this question in the one-matrix model and to  establish precisely the class of physical solution singled out by such bootstrap.

As it was noticed in~\cite{2017NuPhB.921..702A,2020JHEP...06..090L,1982PhLB..108..407S}  the numerical bootstrap for  large \(N\) multi-matrix models presents a higher degree of difficulty than the bootstrap of the large \(N\) one-matrix model. The bootstrap study of   multi-matrix models was restricted to low orders in the length cut-off for the  moments (or ``words'' forming them) and consequently it provided us with a limited precision. The main reason for this inefficiency is the non-convex formulation of the problem. Our main task in this work was to overcome these drawbacks of the matrix bootstrap.

Compared to  the  cited above papers we achieved a better understanding and efficiency of the matrix bootstrap  in the following aspects:

\begin{itemize}
    \item 
In the case of the large \(N\) Hermitian one-matrix model, we managed to prove  that the bootstrap constraints pick up precisely the exact ``physical'' solutions, with positive measure for the distribution of the eigenvalues (the corresponding resolvent has only the cuts on the real axis). In other words, we established  the exact analytic solutions of the bootstrap conditions for the one-matrix model, thus justifying  the numerical bootstrap techniques. We don't have yet the generalization of such analytic argument for the multi-matrix models, which is an interesting question to address.
\item
Then for the multi-matrix models, we developed the relaxation bootstrap method to overcome the crucial obstacle of non-convexity of the original problem. We demonstrated that this relaxation method was a systematic approach, capable to provide  the  numerically viable procedure for the  large \(N\) multi-matrix models. We tested this method on a model that is genuinely analytically  unsolvable  (unlike Lin's 2-matrix  model with cubic interactions in~\cite{2020JHEP...06..090L}).  For particular parameters, when the analytic solution is known, our numerics  reproduces extremely well the  analytic results. For generic values of these parameters, we bootstrap the physical values with a remarkable precision (6 digits).
\item
This method is also able to detect  critical behaviors,  though the precision gets less impressive in the vicinity of critical lines.
\item
Remarkably, our bootstrap method is also applicable for  bootstrapping  the symmetry-breaking solutions and transform the non-convex problem to a convex one.

\end{itemize}

Here we make several comments on the bootstrap method proposed here  and sketch out some further directions:
\begin{itemize}
    \item All the numerical results in this study can be, in principle,  reproduced on a single decent laptop in a decent time laps. So it looks very promising to implement it on a big cluster with parallelization. The main technical difficulty of the method is that we used the precision bigger than the machine precision (double-double or quad-double) in our current work. We believe that this is mainly due to the fact that our problem is badly-scaled as an SDP: the involved variables can have very different orders of magnitude. It would be good to find  a systematic approach to scale appropriately the variables for very large-scale problems.
    \item The positivity of the correlation matrix in  the matrix  bootstrap method must be satisfied for any multi-matrix integrals with a reasonably converging  positive measure. It follows from the fact that the integral of a positive function against a positive measure is positive. Contrary to the conformal bootstrap and $S$-matrix bootstrap where unitarity is one of the most important conditions, we don't know whether the unitarity or reflection positivity can be imposed in the bootstrap method for the  matrix models. 
    We also hope that our method can be generalized  for bootstrapping  non-unitary quantum field theories.
    \item 
    We expect that the correlation matrix contains a lot of redundancies, i.e. very few of its minors  may contain \(99\%\) of the information of the whole correlation matrix. This is reminiscent of a similar feature of the conformal bootstrap: we don't impose the positivity condition on all spin channels, rather a very limited number of spin channels are good enough to make the algorithm to converge~\cite{Rattazzi:2008pe}. At the moment we don't have any scheme to isolate the minors of the correlation matrix that are more important than the other,  which  would be very beneficial when considering large-scale problems.
    \item
    It would be interesting to apply our methods to the Matrix Quantum Mechanics, in the spirit of the work~\cite{2020PhRvL.125d1601H}, including for the non-singlet states there. Another interesting two-matrix model to study by bootstrap would be the generalization of \eqref{2MMcom2} by taking the $q$-deformed version of interaction: $\tr[A,B]_q^2=\tr(qAB-q^{-1}BA)^2$. This model interpolates between the solvable cases with $\tr(A^2B^2)$ or $\tr(BABA)$ interactions~\cite{Kazakov:1998qw}.   
    \item
   An obvious, and one of the most ambitious possible applications of our relaxation bootstrap methods is the lattice Yang-Mills theory. We have thus good chances  to significantly improve on this way the very preliminary results of~\cite{2017NuPhB.921..702A}.  A method alternative to the wide-spread Monte-Carlo simulations, even at large $N$~\cite{Teper:2008yi}, would be extremely welcome for the study and a deeper understanding  of  QCD. Obvious advantages of the bootstrap method based on Migdal-Makeenko loop equations~\cite{Makeenko:1979pb} w.r.t. Monte-Carlo are: i) Exact inequalities on loop averages, no statistical error; ii) absence of finite boundary conditions (the lattice is infinite); iii)  One gets some information on all loops at once up to a given length, although with better precision for short loops. That gives access to more of the physical quantities. The obvious drawback is the limited length of Wilson loops. We hope to  establish  by the future  numerical work  whether this drawback is crucial indeed. 
   
\end{itemize}

\section*{Appendices}

\appendix
\section{Analytic solvability of two-matrix model with cubic interactions and arbitrary potentials}\label{sec:linsolve}

The  Hermitian 2-matrix model with  the general cubic interactions between two matrices and general potentials  in the action
\begin{equation}\label{Zfactor}
S=\tr\bigg(h(AB^2+BA^{2})+W(A)+\tilde W(B)\bigg)\,,
\end{equation}
has been  studied in~\cite{2020JHEP...06..090L}  by numerical bootstrap method,   as an example of bootstrap approach to an analytically  ``unsolvable'' matrix integral. Here we show that this matrix model is in fact analytically solvable for generic potentials \(W\) and \(\tilde W\), in the sense that the matrix integral can be explicitly reduced to \(\sim N \)  amount of variables, instead of the original \(\sim N^2\) matrix variables, which in principal allows the application of the saddle point method at large \(N\). Our derivation will be schematic and we will repeatedly neglect the non-dynamical factors before the integral of partition function.  It is unclear whether this integrability influences the efficiency  of Lin's  bootstrap method but this is our motivation to choose a different, truly ``unsolvable'' 2-matrix integral, with the interaction \(\tr[A,B]^2\),  as the main example of application of bootstrap in this paper. 

Since \(\tr(A+B)^3=\tr(A^3+ 3A^2B+3AB^2+B^3)\) we can always rewrite it as
\begin{equation}\label{cubicMM}
S=\tr\bigg((A+B)^3/3-V(A)-\tilde{V}(B)\bigg)
\end{equation}
where \(V(A)=A^3/3-W(A)\) and \(\tilde{V}(B)=B^3/3-\tilde W(B)\) and we have set wit1out loss of generality \(h=1\). 

We can always reduce it  to Itzykson-Zuber-Charish-Chandra (IZC) integral by an extra matrix integration, following the trick similar to proposed in~\cite{Kazakov:1987qg} in the context of solution of the Potts model on dynamical planar Feynman graphs. Namely, represent the first term in \eqref{cubicMM}  in terms of an extra hermitian  matrix integral over \(X\)~\footnote{We drop here and further all inessential overall factors}
\begin{align}
e^{\tr \,C^3/3 }=\int d^{N^2}X e^{\tr (iXC+F(X))},\qquad C=A+B
\end{align}     
where the function \(F(X)\) is defined as the inverse matrix Fourier transform:
\begin{align}\label{Fint}
e^{\tr F(X)}=\int d^{N^2}C e^{\tr (-iXC+C^3/3)}=\int \prod_{j}dc_j\,e^{c_j^3/3}\,\,\Delta^2(c)\frac{\det_{j,k} e^{-ix_jc_k}}{\Delta(c)\Delta(x)}\,,
\end{align} 
i.e. it represents the matrix Airy function. In the last equality we applied the IZC integral. Then we  rotate the contour of \(X\) integration, change variable \(C\to -C\) and  write the partition function of cubic 2MM model in the form 
\begin{align}\label{Zfactor}
Z=\int d^{N^2}X\int d^{N^2}C e^{\tr (XC-C^3/3)}\int d^{N^2}A\,e^{\tr  [XA-V(A)]}\,\int d^{N^2}B\,e^{\,\tr  [XB-\tilde{V}(B)]}.
\end{align}

 We can  compute the function \(F(X)   \)  in terms of the eigenvalues of \(X=\Omega^{\dagger}x\Omega \) where \(x=\mathrm{diag}\{x_0,x_1,\dots,x_{N-1}\}\). We compute the angular integral in  \eqref{Fint} via IZC integrals: 
 \begin{align}\label{Fint}
e^{\tr F(X)}=\Delta^{-1}(x)\int \prod_{j}dc_j\,e^{x_jc_j-c_j^3/3}\,\,\Delta(c)=\frac{W[-\mathrm{Ai}(x_0),\dots -\mathrm{Ai}(x_{N-1})]}{\Delta(x)}\,
\end{align}

where in the denominator we have the \(N\times N\) ``Wronskian'' of Airy-type functions:
 \begin{align}
\mathrm{Ai}(x)=\int_{{\cal C}} dc\,e^{-xc+\frac{1}{3}c^3},\,\,
\end{align}
and:
\begin{equation}
    W[f_0(x_0),\dots f_{N-1}(x_{N-1})]=\sum_{\sigma \in S_n}\mathrm{sgn}(\sigma)\prod_{i=0}^{N-1}\partial_{x_i}^{\sigma(i)}f_i(x_i).
\end{equation}
Here the complex  contour ${\cal C}$  is usually chosen so that it goes from  infinity with the slope $-\pi/3$ and ends up at infinity with the slope   $\pi/3$.  However, when we study the limit $N\to\infty$ the saddle point configuration of the eigenvalues will adjust itself to the relevant distribution on the real axis given by a solution of the integral saddle point equation. Similarly, the bootstrap numerical procedure should single out such solutions.

Then we treat similarly the other two integrals in \eqref{Zfactor} and represent them also in terms of Wronskians of \begin{align}
f(x)=\int da\,e^{xa-V(a)}\,,\,\qquad \tilde f(x)=\int db\,e^{xb-\tilde V(b)}.
\end{align}

In this way, we managed to re-wrtite the  cubic 2MM entirely in terms of eigenvalue integral: 
\begin{align}\label{Zev}
Z=\int\prod_{j}dx_j\,\frac{W[-\mathrm{Ai}(x_0),\dots ]\,\,W[f(x_0),\dots ]\,\,W[\tilde f(x_0),\dots]\,\,}{\Delta(x)}{}.\,\,
\end{align}

Hence we reduced the cubic two-matrix integral \eqref{cubicMM} to an explicit integral over \(N\) eigenvalues of an auxiliary matrix \(X\).
 We treat  such a matrix model as ''solvable`` though the further details of  the explicit solution can be rather involved.
   Instead of studying the saddle point in terms of wronskians
it is better  to apply the method (inspired by~\cite{Brezin:1980rk}) which was proposed by V.Kazakov and I.Kostov for  solution of Potts model on random planar graphs~\cite{Kostov:1988pe}; it is well presented in ~\cite{Daul:1994qy}.  We will not pursue here this route and we leave it for the future work. 

\section{Solving the positivity condition of the resolvent}\label{Quartic}

We saw in Section~\ref{sec:Hamburger} that the positivity of correlation matrix is equivalent to the positivity of the resolvent. As it was noticed there, this equivalence enables us to analytically solve the bootstrap condition. Here we propose a general method to solve the lower moments from the positivity condition of resolvent. This finishes our analytic solution of the bootstrap problem corresponding to the Hermitian one-matrix model. As a specific  example, we also apply this method to one-matrix model  with quartic potential, for which the results were  summarized in Sec.~\ref{sec:ana}.

\subsection{Cuts and zeros}\label{sec:cut}

Finding the lower moments from the positivity of the resolvent is a well-posed problem in complex analysis, and abundant mathematical tools can be employed to solve it. Here we  study the configuration of cuts and zeros of the cut function defined in the main text~\eqref{eq:cutfunc}. Due to the polynomiality of the discriminant $D(x)=C(x)^2$  we can give the full classification of all possible configurations of cuts and zeros of the cut function $C(x)$ on the real line. It turns out that the properties of these configurations provide not only necessary but also sufficient condition for the positivity of the resolvent. Generally, if we know a configuration of the cuts and zeros, we are able to fix a few lowest   moments  which we want to find the solution. This is how we solve the positivity condition of the resolvent. In the following we will be considering a Hermitian one-matrix model with general polynomial potential \(V(x)\).

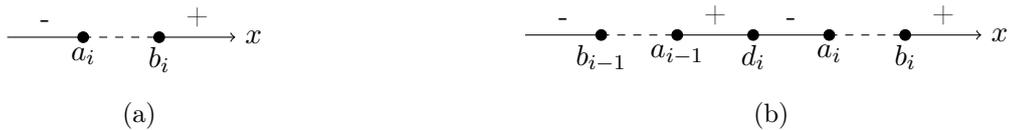
\begin{figure}[ht]
     \begin{subfigure}[b]{0.4\textwidth}
\begin{tikzpicture}
  \draw[->] (1, 0) -- (2, 0) node[right] {$x$} node[pos=0.5, anchor=south]{+};
  \draw[dashed] (0,0) --(1,0) ;
  \draw (-1,0) --(0,0) node[pos=0.5, anchor=south]{-};
  \filldraw[black] (0,0) circle (2pt) node[anchor=north] {$a_i$};
  \filldraw[black] (1,0) circle (2pt) node[anchor=north] {$b_i$};
    \end{tikzpicture}
    \centering
    \caption{}
    \label{fig:cut}
     \end{subfigure}
     \hfill
     \begin{subfigure}[b]{0.5\textwidth}
    \begin{tikzpicture}\label{cut}
  \draw[->] (1, 0) -- (2, 0) node[right] {$x$} node[pos=0.5, anchor=south]{+};
  \draw[dashed] (0,0) --(1,0) ;
  \draw (-1,0) --(0,0) node[pos=0.5, anchor=south]{-};
  \filldraw[black] (0,0) circle (2pt) node[anchor=north] {$a_i$};
  \filldraw[black] (1,0) circle (2pt) node[anchor=north] {$b_i$};
  \draw (-2, 0) -- (-1, 0) node[pos=0.5, anchor=south]{+};
  \draw[dashed] (-3,0) --(-2,0) ;
  \draw (-4,0) --(-3,0) node[pos=0.5, anchor=south]{-};
  \filldraw[black] (-2,0) circle (2pt) node[anchor=north] {$a_{i-1}$};
  \filldraw[black] (-3,0) circle (2pt) node[anchor=north] {$b_{i-1}$};
  \filldraw[black] (-1,0) circle (2pt) node[anchor=north] {$d_{i}$};
    \end{tikzpicture}
    \centering
    \caption{}
    \label{fig:my_label}
     \end{subfigure}
        \caption{The left plot shows the sign of $C(x)$ in the neighborhood of a positive cut. The right plot shows there must be at least one zero, or generally odd number of zeros, between two cuts to fix the sign.}
        \label{fig:cuts}
\end{figure}

First let us pick a single cut of $C(x)$ on the real axis, namely $[a_i,b_i]$, as shown in Fig~\ref{fig:cuts}. The positivity condition of the resolvent implies that for \(x \in [a_i,b_i]\) we have \(\mathrm{Im}C(x+i0)\propto \rho (x)\geq 0\). From the definition of the cut function, we must have $C(x)>0$ in the right neighborhood of $b_i$, and $C(x)<0$ in the left neighborhood of $a_i$. A direct consequence is that we must have at least one zero, or generally odd number of zeros, between two positive cuts to fix the sign. This is also illustrated in Fig~\ref{fig:cuts}.

There is yet another constraint on the zeros. We notice that $V'(x)$ and $C(x)$ get unbounded at infinity, but $G(z)\underset{z\to\infty}{\to} \frac{1}{z}$, i.e. it is analytic there. As a result, asymptotic behavior of $C(x)$ must match the asymptotics of $V'(x)$. For example, if $\lim_{x\rightarrow \infty}V'(x)<0$, we must add another zero to the right of all the cuts to fix the sign of $C(x)$, preserving both positivity of the resolvent and the asymptotic behavior. 

For the zeros of \(D(x)\) which are not located on the real axis, there exist roots with even multiplicities since there shouldn't exist complex cuts for \(C(x)\). We also note that since \(D(x)\) is a polynomial with real coefficients, all its complex roots must come in pairs.

The above analysis  gives the way to count the the possible number of cuts. Here we list the maximum number of cuts $m$ when the degree of the potential is $d+1$, under the asymptotic behavior $(\pm,\pm)$\footnote{For example, $(-,+)$ means $\lim_{x\rightarrow -\infty}V'(x)<0$ and $\lim_{x\rightarrow \infty}V'(x)>0$}:
\begin{itemize}
    \item $(-,+)$: $2m+2(m-1)\leq 2d\Rightarrow m\leq (d+1)/2$,
    \item $(+,-)$: $4+2m+2(m-1)\leq 2d\Rightarrow m\leq (d-1)/2$,
    \item $(+,+)$ and $(-,-)$: $2+2m+2(m-1)\leq 2d\Rightarrow m\leq d/2$.
\end{itemize}

\subsection{A working example}\label{minus2}
\begin{figure}[ht]
    \centering
    \begin{tikzpicture}
  \draw[->] (1, 0) -- (2, 0) node[right] {$x$} node[pos=0.5, anchor=south]{-};
  \draw (0, 0) -- (1, 0) node[pos=0.5, anchor=south]{+};
  \draw[dashed] (-1,0) --(0,0) ;
  \draw (-2, 0) -- (-1, 0) node[pos=0.5, anchor=south]{-};
  \draw (-3, 0) -- (-2, 0) node[pos=0.5, anchor=south]{+};
  
  \filldraw[black] (1,0) circle (2pt) node[anchor=north] {$d_1$};
  \filldraw[black] (0,0) circle (2pt) node[anchor=north] {$b_1$};
  \filldraw[black] (-1,0) circle (2pt) node[anchor=north] {$a_1$};
  \filldraw[black] (-2,0) circle (2pt) node[anchor=north] {$d_2$};
    \end{tikzpicture}
    \caption{The only possible cut configuration for asymptotic $(+,-)$}
    \label{fig:gneg}
\end{figure}
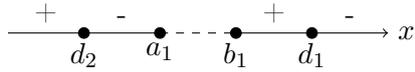

In this part, we will use these cuts and zeros considerations solve the bootstrap condition of the model with potential:
\begin{equation}\label{eq:pot1}
    V(x)=\frac{1}{2} x^2+\frac{1}{4} g x^4,\quad g<0.
\end{equation}
This potential has a $(+,-)$ asymptotic, so in the minimal case, we must have a positive cut in the middle, and two zeros, one placed on the right and another one on the left of the cut, to fix the asymptotics. Since the polynomial $D(x)=V'(x)^2-4P(x)$ is of 6th degree, we have already reached the maximum number of zeros. The conclusion is that the cut configuration in Fig~\ref{fig:gneg} is the only possibility for such asymptotic behavior.

\begin{figure}[ht]
     \begin{subfigure}[b]{0.5\textwidth}
         \centering
         \includegraphics[width=\textwidth]{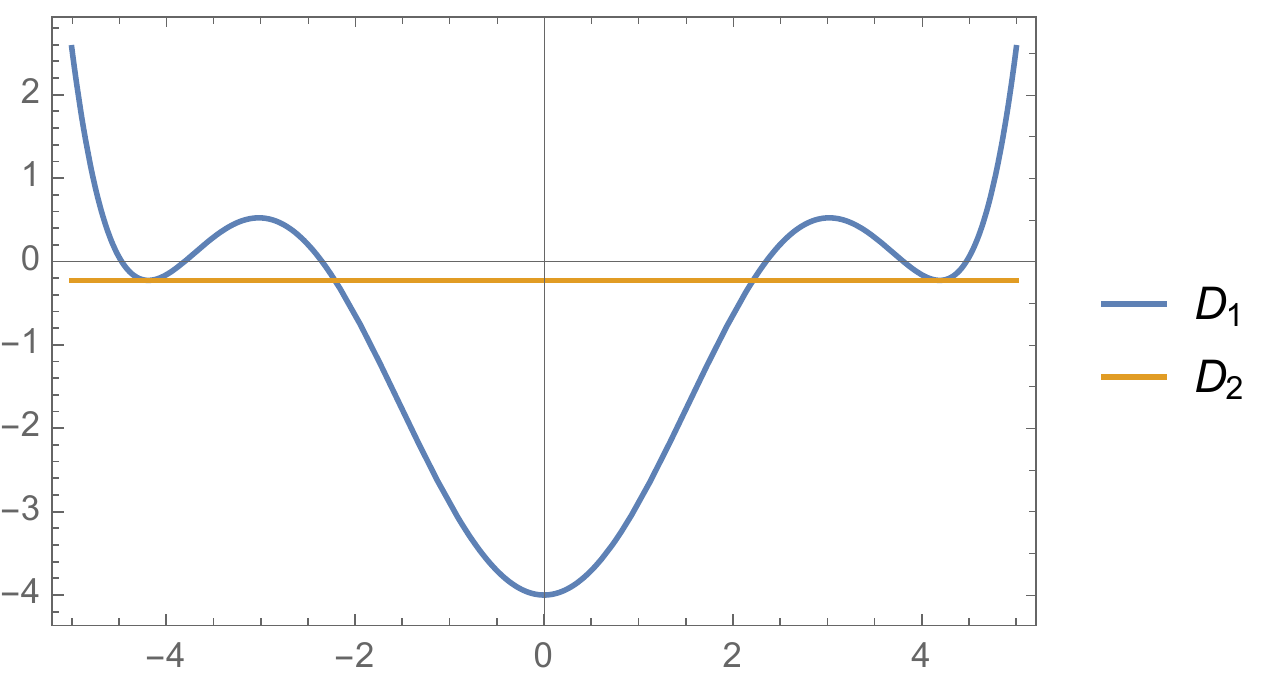}
         \caption{$g=-1/20$}
         
     \end{subfigure}
     \hfill
     \begin{subfigure}[b]{0.5\textwidth}
         \centering
         \includegraphics[width=\textwidth]{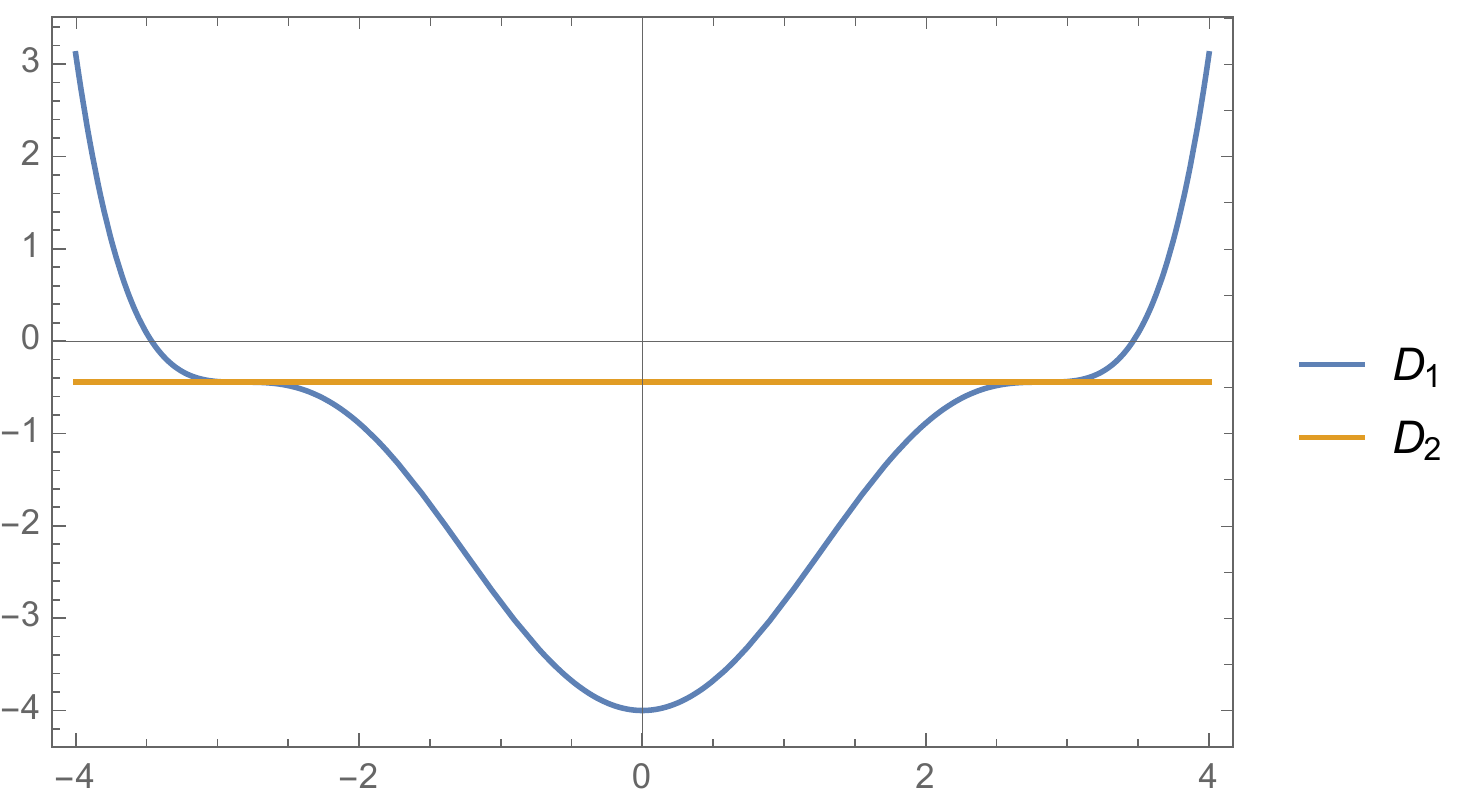}
         \caption{$g=-1/12$}
         
     \end{subfigure}
        \caption{The only possible cut configuration for $\mu=1$ and $g<0$. The critical case is $g=-\frac{1}{12}$, below that, there is no more bootstrap solution.}
        \label{fig:plusminus}
\end{figure}

To see when this cut configuration is possible, we note that:
\begin{equation}
    D(x)=V'(x)^2-4P(x)=\left(g x^3-x\right)^2-4gx^2+4g-4g \left( \mathcal{W}_1 x+\mathcal{W}_2\right)=D_1(x)-D_2(x).
\end{equation}
Here we split the discriminant into the part depending on $\mathcal{W}_1$ and $\mathcal{W}_2$: 
\begin{equation}
    D_1(x)=\left(g x^3-x\right)^2-4gx^2+4g
\end{equation}
and and the part depending only on $g$:
\begin{equation}
    D_2(x)=4g \left( \mathcal{W}_1 x+\mathcal{W}_2\right)\,.
\end{equation}
$D_2(x)$ is a straight line with negative intercept. As depicted in Fig~\ref{fig:plusminus}, it is possible that a straight line crosses $D_1(x)$ with sufficient number of intersections, but only for \(g\geq g_c=-\frac{1}{12}\). For  $g<g_c$  the qualitative shape of the graph of $D_1(x)$ disqualifies the only possible cut configuration Fig~\ref{fig:gneg}, excluding the existence of any bootstrap solution.

From the Fig~\ref{fig:plusminus}, it is obvious that  $\mathcal{W}_1=0$. Otherwise we won't have a resolvent satisfying the positivity condition. We can find $\mathcal{W}_2$ from the vanishing of the discriminant of the polynomial $D(x)$, namely:
\begin{equation}\label{BIPZ}
    \mathcal{W}_2=\frac{(12 g+1)^{3/2}-18 g-1}{54 g^2}
\end{equation}
which corresponds to the standard one-cut IBPZ solution~\cite{Brezin:1977sv}.

To make our intuitive arguments above more systematic, and applicable to higher degree potentials,  let us list the conditions by which the positivity of the resolvent translates into the properties of the polynomial $D(x)$:
\begin{enumerate}
     \item $D(x)$ has 6 real roots counting multiplicity, two of them are double roots. Two of them are single roots.
    \item The simple roots of $D(x)$ lie in between the double roots on the real line.
\end{enumerate}

\begin{figure}[ht]
    \centering
    \includegraphics[width=.5\textwidth]{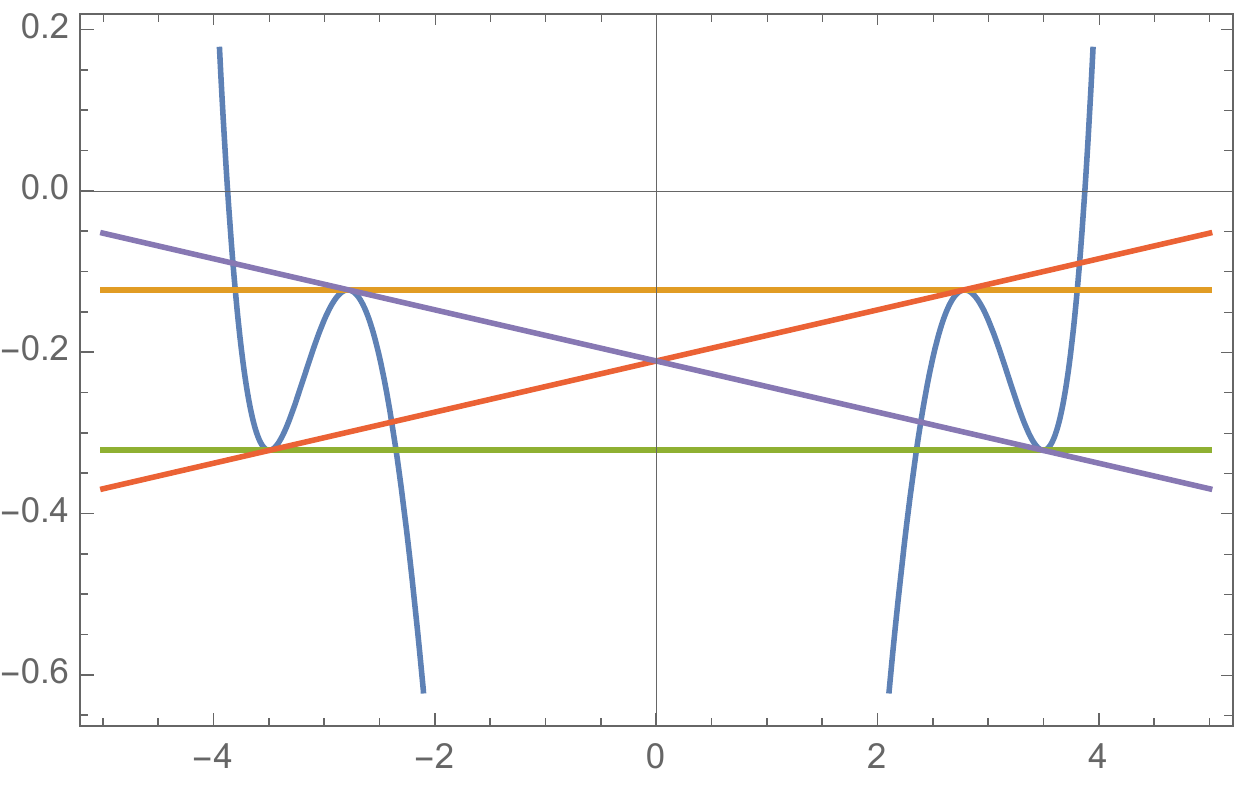}
    \caption{4 solutions under the condition 1 when $g=-1/15$. The blue line is the function\(D_1(x)\) which is independent of moments \(\mathcal{W}_1,\,\mathcal{W}_2\), whereas other color straight lines are different solution of \(D_2(x)\). We see only the green line has the correct configuration of zeros depicted in Fig~\ref{fig:plusminus}.}
    \label{fig:wrong}
\end{figure}

The property 1 boils down to the condition that the degree of the greatest common divisor of $(D(x),D'(x))$ is 2. We apply the Euclidean algorithm to $D(x)$ and $D'(x)$ to rewrite them in the form:
\begin{equation}
\begin{cases} D=q_1 D'+r_1\\ D'=q_2 r_1+r_2 \\ r_1=q_3 r_2+r_3\\ r_2=q_4 r_3+r_4 \end{cases}
\end{equation}
The remainder $r_4$ is a polynomial of degree $1$ in $x$ and we set all its coefficients to zero. In this way we get 2 algebraic equations which, in principle, fix $\mathcal{W}_1$ and $\mathcal{W}_2$ in terms of $g$. For the reality of the roots, we note that this means the discriminant of $(D(x),D'(x))=r_3$  and the discriminant of $D(x)/r_3^2$ is non-negative. These are already a lot of conditions. Luckily the mathematica \textbf{Reduce} function can treat these conditions efficiently, leaving us with four solutions. The four solutions correspond to different orders of single zeros and double zeros. We show the corresponding solutions in Fig~\ref{fig:wrong}. We can take advantage of the condition 2 to select the only physical solution for this case:
\begin{equation}
\begin{cases} -\frac{1}{12}\leq g<0:\, \mathcal{W}_1=0, \,\mathcal{W}_2=\frac{(12 g+1)^{3/2}-18 g-1}{54 g^2}.\\ 
g<-\frac{1}{12}:\, \text{No bootstrap solution}.
\end{cases}
\end{equation}

\section{Dual formulation and relaxation}\label{rdual}
In this appendix, we review some basic facts about dual formulation in optimization theory and clarify the relationship between our relaxation method introduced in Section~\ref{sec:relax} and the dual formulation. For the readers interested in more details about the optimization theory, the book \cite{boyd_vandenberghe_2004} is a good starting point.
\subsection{Dual problem of general optimization problem}
Consider a general optimization problem of the form:
\begin{equation}\label{general}
\begin{aligned}
\min \quad & c(x)\\
\textrm{subject\,\,to} \quad & f_i(x)\leq 0, \quad i=1,...,m\\
  & h_j(x)=0,\quad j=1,...,p    \\
  & x\in \mathbb{R}^n.
\end{aligned}
\end{equation}
For convenience, we denote the optimal value for \eqref{general} by $p^\star$. Whenever $c(x)$, $f_i(x)$ are convex functions and $h_j(x)$ are linear functions the problem \eqref{general} is defined to be convex, otherwise it is non-convex. \footnote{It turns out that usually we can only solve the convex optimization problem efficiently, and non-convex problems are generally NP-hard.}

For a general problem of the form \eqref{general}, convex or not, we can construct its dual problem starting with the Lagrangian defined by:
\begin{equation}\label{Lag}
\mathcal{L}(x,u,v)=c(x)+\sum_{i=0}^m u_i f_i(x) +\sum_{j=0}^p v_j h_j(x), \quad u_i\geq 0
\end{equation}
and  minimize over $x$\footnote{We stress here that the minimization is the unconstrainted minimization, i.e. \(x\in \mathbb{R}^n\)}:
\begin{equation}
g(u,v)=\min_{x\in \mathbb{R}^n} \mathcal{L}(x,u,v) \,.
\end{equation}
It is a simple exercise to show that $g(u,v)$ is concave, since it is a minimization over a family of linear functions in $u,v$. For all $u,v$ under the constraints $u_i\geq 0$ , we have:
\begin{equation}\label{dp1}
g(u,v)\leq \max g(u,v)\leq p^\star.
\end{equation}
This indicates us the formulation of the dual problem:
\begin{equation}\label{dual}
\begin{aligned}
\max \quad & g(u,v)\\
\textrm{s.t.} \quad & u_i\geq 0 \quad i=1,...,m.
\end{aligned}
\end{equation}
We denote the optimal solution of this problem to be $d^\star$. This optimization problem is guaranteed to be convex, since it is maximizing a concave function $g(u,v)$. In this sense the dual problem \eqref{dual} is simpler when the primal problem is non-convex. Of course we always have:
\begin{equation}\label{dp2}
    d^\star\leq p^\star
\end{equation}
due to \eqref{dp1}. The inequality~\eqref{dp2} is conventionally called weak duality. It would be actually nice to have the equality, i.e. when \textit{strong duality} holds. For that case we have the well-known Slater's condition:

\textit{The strong duality holds if the primal problem is convex and it has a strictly feasible solution.}

By definition, a solution $x^*$ is strictly feasible when:
\begin{equation}\label{strict}
\begin{aligned}
& f_i(x^*)< 0 \quad i=1,...,m\\
  & h_j(x^*)=0,\quad j=1,...,p.    
\end{aligned}
\end{equation}

In general, we don't have strong duality for non-convex problem.

\subsection{Relaxation problem and dual problem.}

In this section, for  completeness we present a proof that the relaxation problem introduced in \eqref{relaxp} is the dual of the dual problem of the original problem \eqref{NSDP}. The proof is actually trivial but lengthy, so the reader could treat it as an implementation example of the dual formulation introduced in the last section.

First we transform the original problem \eqref{NSDP} to the following form:
\begin{equation}\label{NSDP1}
\begin{aligned}
\min \quad & c^{\mathrm{T}}x\\
\textrm{s.t.} \quad & \tr X A_i +b_i^{\mathrm{T}} x+a_i=0,\\
  & M_0+\sum_{j=1}^L M_j x_j\succeq 0    ,\\& X=xx^{\mathrm{T}}.
\end{aligned}
\end{equation}
To take the dual problem of it, we write down its Lagrangian:
\begin{equation}
    \mathcal{L}(x, X, \lambda, Y, Z)=c^{\mathrm{T}}x+\sum_{i} \lambda_i (\tr X A_i +b_i^{\mathrm{T}} x+a_i) -\tr (Y(M_0+\sum_{j=1}^L M_j x_j))+\tr(Z(xx^{\mathrm{T}}-X)).
\end{equation}
The matrices $Y$ and $Z$ introduced here are real symmetric matrix variables, satisfying $Y\succeq 0$ due to the inequality condition in \eqref{NSDP1}. To minimize the Lagrangian, we collect all the terms involving primal variables $x$ and $X$:
\begin{equation}
\mathcal{L}(x, X, \lambda, Y, Z)=\tr((\sum_{i}\lambda_i A_i-Z) X) +x^{\mathrm{T}} Z x +(c+d+\sum_{i}\lambda_i b_i)^{\mathrm{T}} x +\sum_{i} \lambda_i a_i -\tr (Y M_0)
\end{equation}
where we introduced the vector variable $d_k=-\Tr (Y M_k)$ to make the formula more compact. We get $g(\lambda, Y, Z)$ by taking the minimization over $x,X$:
\begin{equation}\label{eq:d1}
\begin{medsize}
\begin{aligned}
&g(\lambda, Y, Z)=\sum_{i} \lambda_i a_i -\tr (Y M_0) -\frac{1}{4} (c+d+\sum_{i}\lambda_i b_i)^{\mathrm{T}} Z^\dagger (c+d+\sum_{i}\lambda_i b_i),\\
    &\mathrm{if}\quad \sum_{i}\lambda_i A_i-Z=0,\quad Z\succeq0,\quad (I-ZZ^{\dagger})(c+d+\sum_{i}\lambda_i b_i)=0. \\
\end{aligned}
\end{medsize}
\end{equation}
The reader can verify that if the conditions in the second line of~\eqref{eq:d1} are not saturated, the minimal value $g(\lambda, Y, Z)$ is $-\infty$, which is irrelevant since we are only interested in its the maximal. Note that $Z^\dagger$ here denotes the pseudo-inverse of the matrix $Z$:
\begin{equation}
    Z^\dagger=\lim_{\epsilon\rightarrow 0^+} (Z^{\mathrm{T}} Z +\epsilon I)^{-1} Z^{\mathrm{T}}.
\end{equation}
Introducing an auxiliary variable $\gamma$ by Schur complement, we can formulate the dual problem in a more compact form:
\begin{equation}\label{dual1}
\begin{aligned}
\max \quad & \gamma\\
\textrm{s.t.} \quad & \sum_{i}\lambda_i A_i-Z=0,\quad Y\succeq 0,\\
  & \begin{pmatrix}
\sum_{i} \lambda_i a_i -\tr (Y M_0)-\gamma\quad & (c+d+\sum_{i}\lambda_i b_i)^{\mathrm{T}}/2\\
(c+d+\sum_{i}\lambda_i b_i)/2 & Z
\end{pmatrix}\succeq 0.
\end{aligned}
\end{equation}
To do the second dualization, we introduce again the Lagrangian:
\begin{equation}
\begin{aligned}
&\mathcal{L}_2(x, X, \delta, W, S, \gamma, \lambda, Y, Z)=-\gamma-\tr(S Y)+\tr(W(\sum_{i}\lambda_i A_i-Z))\\
&-\tr\left(\begin{pmatrix}
\delta & x^{\mathrm{T}}\\
x & X
\end{pmatrix}\begin{pmatrix}
\sum_{i} \lambda_i a_i -\tr (Y M_0)-\gamma & (c+d+\sum_{i}\lambda_i b_i)^{\mathrm{T}}/2\\
(c+d+\sum_{i}\lambda_i b_i)/2 & Z
\end{pmatrix}\right)
\end{aligned}
\end{equation}
Here we slightly abuse the notation by introducing a dual variable contains $x$ and $X$, which satisfies:
\begin{equation}
    \begin{pmatrix}
\delta & x^{\mathrm{T}}\\
x & X
\end{pmatrix}\succeq 0.
\end{equation}
We also note that $S\succeq 0$. The dual problem becomes apparent when we collect the coefficients of $\gamma,\, Y,\, Z,\, \gamma$:
\begin{equation}
\begin{aligned}
&\mathcal{L}_2(x, X, \delta, W, S, \gamma, \lambda, Y, Z)=(\delta-1)\gamma+\tr( ( M_0 \delta-S +\sum_k M_k x_k)Y)-\tr((W+X)Z)\\
&+\sum_i \lambda_i (\tr(W A_i)- a_i \delta -b_i^{\mathrm{T}} x)-c^{\mathrm{T}} x.
\end{aligned}
\end{equation}
Again, all the coefficients of the linear terms must vanish, otherwise the minimal value of ${\cal L}_2$ will be $-\infty$ when we take the minimization. So this time  we can write down the dual problem as:
\begin{equation}\label{dual2}
\begin{aligned}
\min \quad & c^{\mathrm{T}} x,\\
\textrm{subject to} \quad & M_0+\sum_k M_k x_k\succeq 0\\&\tr (X A_i) +b_i^{\mathrm{T}} x +a_i=0,\\&\begin{pmatrix}
1 & x^{\mathrm{T}}\\
x & X
\end{pmatrix}\succeq 0
\end{aligned}
\end{equation}
which is exactly our original  problem with relaxation.

\section{Implementation in SDPA}\label{imple}
Here we gather the details on the  implementation of our methods which are not covered in Section~\ref{QCSM}. We stress that although we are quite satisfied by by the current efficiency of the method, we cannot guarantee that the choices we made are the best possible. In Appendix~\ref{example}, we will demonstrate explicitly the importance of all these details on a simple example.

\subsection{Choice of SDP solvers}

The SDP solver for all of the numerical results in Section~\ref{QCSM} is SDPA\cite{eca870fa006646ae8352bf7a66ad292e}. In the search for a better performance, we tried a lot of other solvers, including PENLAB, SDPB, SDPA, MOSEK, SDPT3 and SeDuMi. In the early stage of this work, we were mainly working with SDPB\cite{2015JHEP...06..174S}. Our experience shows that SDPB gives the most stable convergence in all situations, even when we set the SPDB precision to $64$. All other solvers sometimes fail to converge. The disadvantage of SDPB is that it is much slower and uses much more memory compared to all  other solvers. This is due to the fact that SDPB is an arbitrary precision solver relying on GMP, which potentially slows the program by more than $1000$ times. Due to this reason, when we try to deal with large-scale problems, we switch to SDPA which, according to our limited tests, converges better than all other SDP solvers on large-scale problems.

\subsection{Generation of the loop equations}

For loop equations of a given length $l$, the set of all loop equations could be generated by exhaustive method\footnote{Here exhaustive method means that we generate all the loop equations by taking matrix derivatives on all possible positions of all possible words in Schwinger-Dyson equations.}. This is conceptually simple and numerically easy to implement. In fact, a slight optimization of Lin's code in \cite{2020JHEP...06..090L} is efficient enough to generate loop equations up to $\Lambda=11$ in our work. The tricky part here is to determine what is the sufficient length cutoff of loop equations for our numerics.

Let's take the model~\eqref{2MMcom21} as an example again. Suppose we are considering a cutoff $\Lambda$, which means that the longest operator in the correlation matrix is $2\Lambda$. We recall that in loop equation of length $l$ in this model, we have the operator expectations of lengths $l-1,l+1,l+3$\footnote{Here we define the length of quadratic terms in loop equations as the sum of the lengths of their factors.}. We can  consider only the loop equations up to the length $l=2\Lambda-3$, since they contain only the operators we are interested in. But higher length loop equations could also generate non-trivial relations on the operator expectations of lengths within the cutoff. Namely, as loop equations of higher length contain lower operators that we are interested in, we could eliminate higher operators and get new,  generally linear independent  loop equations containing only operators up to the length $l=2\Lambda$. 
However,  empirically for the model~\eqref{2MMcom21}  this doesn't happen: the procedure described above does not produce any new independent loop equations, in addition to the standard ones, constraining the operators of lengths $l\le 2\Lambda-3$. One exception is the case $h=0$ and $g=0$, i.e. when the model is solvable. Then we need to consider the loop equations up to the lengths $l=2\Lambda+1$ and $l=2\Lambda-1$ to generate all the equations between operators we are interested in. We don't have any proof why this is true. But this is not hard to understand intuitively. For example, when $g=0$ and we consider the loop equations only up to $l=2\Lambda-3$, so we have a smaller number of length $2\Lambda$ operators comparing to the case $g\neq 0$ in the loop equations.
But we also note that even for $h=0$ or $g=0$, when we  consider only the loop equation up to length $l=2\Lambda-3$ in our bootstrap assumptions, we still get a bound very close to the one using the higher length loop equations as well. This example supports  our general intuition that higher loop equations are expected to have a relatively limited influence on small length operators. 

In conclusion, for the model~\eqref{2MMcom21} with cutoff \(2\Lambda\), it is enough to take into account the loop equations up to the length $l=2\Lambda-3$.

\subsection{Solving the loop equations}

Unlike other solvers, the input form of SDPA is rather demanding. It only admits an input problem strictly of the \eqref{SDPp} form. As a consequence, we need to solve the loop equation and substitute the solution into the correlation matrix and relaxation matrix. More precisely, by solving the loop equations we mean that there is a reduced set of operator expectations, such that all other operator expectations and all the quadratic terms $\mathcal{R}_{ij}$ can be solved by loop equations as linear combinations of them. For later convenience, we denote the vector of this reduced set by $x_{\mathrm{red}}$.

A bad news is that the loop equations generated by exhaustive method are not linearly independent, so solving it with floating point numbers will cause some numerical instabilities. One way out of this difficulty is to remove the linear redundancies in loop equations. Another way is to use exact numbers in Mathematica. The first is hard since we found that for some specific values of $h$ and $g$, there will be coincident degeneracies. So we doubt we could have a systematic way to select the linear independent loop equations. Actually we are using the second way, i.e. only using the exact number computation when generating and solving loop equations. The exact number arithmetics consumes a lot of memory and makes the solving procedure much slower, but up to the most involved case $\Lambda=11$ in this article, we still have enough of memory(around $200$ Gigabytes) and the solving functions are efficient enough.

\subsection{Relaxation matrix}\label{sec:implerelax}

There are some technical details concerning generating the relaxation matrix constraints. First we note that if we are considering loop equations up to $l=2\Lambda-3$, they contain quadratic terms only up to length $2\Lambda-4$. The operators in the bottom right corner of the relaxation matrix \(\mathcal{R}\) are neither constrained at all, nor contained in the objective function (cf~\eqref{relaxm}). So they are actually irrelevant. Formally these can be shown by Schur complement\footnote{Numerically there is no difference between strict inequalities and non-strict ones.}:
\begin{equation}
    \mathcal{R}=\begin{pmatrix}
    X_1 & B_1\\ B_1^{\mathrm{T}} & X_2
    \end{pmatrix}\succ 0 \Leftrightarrow X_1\succ 0\wedge X_2-B_1^{\mathrm{T}} X_1^{-1} B_1\succ 0.
\end{equation}
Here $X_1$ is the sub-matrix contain the product of $\langle \Tr\mathcal{O}_i\rangle \langle \Tr\mathcal{O}_j\rangle$, where the length of $\mathcal{O}_i$ plus $\mathcal{O}_j$ is less or equal than $\Lambda-2$. Since $X_2$ is not constrained at all, the second inequality can always be satisfied. In this way, the positivity condition of $\mathcal{R}$ is equivalent to the positivity of the $X_1$.

The matrix $X_1$ needs to be further reduced. Since we know that the operator expectations are linearly dependent on the loop equations, it is not surprising that the quadratic terms are linearly dependent. We must reduce this linear dependency, otherwise it will induce numerical instabilities for SDPA. To solve this potential problem, we consider the minor $X_{1\mathrm{red}}$ in the form:
\begin{equation}
    X_{1\mathrm{red}}=x_{1\mathrm{red}}x_{1\mathrm{red}}^{\mathrm{T}}.
\end{equation}
The vector \(x_{1\mathrm{re}}\) consists of operator expectations $\langle \Tr\mathcal{O}_i\rangle$, such that not only the lengths of $\mathcal{O}_i$ are equal or less than $\Lambda-2$ but also they are in the set $x_\mathrm{red}$ derived by solving the loop equations. Then we impose the positivity condition on the reduced relaxation matrix:
\begin{equation}
    \mathcal{R}_\mathrm{1red}=\begin{pmatrix}
1 & x_\mathrm{1red}^{\mathrm{T}}\\
x_\mathrm{1red} & X_\mathrm{1red}
\end{pmatrix}\succeq 0\,.
\end{equation}

\subsection{Feasibility}

In Section~\ref{sec:phase}, in the bisection process we need to test whether the SDP is feasible. However, SDPA doesn't have a built-in option to test the feasibility of a SDP. To deal with this difficulty, we introduce a slack variable \(\mu\) in addition to the original variables in the SDP, to transform the original problem~\eqref{relaxp} into the form:
\begin{equation}
\begin{aligned}
\min \quad\mu\\
\textrm{subject to} \quad & M_0+\sum_k M_k x_k+\mu I_1\succeq 0,\\&\mathrm{and }\quad\tr (X A_i) +b_i^{\mathrm{T}} x +a_i=0,\\&\mathrm{and }\,\begin{pmatrix}
1 & x^{\mathrm{T}}\\
x & X
\end{pmatrix}+\mu I_2\succeq 0.
\end{aligned}
\end{equation}
Here \(I_1\) and \(I_2\) are the identity matrices of the appropriate size. If the optimal value is negative, then the the original problem~\eqref{relaxp} is feasible, otherwise it is infeasible.

\subsection{Normalization}
When we are dealing with a large-scale SDP, a common problem is that our input data is badly-scaled. We usually need to adjust our normalization to make all the numbers in the optimization problem to be of similar order of magnitude. In our problem, we make the substitution:
\begin{equation}
    \mathcal{O}_l^{'}=g_s^{l/2}\mathcal{O}_l
\end{equation}
and then we bootstrap the primed operator instead of the original one. Here $l$ is the length of the operator. When $g$ and $h$ are not vanishing we put $g_s=\mathrm{min}\{\abs{g},2\abs{h}\}$. When it is vanishing, we take $g_s$ to be the absolute value of the non-vanishing coupling. 

This naive  choice to scale  is not guaranteed to be the best for the problem, though it appears to lead   to decent results for our range of cutoffs. It may be more advantageous  to scale this problem according to the asymptotic behavior of large length operator expectations, which would be interesting to study analytically.

\subsection{Precision of the solver}

Finally, we want to make some comments on the numerical precision. Most of the SDP solvers including SDPA are machine precision solvers. Numerical instabilities may happen when we are too ambitious about the precision of the results or the input data is badly scaled. In this situation, one could try arbitrary precision solvers like SDPB and SDPA-GMP, or other solvers in SDPA family called SDPA-DD and SDPA-QD, which is actually our recommendation. SDPA-DD and SDPA-QD is based on double-double and quad-double data type in QD library. From our tests, 32 digits or 64 digits are always enough for our purpose, and they are still efficient enough to solve the problem. 

\section{\texorpdfstring{$\Lambda=4$}{lambda=4} example}\label{example}

In this appendix, we demonstrate the numerical implementation of SDPA introduced in the Appendix~\ref{imple} on the simplest non-linear case $\Lambda=4$ of the model~\eqref{2MMcom21}, under the assumption of $\mathbb{Z}_2^3$ symmetry.

\subsection{Operators and Loop equations}

Before generating the loop equations, one of the preparatory work is to generate all the nonequivalent operators and possible quadratic terms up to the cutoff $\Lambda=4$. As already discussed in Section~\ref{QCSM}, we have only very few operators that are non-equivalent up to identifications due to the symmetries of the problem and of the class of solutions considered here\footnote{We slightly abuse the notations in this appendix: all  single trace operators in this section actually mean their expectation values. For example $\Tr A^2$ is actually $\langle \Tr A^2 \rangle$}.  There are 20 operators with the length smaller or equal to $8$:
\begin{equation}
\begin{aligned}
&\mathrm{Tr}A^2,\, \mathrm{Tr}A^4,\, \mathrm{Tr}A^2B^2,\, \mathrm{Tr}ABAB,\, \mathrm{Tr}A^6,\, \mathrm{Tr}A^4B^2,\, \mathrm{Tr}A^3BAB,\, \mathrm{Tr}A^2BA^2B,\, \mathrm{Tr}A^8,\,\\
&\mathrm{Tr}A^6B^2,\, \mathrm{Tr}A^5BAB,\, \mathrm{Tr}A^4BA^2B,\, \mathrm{Tr}A^4B^4,\, \mathrm{Tr}A^3BA^3B,\, \mathrm{Tr}A^3BAB^3,\, \mathrm{Tr}A^3B^2AB^2,\,\\ &\mathrm{Tr}A^2BABAB^2,\, \mathrm{Tr}A^2BAB^2AB,\, \mathrm{Tr}A^2B^2A^2B^2,\, \mathrm{Tr}ABABABAB\,.
\end{aligned}
\end{equation}
All other operators are identical to  those in this list or they  vanish under the $\mathbb{Z}_2^3$ symmetry assumption. There is only one quadratic term left under this assumption. For conciseness, we denote it by $\beta$ in this appendix:
\begin{equation}
    \beta=(\mathrm{Tr}A^2)^2=(\mathrm{Tr}B^2)^2=\mathrm{Tr}A^2\mathrm{Tr}B^2.
\end{equation}

To generate the loop equations, we simply apply the exhaustive method and then delete the duplicates. For general couplings $h$ and $g$, we have $14$ loop equations left:

\begin{equation}
\begin{medsize}
    \begin{array}{lcl}
    1 =  \mathrm{Tr}A^2 + g \mathrm{Tr}A^4 - h (-2 \mathrm{Tr}A^2B^2 + 2 \mathrm{Tr}ABAB)\\ 0 = -2 \mathrm{Tr}A^2 + \mathrm{Tr}A^4 - h (2 \mathrm{Tr}A^3BAB - 2 \mathrm{Tr}A^4B^2) + g \mathrm{Tr}A^6\\ 0 = -\mathrm{Tr}A^2 + \mathrm{Tr}A^2B^2 - h (-\mathrm{Tr}A^2BA^2B + 2 \mathrm{Tr}A^3BAB - \mathrm{Tr}A^4B^2) + g \mathrm{Tr}A^4B^2\\ 0 = -h (2 \mathrm{Tr}A^2BA^2B - 2 \mathrm{Tr}A^3BAB) + g \mathrm{Tr}A^3BAB + \mathrm{Tr}ABAB\\ \beta =  - 2 \mathrm{Tr}A^4 + \mathrm{Tr}A^6 - h (2 \mathrm{Tr}A^5BAB - 2 \mathrm{Tr}A^6B^2) + g \mathrm{Tr}A^8\\ \beta =  - \mathrm{Tr}A^2B^2 + \mathrm{Tr}A^4B^2 - h (-\mathrm{Tr}A^3B^2AB^2 + 2 \mathrm{Tr}A^3BAB^3 - \mathrm{Tr}A^4B^4) + g \mathrm{Tr}A^6B^2\\ 0 = -2 \mathrm{Tr}A^2B^2 - h (-\mathrm{Tr}A^2B^2A^2B^2 + 2 \mathrm{Tr}A^2BABAB^2 - \mathrm{Tr}A^3B^2AB^2) + \mathrm{Tr}A^4B^2 + g \mathrm{Tr}A^6B^2\\ 0 = -\mathrm{Tr}A^4 + \mathrm{Tr}A^4B^2 + g \mathrm{Tr}A^4B^4 - h (-\mathrm{Tr}A^4BA^2B + 2 \mathrm{Tr}A^5BAB - \mathrm{Tr}A^6B^2)\\ 0 = \mathrm{Tr}A^3BAB - h (2 \mathrm{Tr}A^2BAB^2AB - \mathrm{Tr}A^2BABAB^2 - \mathrm{Tr}A^3BAB^3) + g \mathrm{Tr}A^5BAB - \mathrm{Tr}ABAB\\ 0 = \mathrm{Tr}A^3BAB + g \mathrm{Tr}A^5BAB - 2 \mathrm{Tr}ABAB - h (-2 \mathrm{Tr}A^2BABAB^2 + 2 \mathrm{Tr}ABABABAB)\\ 0 = \mathrm{Tr}A^3BAB + g \mathrm{Tr}A^3BAB^3 - h (-\mathrm{Tr}A^3BA^3B + 2 \mathrm{Tr}A^4BA^2B - \mathrm{Tr}A^5BAB)\\ 0 = g \mathrm{Tr}A^3BA^3B + \mathrm{Tr}A^3BAB - h (2 \mathrm{Tr}A^3B^2AB^2 - 2 \mathrm{Tr}A^3BAB^3)\\ 0 = -\mathrm{Tr}A^2B^2 + \mathrm{Tr}A^2BA^2B - h (-\mathrm{Tr}A^2BAB^2AB + 2 \mathrm{Tr}A^2BABAB^2 - \mathrm{Tr}A^3B^2AB^2) + g \mathrm{Tr}A^4BA^2B\\ \beta =  \mathrm{Tr}A^2BA^2B + g \mathrm{Tr}A^3B^2AB^2 - h (2 \mathrm{Tr}A^3BA^3B - 2 \mathrm{Tr}A^4BA^2B).
    \end{array}
    \end{medsize}
\end{equation}
This is a system  of $14$ linear  equations for $21$ variables. For generic values of $h$ and $g$, like the one we often chose in this paper $h=g=1$, they are all linearly independent. So we can express $14$ variables including $\beta$ through seven variables of shortest lengths. These seven variables form the subset $x_\mathrm{red}$ introduced in Appendix~\ref{sec:implerelax}. For $h=g=1$, we can take it  as:
\begin{equation}
    x_\mathrm{red}=(\mathrm{Tr}A^2,\, \mathrm{Tr}A^4,\, \mathrm{Tr}A^2B^2,\, \mathrm{Tr}A^6,\, \mathrm{Tr}A^8,\, \mathrm{Tr}A^6B^2,\, \mathrm{Tr}A^5BAB)^{\mathrm{T}}.
\end{equation}
The other operators, including $\beta$, can be expressed as linear combinations of these variables:
\begin{equation}\label{rule}
    \begin{medsize}
        \begin{array}{lcl} \mathrm{Tr}ABAB=\frac{1}{2} \mathrm{Tr}A^2+\frac{1}{2} \mathrm{Tr}A^4+\mathrm{Tr}A^2B^2-\frac{1}{2}\\\mathrm{Tr}A^4B^2=\frac{1}{6} \mathrm{Tr}A^2-\mathrm{Tr}A^2B^2+\frac{1}{6} \mathrm{Tr}A^6+\frac{1}{6}\\\mathrm{Tr}A^3BAB=-\frac{5}{6} \mathrm{Tr}A^2+\frac{1}{2} \mathrm{Tr}A^4-\mathrm{Tr}A^2B^2+\frac{2}{3} \mathrm{Tr}A^6+\frac{1}{6}\\\mathrm{Tr}A^2BA^2B=-\mathrm{Tr}A^2+\mathrm{Tr}A^4-\mathrm{Tr}A^2B^2+\mathrm{Tr}A^6\\\mathrm{Tr}A^4BA^2B=-\frac{8}{3} \mathrm{Tr}A^2+9 \mathrm{Tr}A^4-\frac{14}{3} \mathrm{Tr}A^2B^2-\frac{1}{3} \mathrm{Tr}A^6-\frac{8}{3} \mathrm{Tr}A^8-\frac{16}{3} \mathrm{Tr}A^6B^2+\frac{28}{3} \mathrm{Tr}A^5BAB+\frac{1}{3}\\\mathrm{Tr}A^4B^4=\frac{5}{2} \mathrm{Tr}A^2-8 \mathrm{Tr}A^4+\frac{17}{3} \mathrm{Tr}A^2B^2+\frac{1}{6} \mathrm{Tr}A^6+\frac{8}{3} \mathrm{Tr}A^8+\frac{13}{3} \mathrm{Tr}A^6B^2-\frac{22}{3} \mathrm{Tr}A^5BAB-\frac{1}{2}\\\mathrm{Tr}A^3BA^3B=-\frac{9}{2} \mathrm{Tr}A^2+\frac{31}{2} \mathrm{Tr}A^4-\frac{23}{3} \mathrm{Tr}A^2B^2-\frac{2}{3} \mathrm{Tr}A^6-\frac{14}{3} \mathrm{Tr}A^8-\frac{28}{3} \mathrm{Tr}A^6B^2+\frac{46}{3} \mathrm{Tr}A^5BAB+\frac{1}{2}\\\mathrm{Tr}A^3BAB^3=2 \mathrm{Tr}A^4-\frac{2}{3} \mathrm{Tr}A^2B^2-\frac{2}{3} \mathrm{Tr}A^6-\frac{2}{3} \mathrm{Tr}A^8-\frac{4}{3} \mathrm{Tr}A^6B^2+\frac{7}{3} \mathrm{Tr}A^5BAB\\\mathrm{Tr}A^3B^2AB^2=-\frac{8}{3} \mathrm{Tr}A^2+10 \mathrm{Tr}A^4-5 \mathrm{Tr}A^2B^2-\frac{2}{3} \mathrm{Tr}A^6-3 \mathrm{Tr}A^8-6 \mathrm{Tr}A^6B^2+10 \mathrm{Tr}A^5BAB+\frac{1}{3}\\\mathrm{Tr}A^2BABAB^2=-\frac{14}{3} \mathrm{Tr}A^2+14 \mathrm{Tr}A^4-\frac{26}{3} \mathrm{Tr}A^2B^2-4 \mathrm{Tr}A^8-8 \mathrm{Tr}A^6B^2+14 \mathrm{Tr}A^5BAB+\frac{2}{3}\\\mathrm{Tr}A^2BAB^2AB=-3 \mathrm{Tr}A^2+8 \mathrm{Tr}A^4-\frac{17}{3} \mathrm{Tr}A^2B^2-\frac{7}{3} \mathrm{Tr}A^8-\frac{14}{3} \mathrm{Tr}A^6B^2+\frac{26}{3} \mathrm{Tr}A^5BAB+\frac{2}{3}\\\mathrm{Tr}A^2B^2A^2B^2=-\frac{41}{6} \mathrm{Tr}A^2+18 \mathrm{Tr}A^4-\frac{28}{3} \mathrm{Tr}A^2B^2+\frac{1}{2} \mathrm{Tr}A^6-5 \mathrm{Tr}A^8-11 \mathrm{Tr}A^6B^2+18 \mathrm{Tr}A^5BAB+\frac{5}{6}\\\mathrm{Tr}ABABABAB=-\frac{67}{12} \mathrm{Tr}A^2+\frac{55}{4} \mathrm{Tr}A^4-\frac{61}{6} \mathrm{Tr}A^2B^2+\frac{1}{3} \mathrm{Tr}A^6-4 \mathrm{Tr}A^8-8 \mathrm{Tr}A^6B^2+\frac{29}{2} \mathrm{Tr}A^5BAB+\frac{5}{4}\\\beta=-2 \mathrm{Tr}A^4+\mathrm{Tr}A^6+\mathrm{Tr}A^8+2 \mathrm{Tr}A^6B^2-2 \mathrm{Tr}A^5BAB.
        \end{array}
    \end{medsize}
\end{equation}

\subsection{Correlation matrix and relaxation matrix}

As we discussed in Section~\ref{QCSM}, under the $\mathbb{Z}_2^3$ symmetry our correlation matrix decouples into a block-diagonal matrix with three blocks. They are, respectively, the inner product\footnote{Here inner product of \(\mathcal{O}_1\) and \(\mathcal{O}_2\) is defined to be \(\langle\Tr \mathcal{O}_1^\dagger \mathcal{O}_2\rangle\).} matrix of even-even words:
\begin{equation}
    I, AA, BB, AAAA, AABB, ABAB, ABBA, BAAB, BABA, BBAA, BBBB
\end{equation}
odd-odd words:
\begin{equation}
    AB, BA, AAAB, AABA, ABAA, ABBB, BAAA, BABB, BBAB, BBBA
\end{equation}
and even-odd words:
\begin{equation}
    B, AAB, ABA, BAA, BBB\,.
\end{equation}
For example, the block for the even-odd words reads:
\begin{equation}
    \left(
\begin{array}{ccccc}
 \mathrm{Tr}A^2 & \mathrm{Tr}A^4 & \mathrm{Tr}A^2B^2 & \mathrm{Tr}ABAB & \mathrm{Tr}A^2B^2 \\
 \mathrm{Tr}A^4 & \mathrm{Tr}A^6 & \mathrm{Tr}A^4B^2 & \mathrm{Tr}A^3BAB & \mathrm{Tr}A^4B^2 \\
 \mathrm{Tr}A^2B^2 & \mathrm{Tr}A^4B^2 & \mathrm{Tr}A^4B^2 & \mathrm{Tr}A^3BAB & \mathrm{Tr}A^2BA^2B \\
 \mathrm{Tr}ABAB & \mathrm{Tr}A^3BAB & \mathrm{Tr}A^3BAB & \mathrm{Tr}A^2BA^2B & \mathrm{Tr}A^3BAB \\
 \mathrm{Tr}A^2B^2 & \mathrm{Tr}A^4B^2 & \mathrm{Tr}A^2BA^2B & \mathrm{Tr}A^3BAB & \mathrm{Tr}A^4B^2 \\
\end{array}
\right)
\end{equation}

It is easy to construct the relaxation matrix  for this example using the  explanations of Appendix~\ref{sec:implerelax}. It is: 
\begin{equation}
    \begin{pmatrix}
    1& \Tr A^2\\
    \Tr A^2 & \beta
    \end{pmatrix}\succeq 0.
\end{equation}
We note that in our current setting the vector \(x_\mathrm{1red}\) is a single component vector with the component \(\Tr A^2\).

To turn the original problem into the form~\eqref{SDPp}, we substitute the ``solution'' of the loop equations~\eqref{rule} into the correlation matrix and the relaxation matrix. As for the objective function, we choose it to  minimize $\Tr A^2$ and $-\Tr A^2$ to find the minimal and the maximal value of $\Tr A^2$, respectively. In this way, we can get the allowed region of $\Tr A^2$. At the next step we generate the input file for SDPA and solve it. With the appropriate setup, the entire time consumed for generating the input and solving it should last less than 0.1s CPU time. The result for \(\Lambda=4\) bootstrap is then:
\begin{equation}
    0.393566\leq\Tr A^2\leq0.431148\,.
\end{equation}

\section{Structure of loop equations and solvable 2-matrix model}\label{sec:structure}

Generalizing the results derived in Section~\ref{sec:Hamburger} to multi-matrix model is far from straightforward. In one-matrix model, all the information contained in the loop equations and the positivity conditions can be encoded compactly into the resolvent function. On the contrary, for the multi-matrix model, as the loop equations and the correlation matrix both get much more involved, we don't expect to have such an analytic function enclosing the information of all the moments. Due to this complication,  the bootstrap problem for a general multi-matrix model is generally not exactly solvable. In this appendix, we will discuss the nature of these complications in the structure of the loop equations and remind, from this point of view,  an old result for the simplest 2-mattrix model with $\tr(AB)$ interaction when the loop equations  greatly simplify~\cite{Staudacher:1993xy,Eynard:2002kg}.

\subsection{Base moments}\label{base}
As demonstrated in Section~\ref{sec:review} by the loop equations, all higher moments of one-matrix model are fully determined by a fixed number of lower moments (which we will call the \emph{base moments}). But this is not generally true in multi-matrix model: namely, the number of such base moments generally grows with the increase of the cutoff $\Lambda$. Let us take the model d~\eqref{2MMcom21} as an example. The following results can be observed from our numerical investigation~\footnote{  at some finite but high cutoffs,  but we strongly believe that it holds also for arbitrarily high cutoff.}:
\begin{enumerate}
    \item In the simplest case \(h=0\) when the model   effectively factorizes into two decoupled one-matrix models, all the moments can be expressed by a polynomial of \(t_2=\langle \Tr A^2\rangle\) and \(g\). The number of base moments is \(1\) here.\footnote{We assumed the global symmetry.}
    \item For \(g=0\) which is also solvable, all the moments are fixed by \(t_{2k}=\langle \Tr A^{2k}\rangle\), \(k=1,2,...\). This is very different from the one-matrix model since here we have to specify the value of infinite number of moments to determine the value of the remaining moments. If we set a finite cutoff \(2\Lambda\) to the length of the moments, we have a set of truncated set of base moments of the size \(\Lambda\). 
    \item For the general parameters \((g\neq 0,\, h\neq 0)\), there are much more base moments than the case \(g=0\).
\end{enumerate}
The intuition here is that, for a given multi-matrix model, the number of the base moments is negatively related with the solvability of the multi-matrix model. For a given cutoff, the number of the base moments are relatively easy to calculate. So one of the possible application of this intuitive observation is that one can use the number of the base moments for a finite but high cutoff to predict the solvability of the model. 
\subsection{Closed subset of loop equations}\label{solve}

Sometimes,  the loop equations can be closed on a proper, much reduced subset of all moments. This usually leads to a great simplification of the system of loop equations and potentially makes the bootstrap problem for the multi-matrix model exactly solvable. we demonstrate this on the simplest solvable 2-matrix   model, with a long history of stydy and applications~\cite{Itzykson:1979fi,Mehta:1981xt,Kazakov:1986hy} and used in~\cite{2020JHEP...06..090L} to demonstrate the matrix bootstrap:
\begin{equation}\label{eq:2MMsolv}
    Z=\lim_{N\rightarrow \infty}\int d^{N^2}A\,d^{N^2}B\,\e^{-N\tr\left( -AB+V(A)+V(B)\right)},\qquad V(x)=g_2 x^2/2+g_3 x^3/3.
\end{equation}
We notice that the following subset of loop equations:
\begin{equation}\label{eq:isingloop}
      \begin{split}
      -g_3 t_{n+1}&=g_2 t_n-t_{n-1,1}-\sum_{j=0}^{n-2} t_j t_{n-2-j},\\
      -g_3 t_{n+1,1}&=g_2 t_{n,1}-t_{n-1,2}-\sum_{j=0}^{n-2} t_j t_{n-2-j,1},\\
      -g_3 t_{n-1,2}&=g_2 t_{n-1,1}-t_n,
      \end{split}
\end{equation}      
are closed among the the operators \(t_n,\, t_{n,1},\,t_{n,2}\), here:
\begin{equation}
t_n=\langle \mathrm{Tr} A^{n} \rangle,\, t_{n,m}=\langle \Tr A^n B^m\rangle.
\end{equation}

Summing over the equations by the way in Section~\ref{sec:revisit}, we can get the \textit{Master loop equation} for model~\eqref{eq:2MMsolv}~\cite{Eynard:2002kg}:
\begin{equation}
    (Y(z)-V'(z))(z-V'(Y(z)))+P(Y(z),z)=0,
\end{equation}
where
\begin{equation}
\begin{split}
    &P(x,y)=-\langle \mathrm{Tr}\frac{V'(x)-V'(A)}{x-A}\frac{V'(y)-V'(B)}{y-B} \rangle+1,\\
    &G(z)=\langle \mathrm{Tr} \frac{1}{z-A} \rangle=\sum_{i=0}^{\infty}z^{-i-1} t_i,\\
    &Y(z)=V^{'}(z)-G(z).
\end{split}
\end{equation}

Since this is a closed subset of loop equations, we assume that the sub-correlation matrix defined by \(T_{i,j}=t_{i+j-2}\) is positive semi-definite, which is  the positivity condition for the minor of the whole correlation matrix consisting of the elements \(t_n\). This brings us back to the one-matrix type bootstrap problem considered in Section~\ref{sec:ana}. Here the positivity condition is equivalent to that the eigenvalue distribution corresponding to \(G(z)\) is real and positive, or \(Y(z)\) has a negative cut. In principle, this problem is analytically solvable, the complication compared with one-matrix model is that we have a cubic equation instead of a quadratic one (or an equation of $n$th degree for the potentials of order $n$). The solution of these loop equations has been found in  \cite{Kazakov:2002yh} in terms of an algebraic curve depending on the base moments~\cite{Kazakov:2004du}.

\acknowledgments

We thank M.Paulos for a useful discussion. This work benifits a lot from the help from Walter Landry on SDPB in the early stage.


\bibliography{MatrixBootstrap}

\providecommand{\href}[2]{#2}\begingroup\raggedright\begin{thebibliography}{10}

\bibitem{tHooft:1973alw}
G.~'t~Hooft, \emph{{A Planar Diagram Theory for Strong Interactions}},
  \href{https://doi.org/10.1016/0550-3213(74)90154-0}{\emph{Nucl. Phys. B}
  {\bfseries 72} (1974) 461}.

\bibitem{Migdal:1983qrz}
A.A.~Migdal, \emph{{Loop Equations and 1/N Expansion}},
  \href{https://doi.org/10.1016/0370-1573(83)90076-5}{\emph{Phys. Rept.}
  {\bfseries 102} (1983) 199}.

\bibitem{David:1984tx}
F.~David, \emph{{Planar Diagrams, Two-Dimensional Lattice Gravity and Surface
  Models}}, \href{https://doi.org/10.1016/0550-3213(85)90335-9}{\emph{Nucl.
  Phys. B} {\bfseries 257} (1985) 45}.

\bibitem{Kazakov:1985ea}
V.A.~Kazakov, A.A.~Migdal and I.K.~Kostov, \emph{{Critical Properties of
  Randomly Triangulated Planar Random Surfaces}},
  \href{https://doi.org/10.1016/0370-2693(85)90669-0}{\emph{Phys. Lett. B}
  {\bfseries 157} (1985) 295}.

\bibitem{Kazakov:1985ds}
V.A.~Kazakov, \emph{{Bilocal Regularization of Models of Random Surfaces}},
  \href{https://doi.org/10.1016/0370-2693(85)91011-1}{\emph{Phys. Lett. B}
  {\bfseries 150} (1985) 282}.

\bibitem{Kazakov:1987qg}
V.A.~Kazakov, \emph{{EXACTLY SOLVABLE POTTS MODELS, BOND AND TREE LIKE
  PERCOLATION ON DYNAMICAL (RANDOM) PLANAR}},  in \emph{{International
  Symposium on Field Theory of the Lattice}}, 12, 1987.

\bibitem{PhysRevLett.52.1}
O.~Bohigas, M.J.~Giannoni and C.~Schmit, \emph{Characterization of chaotic
  quantum spectra and universality of level fluctuation laws},
  \href{https://doi.org/10.1103/PhysRevLett.52.1}{\emph{Phys. Rev. Lett.}
  {\bfseries 52} (1984) 1}.

\bibitem{Dijkgraaf:2002fc}
R.~Dijkgraaf and C.~Vafa, \emph{{Matrix models, topological strings, and
  supersymmetric gauge theories}},
  \href{https://doi.org/10.1016/S0550-3213(02)00766-6}{\emph{Nucl. Phys. B}
  {\bfseries 644} (2002) 3}
  [\href{https://arxiv.org/abs/hep-th/0206255}{{\ttfamily hep-th/0206255}}].

\bibitem{Dijkgraaf:2002pp}
R.~Dijkgraaf, S.~Gukov, V.A.~Kazakov and C.~Vafa, \emph{{Perturbative analysis
  of gauged matrix models}},
  \href{https://doi.org/10.1103/PhysRevD.68.045007}{\emph{Phys. Rev. D}
  {\bfseries 68} (2003) 045007}
  [\href{https://arxiv.org/abs/hep-th/0210238}{{\ttfamily hep-th/0210238}}].

\bibitem{Eynard:2007kz}
B.~Eynard and N.~Orantin, \emph{{Invariants of algebraic curves and topological
  expansion}}, \href{https://doi.org/10.4310/CNTP.2007.v1.n2.a4}{\emph{Commun.
  Num. Theor. Phys.} {\bfseries 1} (2007) 347}
  [\href{https://arxiv.org/abs/math-ph/0702045}{{\ttfamily math-ph/0702045}}].

\bibitem{Kontsevich:1992ti}
M.~Kontsevich, \emph{{Intersection theory on the moduli space of curves and the
  matrix Airy function}},
  \href{https://doi.org/10.1007/BF02099526}{\emph{Commun. Math. Phys.}
  {\bfseries 147} (1992) 1}.

\bibitem{montgomery1973pair}
H.L.~Montgomery, \emph{The pair correlation of zeros of the zeta function},  in
  \emph{Proc. Symp. Pure Math}, vol.~24, pp.~181--193, 1973.

\bibitem{Eguchi:1982nm}
T.~Eguchi and H.~Kawai, \emph{{Reduction of Dynamical Degrees of Freedom in the
  Large N Gauge Theory}},
  \href{https://doi.org/10.1103/PhysRevLett.48.1063}{\emph{Phys. Rev. Lett.}
  {\bfseries 48} (1982) 1063}.

\bibitem{Brezin:1977sv}
E.~Brezin, C.~Itzykson, G.~Parisi and J.B.~Zuber, \emph{{Planar Diagrams}},
  \href{https://doi.org/10.1007/BF01614153}{\emph{Commun. Math. Phys.}
  {\bfseries 59} (1978) 35}.

\bibitem{Itzykson:1979fi}
C.~Itzykson and J.B.~Zuber, \emph{{The Planar Approximation. 2.}},
  \href{https://doi.org/10.1063/1.524438}{\emph{J. Math. Phys.} {\bfseries 21}
  (1980) 411}.

\bibitem{Mehta:1981xt}
M.L.~Mehta, \emph{{A Method of Integration Over Matrix Variables}},
  \href{https://doi.org/10.1007/BF01208498}{\emph{Commun. Math. Phys.}
  {\bfseries 79} (1981) 327}.

\bibitem{Kazakov:1986hy}
V.A.~Kazakov, \emph{{Exact Solution of the Ising Model on a Random
  Two-dimensional Lattice}}, {\emph{JETP Lett.} {\bfseries 44} (1986) 133}.

\bibitem{Boulatov:1986sb}
D.V.~Boulatov and V.A.~Kazakov, \emph{{The Ising Model on Random Planar
  Lattice: The Structure of Phase Transition and the Exact Critical
  Exponents}}, \href{https://doi.org/10.1016/0370-2693(87)90312-1}{\emph{Phys.
  Lett. B} {\bfseries 186} (1987) 379}.

\bibitem{Kostov:1988fy}
I.K.~Kostov, \emph{{O($n$) Vector Model on a Planar Random Lattice: Spectrum of
  Anomalous Dimensions}},
  \href{https://doi.org/10.1142/S0217732389000289}{\emph{Mod. Phys. Lett. A}
  {\bfseries 4} (1989) 217}.

\bibitem{Daul:1994qy}
J.-M.~Daul, \emph{{Q states Potts model on a random planar lattice}},
  \href{https://arxiv.org/abs/hep-th/9502014}{{\ttfamily hep-th/9502014}}.

\bibitem{Kazakov:1988ch}
V.A.~Kazakov and A.A.~Migdal, \emph{{Recent Progress in the Theory of
  Noncritical Strings}},
  \href{https://doi.org/10.1016/0550-3213(88)90146-0}{\emph{Nucl. Phys. B}
  {\bfseries 311} (1988) 171}.

\bibitem{Kazakov:2000aq}
V.A.~Kazakov, \emph{{Solvable matrix models}},  2, 2000
  [\href{https://arxiv.org/abs/hep-th/0003064}{{\ttfamily hep-th/0003064}}].

\bibitem{Jevicki:1982jj}
A.~Jevicki, O.~Karim, J.P.~Rodrigues and H.~Levine, \emph{{Loop Space
  Hamiltonians and Numerical Methods for Large $N$ Gauge Theories}},
  \href{https://doi.org/10.1016/0550-3213(83)90180-3}{\emph{Nucl. Phys. B}
  {\bfseries 213} (1983) 169}.

\bibitem{Jevicki:1983wu}
A.~Jevicki, O.~Karim, J.P.~Rodrigues and H.~Levine, \emph{{Loop Space
  Hamiltonians and Numerical Methods for Large $N$ Gauge Theories. 2.}},
  \href{https://doi.org/10.1016/0550-3213(84)90215-3}{\emph{Nucl. Phys. B}
  {\bfseries 230} (1984) 299}.

\bibitem{Rodrigues:1985aq}
J.P.~Rodrigues, \emph{{Numerical Solution of Lattice Schwinger-dyson Equations
  in the Large $N$ Limit}},
  \href{https://doi.org/10.1016/0550-3213(85)90077-X}{\emph{Nucl. Phys. B}
  {\bfseries 260} (1985) 350}.

\bibitem{2008JHEP...12..031R}
R.~{Rattazzi}, V.S.~{Rychkov}, E.~{Tonni} and A.~{Vichi}, \emph{{Bounding
  scalar operator dimensions in 4D CFT}},
  \href{https://doi.org/10.1088/1126-6708/2008/12/031}{\emph{Journal of High
  Energy Physics} {\bfseries 2008} (2008) 031}
  [\href{https://arxiv.org/abs/0807.0004}{{\ttfamily 0807.0004}}].

\bibitem{2017JHEP...03..086S}
D.~{Simmons-Duffin}, \emph{{The lightcone bootstrap and the spectrum of the 3d
  Ising CFT}}, \href{https://doi.org/10.1007/JHEP03(2017)086}{\emph{Journal of
  High Energy Physics} {\bfseries 2017} (2017) 86}
  [\href{https://arxiv.org/abs/1612.08471}{{\ttfamily 1612.08471}}].

\bibitem{2020JHEP...06..090L}
H.W.~{Lin}, \emph{{Bootstraps to strings: solving random matrix models with
  positivite}}, \href{https://doi.org/10.1007/JHEP06(2020)090}{\emph{Journal of
  High Energy Physics} {\bfseries 2020} (2020) 90}
  [\href{https://arxiv.org/abs/2002.08387}{{\ttfamily 2002.08387}}].

\bibitem{2020PhRvL.125d1601H}
X.~{Han}, S.A.~{Hartnoll} and J.~{Kruthoff}, \emph{{Bootstrapping Matrix
  Quantum Mechanics}},
  \href{https://doi.org/10.1103/PhysRevLett.125.041601}{\emph{\prl} {\bfseries
  125} (2020) 041601} [\href{https://arxiv.org/abs/2004.10212}{{\ttfamily
  2004.10212}}].

\bibitem{2017NuPhB.921..702A}
P.D.~{Anderson} and M.~{Kruczenski}, \emph{{Loop equations and bootstrap
  methods in the lattice}},
  \href{https://doi.org/10.1016/j.nuclphysb.2017.06.009}{\emph{Nuclear Physics
  B} {\bfseries 921} (2017) 702}
  [\href{https://arxiv.org/abs/1612.08140}{{\ttfamily 1612.08140}}].

\bibitem{Makeenko:1979pb}
Y.M.~Makeenko and A.A.~Migdal, \emph{{Exact Equation for the Loop Average in
  Multicolor QCD}},
  \href{https://doi.org/10.1016/0370-2693(79)90131-X}{\emph{Phys. Lett. B}
  {\bfseries 88} (1979) 135}.

\bibitem{2017JHEP...11..143P}
M.F.~{Paulos}, J.~{Penedones}, J.~{Toledo}, B.C.~{van Rees} and P.~{Vieira},
  \emph{{The S-matrix bootstrap II: two dimensional amplitudes}},
  \href{https://doi.org/10.1007/JHEP11(2017)143}{\emph{Journal of High Energy
  Physics} {\bfseries 2017} (2017) 143}
  [\href{https://arxiv.org/abs/1607.06110}{{\ttfamily 1607.06110}}].

\bibitem{2019JHEP...02..162M}
D.~{Maz{\'a}{\v{c}}} and M.F.~{Paulos}, \emph{{The analytic functional
  bootstrap. Part I: 1D CFTs and 2D S-matrices}},
  \href{https://doi.org/10.1007/JHEP02(2019)162}{\emph{Journal of High Energy
  Physics} {\bfseries 2019} (2019) 162}
  [\href{https://arxiv.org/abs/1803.10233}{{\ttfamily 1803.10233}}].

\bibitem{1999NuPhB.557..413K}
V.A.~{Kazakov}, I.K.~{Kostov} and N.~{Nekrasov}, \emph{{D-particles, matrix
  integrals and KP hierarchy}},
  \href{https://doi.org/10.1016/S0550-3213(99)00393-4}{\emph{Nuclear Physics B}
  {\bfseries 557} (1999) 413}
  [\href{https://arxiv.org/abs/hep-th/9810035}{{\ttfamily hep-th/9810035}}].

\bibitem{Kazakov:1998qw}
V.A.~Kazakov and P.~Zinn-Justin, \emph{{Two matrix model with ABAB
  interaction}},
  \href{https://doi.org/10.1016/S0550-3213(99)00015-2}{\emph{Nucl. Phys. B}
  {\bfseries 546} (1999) 647}
  [\href{https://arxiv.org/abs/hep-th/9808043}{{\ttfamily hep-th/9808043}}].

\bibitem{Kostov:1999qx}
I.K.~Kostov, \emph{{Exact solution of the six vertex model on a random
  lattice}}, \href{https://doi.org/10.1016/S0550-3213(00)00060-2}{\emph{Nucl.
  Phys. B} {\bfseries 575} (2000) 513}
  [\href{https://arxiv.org/abs/hep-th/9911023}{{\ttfamily hep-th/9911023}}].

\bibitem{Zinn-Justin:1999chi}
P.~Zinn-Justin, \emph{{The Six vertex model on random lattices}},
  \href{https://doi.org/10.1209/epl/i2000-00229-y}{\emph{Europhys. Lett.}
  {\bfseries 50} (2000) 15}
  [\href{https://arxiv.org/abs/cond-mat/9909250}{{\ttfamily
  cond-mat/9909250}}].

\bibitem{Jha:2021exo}
R.G.~Jha, \emph{{Introduction to Monte Carlo for Matrix Models}},
  \href{https://arxiv.org/abs/2111.02410}{{\ttfamily 2111.02410}}.

\bibitem{Eynard:2004mh}
B.~Eynard, \emph{{Topological expansion for the 1-Hermitian matrix model
  correlation functions}},
  \href{https://doi.org/10.1088/1126-6708/2004/11/031}{\emph{JHEP} {\bfseries
  11} (2004) 031} [\href{https://arxiv.org/abs/hep-th/0407261}{{\ttfamily
  hep-th/0407261}}].

\bibitem{DiFrancesco:1993cyw}
P.~Di~Francesco, P.H.~Ginsparg and J.~Zinn-Justin, \emph{{2-D Gravity and
  random matrices}},
  \href{https://doi.org/10.1016/0370-1573(94)00084-G}{\emph{Phys. Rept.}
  {\bfseries 254} (1995) 1}
  [\href{https://arxiv.org/abs/hep-th/9306153}{{\ttfamily hep-th/9306153}}].

\bibitem{reed1975ii}
M.~Reed and B.~Simon, \emph{II: Fourier Analysis, Self-Adjointness}, vol.~2,
  Elsevier (1975).

\bibitem{Kazakov:1989bc}
V.A.~Kazakov, \emph{{The Appearance of Matter Fields from Quantum Fluctuations
  of 2D Gravity}}, \href{https://doi.org/10.1142/S0217732389002392}{\emph{Mod.
  Phys. Lett. A} {\bfseries 4} (1989) 2125}.

\bibitem{Staudacher:1993xy}
M.~Staudacher, \emph{{Combinatorial solution of the two matrix model}},
  \href{https://doi.org/10.1016/0370-2693(93)91063-S}{\emph{Phys. Lett. B}
  {\bfseries 305} (1993) 332}
  [\href{https://arxiv.org/abs/hep-th/9301038}{{\ttfamily hep-th/9301038}}].

\bibitem{boyd_vandenberghe_2004}
S.~Boyd and L.~Vandenberghe, \emph{Convex Optimization}, Cambridge University
  Press (2004),
  \href{https://doi.org/10.1017/CBO9780511804441}{10.1017/CBO9780511804441}.

\bibitem{1982PhDT........32H}
J.R.~{Hoppe}, \emph{{Quantum Theory of a Massless Relativistic Surface and a
  Two-Dimensional Bound State Problem.}}, Ph.D. thesis, MASSACHUSETTS INSTITUTE
  OF TECHNOLOGY., Jan., 1982.

\bibitem{ElShowk:2012ht}
S.~El-Showk, M.F.~Paulos, D.~Poland, S.~Rychkov, D.~Simmons-Duffin and
  A.~Vichi, \emph{{Solving the 3D Ising Model with the Conformal Bootstrap}},
  \href{https://doi.org/10.1103/PhysRevD.86.025022}{\emph{Phys. Rev. D}
  {\bfseries 86} (2012) 025022}
  [\href{https://arxiv.org/abs/1203.6064}{{\ttfamily 1203.6064}}].

\bibitem{1982PhLB..108..407S}
Y.~{Shimamune}, \emph{{On the phase structure of large N matrix models and
  gauge models}},
  \href{https://doi.org/10.1016/0370-2693(82)91223-0}{\emph{Physics Letters B}
  {\bfseries 108} (1982) 407}.

\bibitem{Rattazzi:2008pe}
R.~Rattazzi, V.S.~Rychkov, E.~Tonni and A.~Vichi, \emph{{Bounding scalar
  operator dimensions in 4D CFT}},
  \href{https://doi.org/10.1088/1126-6708/2008/12/031}{\emph{JHEP} {\bfseries
  12} (2008) 031} [\href{https://arxiv.org/abs/0807.0004}{{\ttfamily
  0807.0004}}].

\bibitem{Teper:2008yi}
M.~Teper, \emph{{Large N}},
  \href{https://doi.org/10.22323/1.066.0022}{\emph{PoS} {\bfseries LATTICE2008}
  (2008) 022} [\href{https://arxiv.org/abs/0812.0085}{{\ttfamily 0812.0085}}].

\bibitem{Brezin:1980rk}
E.~Brezin and D.J.~Gross, \emph{{The External Field Problem in the Large N
  Limit of QCD}},
  \href{https://doi.org/10.1016/0370-2693(80)90562-6}{\emph{Phys. Lett. B}
  {\bfseries 97} (1980) 120}.

\bibitem{Kostov:1988pe}
I.K.~Kostov, \emph{{RANDOM SURFACES, SOLVABLE LATTICE MODELS AND DISCRETE
  QUANTUM GRAVITY IN TWO-DIMENSIONS}},  in \emph{{XIX International Seminar on
  Theoretical Physics: Nonperturbative Aspects of the Standard Model (GIFT
  Seminar)}}, 6, 1988.

\bibitem{eca870fa006646ae8352bf7a66ad292e}
M.~Yamashita, K.~Fujisawa, M.~Fukuda, K.~Kobayashi, K.~Nakata and M.~Nakata,
  \emph{Latest developments in the sdpa family for solving large-scale sdps},
  in \emph{International Series in Operations Research and Management Science},
  International Series in Operations Research and Management Science,
  pp.~687--713, Springer New York LLC (2012),
  \href{https://doi.org/10.1007/978-1-4614-0769-0_24}{DOI}.

\bibitem{2015JHEP...06..174S}
D.~{Simmons-Duffin}, \emph{{A semidefinite program solver for the conformal
  bootstrap}}, \href{https://doi.org/10.1007/JHEP06(2015)174}{\emph{Journal of
  High Energy Physics} {\bfseries 2015} (2015) 174}
  [\href{https://arxiv.org/abs/1502.02033}{{\ttfamily 1502.02033}}].

\bibitem{Eynard:2002kg}
B.~Eynard, \emph{{Large N expansion of the 2 matrix model}},
  \href{https://doi.org/10.1088/1126-6708/2003/01/051}{\emph{JHEP} {\bfseries
  01} (2003) 051} [\href{https://arxiv.org/abs/hep-th/0210047}{{\ttfamily
  hep-th/0210047}}].

\bibitem{Kazakov:2002yh}
V.A.~Kazakov and A.~Marshakov, \emph{{Complex curve of the two matrix model and
  its tau function}},
  \href{https://doi.org/10.1088/0305-4470/36/12/315}{\emph{J. Phys. A}
  {\bfseries 36} (2003) 3107}
  [\href{https://arxiv.org/abs/hep-th/0211236}{{\ttfamily hep-th/0211236}}].

\bibitem{Kazakov:2004du}
V.A.~Kazakov and I.K.~Kostov, \emph{{Instantons in noncritical strings from the
  two matrix model}},  in \emph{{From Fields to Strings: Circumnavigating
  Theoretical Physics: A Conference in Tribute to Ian Kogan}}, 3, 2004,
  \href{https://doi.org/10.1142/9789812775344_0045}{DOI}
  [\href{https://arxiv.org/abs/hep-th/0403152}{{\ttfamily hep-th/0403152}}].

\end{thebibliography}\endgroup
\bibliographystyle{jhep}
\end{document}